\documentclass[reprint,NumberedRefs]{JASAnew_mjb} 

\graphicspath{{figures/}}

\usepackage[hang]{subfigure}
\usepackage{booktabs}

\newcommand{\argmin}{\operatornamewithlimits{argmin}}

\begin{document}

\title[JASA]{Machine learning in acoustics: theory and applications}
\author{Michael J. Bianco} \email{mbianco@ucsd.edu}
\affiliation{Scripps Institution of Oceanography, University of California San Diego, La Jolla, CA 92093, USA}
\author{Peter Gerstoft}
\affiliation{Scripps Institution of Oceanography, University of California San Diego, La Jolla, CA 92093, USA}
\author{James Traer}
\affiliation{Department of Brain and Cognitive Sciences, Massachusetts Institute of Technology, Cambridge, MA 02139}
\author{Emma Ozanich}
\affiliation{Scripps Institution of Oceanography, University of California San Diego, La Jolla, CA 92093, USA}
\author{Marie A. Roch}
\affiliation{Department of Computer Science, San Diego State University, San Diego, CA 92182, USA}
\author{Sharon Gannot}
\affiliation{Faculty of Engineering, Bar-Ilan University, Ramat-Gan 5290002, Israel}
\author{Charles-Alban Deledalle}
\affiliation{Department of Electrical and Computer Engineering, University of California San Diego, La Jolla, CA 92093, USA}

\date{\today}

\begin{abstract}
Acoustic data provide scientific and engineering insights in fields ranging from biology and communications to ocean and Earth science. We survey the recent advances and transformative potential of machine learning (ML), including deep learning, in the field of acoustics. ML is a broad family of techniques, which are often based in statistics, for automatically detecting and utilizing patterns in data. Relative to conventional acoustics and signal processing, ML is data-driven. Given sufficient training data, ML can discover complex relationships between features and desired labels or actions, or between features themselves. With large volumes of training data, ML can discover models describing complex acoustic phenomena such as human speech and reverberation. ML in acoustics is rapidly developing with compelling results and significant future promise. We first introduce ML, then highlight ML developments in four acoustics research areas: source localization in speech processing, source localization in ocean acoustics, bioacoustics, and environmental sounds in everyday scenes.
\end{abstract}


\maketitle

\section{\label{sec:1} Introduction}
Acoustic data provide scientific and engineering insights in a very broad range of fields including machine interpretation of human speech\cite{Gannot2017,vincent2018audio} and animal vocalizations, \cite{mellinger2016signal} ocean source localization,\cite{gemba2019robust,niu2017} and imaging geophysical structures in the ocean.\cite{gerstoft1996,jensen2011computational} In all these fields, data analysis is complicated by a number of challenges, including data corruption, missing or sparse measurements, reverberation, and large data volumes. For example, multiple acoustic arrivals of a single event or utterance make source localization and speech interpretation a difficult task for machines.\cite{traer2016statistics,vincent2018audio} In many cases, such as acoustic tomography and bioacoustics, large volumes of data can be collected. The amount of human effort required to manually identify acoustic features and events rapidly becomes limiting as the size of the data sets increase. Further, patterns may exist in the data that are not easily recognized by human cognition.

Machine learning (ML) techniques\cite{jordan2015machine,lecun2015} have enabled broad advances in automated data processing and pattern recognition capabilities across many fields, including computer vision, image processing, speech processing, and (geo)physical science.\cite{kong2018machine,bergen2019machine} ML in acoustics is a rapidly developing field, with many compelling solutions to the aforementioned acoustics challenges. The potential impact of ML-based techniques in the field of acoustics, and the recent attention they have received, motivates this review.
 
Broadly defined, ML is a family of techniques for automatically detecting and utilizing patterns in data. In ML the patterns are used for example to estimate data labels based on measured attributes, such as the species of an animal or their location based on recordings from acoustic arrays. These measurements and their labels are often uncertain, thus statistical methods are often involved. In this way ML provides a means for machines to gain knowledge, or to `learn'.\cite{bishop2006,murphy2012} ML methods are often divided into two major categories: supervised and unsupervised learning. There is also a third category called reinforcement learning, though it is not discussed in this review. In supervised learning, the goal is to learn a predictive mapping from inputs to outputs given labeled input and output pairs. The labels can be categorical or real-valued scalars for classification and regression, respectively. In unsupervised learning, no labels are given, and the task is to discover interesting or useful structure within the data. An example of unsupervised learning is clustering analysis (e.g. K-means). Supervised and unsupervised modes can also be combined. Namely semi- and weakly supervised learning methods can be used when the labels only give partial or contextual information.

\begin{figure}[t]
\includegraphics[width=\reprintcolumnwidth]{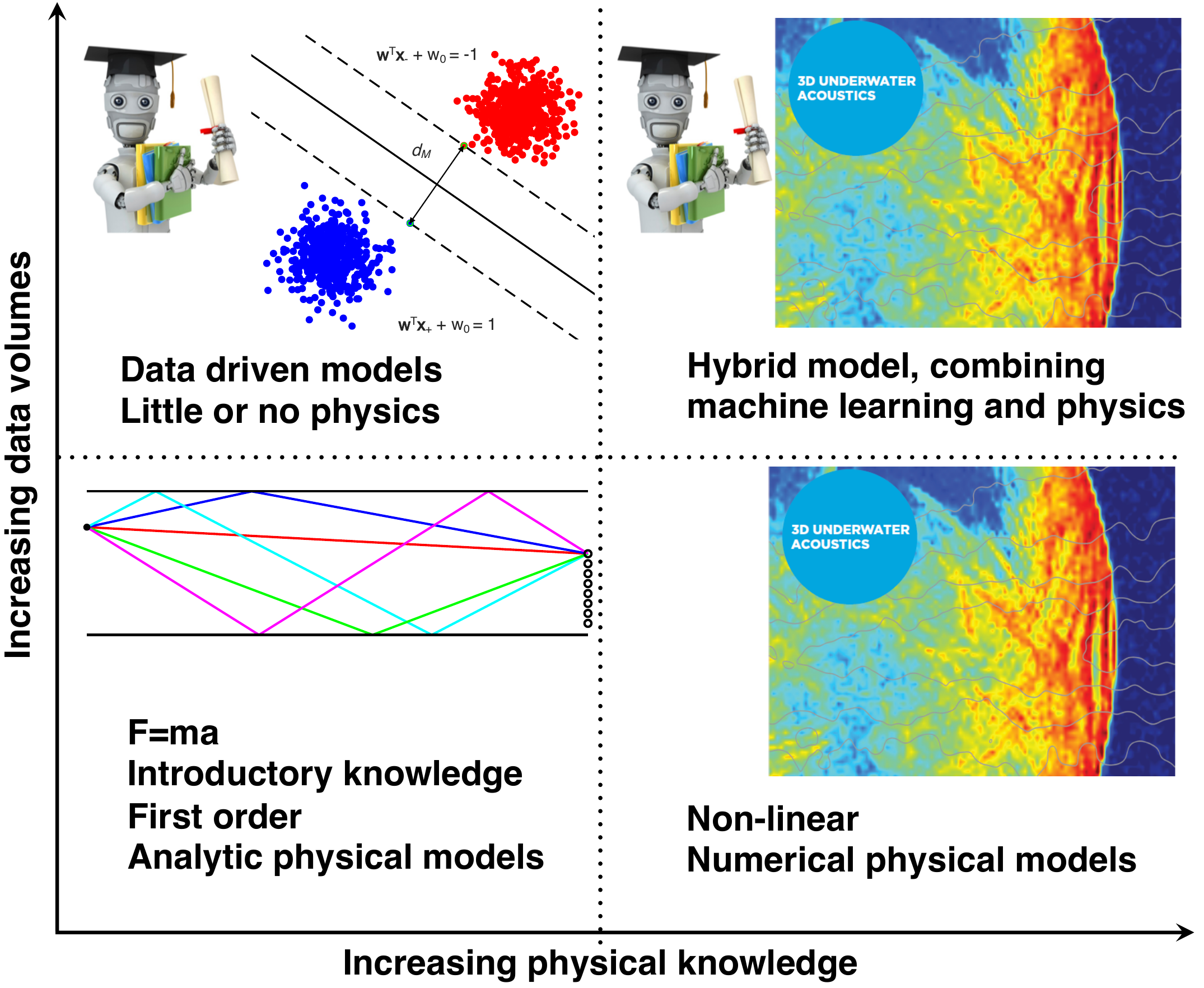}
\caption{\label{fig:FIG1}{(Color Online) Acoustic insight can be improved by leveraging the strengths of both physical and ML-based, data-driven models. Analytic physical models (lower left) give basic insights about physical systems. More sophisticated models, reliant on computational methods (lower right), can model more complex phenomena. Whereas physical models are reliant on rules, which are updated by physical evidence (data), ML is purely data-driven (upper left). By augmenting ML methods with physical models to obtain hybrid models (upper right), a synergy of the strengths of physical intuition and data-driven insights can be obtained.}}
\raggedright
\end{figure}

Research in acoustics has traditionally focused on developing high-level physical models and using these models for inferring properties of the environment and objects in the environment. The complexity of physical principal-based models is indicated by the x-axis in Fig.~\ref{fig:FIG1}. With increasing amounts of data, data-driven approaches have made enormous success. The volume of available data is indicated by the y-axis in Fig.~\ref{fig:FIG1}. It is expected that as more data becomes available in physical sciences that we will be able to better combine advanced acoustic models with ML.

In ML, it is preferred to learn representation models of the data, which provide useful patterns in the data for the ML task at hand, directly from the data rather than by using specific domain knowledge to engineer representations.\cite{bengio_representation_2013} ML can build upon physical models and domain knowledge, improving interpretation by finding representations (e.g. transformations of the features) that are `optimal' for a given task.\cite{goodfellow2016deep} Representations in ML are patterns the input {\it features}, which are particular attributes of the data. Features include spectral characteristics of human speech, or morphological features of a physical environment. Feature inputs to an ML pipeline can be raw measurements of a signal (data) or transformations of the data, e.g., obtained by the classic principal components analysis (PCA) approach. More flexible representations, including Gaussian mixture models (GMMs) are obtained using the expectation-maximization (EM). 
The fundamental concepts of ML are by no means new. For example, linear discriminant analysis (LDA), a fundamental classification model, was developed as early as the 1930's.\cite{fisher1936use} The K-means\cite{macqueen1967some} clustering algorithm and the perceptron\cite{rosenblatt1961principles} algorithm, which was a precursor to modern neural networks (NNs), were developed in the 1960s. Shortly after the perceptron algorithm was published, interest in NNs waned until 1980s when the backpropagation algorithm was developed\cite{rumelhart1986learning}. Currently we are in the midst of a `third-wave' of interest in ML and AI principles.\cite{goodfellow2016deep}

ML in acoustics has made significant progress in recent years. ML-based methods can provide superior performance relative to conventional signal processing methods. However, a clear limitation of ML-based methods is that they are data-driven and thus require large amounts of data for testing and training. Conventional methods  also have the benefit of being more interpretable than many ML models. Particularly in deep learning, ML models can be considered ‘black-boxes’ --- meaning that the intervening operations, between the inputs and outputs of the ML system, are not necessarily physically intuitive. Further, due to the {\it no free-lunch } theorem, models optimized for one task will likely perform worse at others. The intention of this review is to indicate that, despite these challenges, ML has considerable potential in acoustics.

This review focuses on the significant advances ML has already provided in the field of acoustics. We first introduce ML theory, including deep learning (DL). Then we discuss applications and advances of the theory in five acoustics research areas. In Secs.~\ref{sec:theory}--\ref{sec:unsup}, basic ML concepts are introduced, and some fundamental algorithms are developed. In Sec.~\ref{sec:deep_learning}, the field of DL is introduced, and applications to acoustics are discussed. Next, we discuss applications of ML theory to the following fields: speaker localization in reverberant environments (Sec.~\ref{sec:speech}), source localization in ocean acoustics (Sec.~\ref{sec:underwater_acoustics}), bioacoustics (Sec.~\ref{sec:bioacoustics}), and reverberation and environmental sounds in everyday scenes (Sec.~\ref{sec:reverb}). While the list of fields we cover and the treatment of ML theory is not exhaustive, we hope this article can serve as inspiration for future ML research in acoustics. For further reference, we refer readers to several excellent ML and signal processing textbooks, which are useful supplements to the material presented here: Refs.~\onlinecite{murphy2012,hastie2009,bishop2006,goodfellow2016deep,duda2012pattern,cohen2009speech,vincent2018audio,elad2010,mairal2014}

\section{Machine learning principles}\label{sec:theory}

ML is data-driven and can model potentially more complex patterns in the data than conventional methods. Classic signal processing techniques for modeling and predicting data are based on provable performance guarantees. These methods use simplifying assumptions, such as Gaussian independent and identically distributed (iid) variables, and 2\textsuperscript{nd} order statistics (covariance). However, ML methods, and recently DL methods in particular, have shown improved performance in a number of tasks compared to conventional methods.\cite{lecun2015} But, the increased flexibility of the ML models comes with certain difficulties.

Often the complexity of ML models and their training algorithms make guaranteeing their performance difficult and can hinder model interpretation. Further, ML models can require significant amounts of training data, though we note that `vast' quantities of training data are not required to take advantage of ML techniques. Due to the {\it no free lunch } theorem,\cite{wolpert1997no} models whose performance is maximized for one task will likely perform worse at others. Provided high-performance is desired only for a specific task, and there is enough training data, the benefits of ML may outweigh these issues.

\subsection{Inputs and outputs}

In acoustics and signal processing, measurement models explain sets of observations using a set of models. The model explaining the observations is typically called the ``forward'' model. To find the best model parameters, the forward model is ``inverted''. However, ML measurement models are articulated in terms of models relating inputs and outputs, both of which are observed,
\begin{equation}
    \mathbf{y}=f(\mathbf{x})+\epsilon.
    \label{eq:pred}
\end{equation}
Here, $\mathbf{x}\in\mathbb{R}^N$ are $N$ inputs and $\mathbf{y}\in\mathbb{R}^{P}$ are $P$ outputs to the model $f(\mathbf{x})$. $f(\mathbf{x})$ can be a linear or non-linear mapping from input to output. $\epsilon$ is the uncertainty in the estimate $f(\mathbf{x})$ which is due to model limitations and uncertainty in the measurements. Thus, the ML measurement model \eqref{eq:pred} has similarities with the ``inverse'' of the typical ``forward'' model.

Per \eqref{eq:pred}, $\mathbf{x}$ is a single observation of $N$ inputs, called features, from which we would like to estimate a single set of outputs $\mathbf{y}$. For example, in a simple feed-forward NN (Secs.~\ref{subsec:nn},\ref{sec:deep_learning}), the input layer ($\mathbf{x}$) has dimension $N$ and the output layer ($\mathbf{y}$) has dimension $P$. The NN then constitutes a non-linear function $f(\mathbf{x})$ relating the inputs to the outputs. To train the NN (learn $f(\mathbf{x})$) requires many samples of input/output pairs. We define $\mathbf{X}=\mathbf{[x}_1,\ldots,\mathbf{x}_M]^\mathrm{T}\in\mathbb{R}^{M\times N}$ and $\mathbf{Y}=[\mathbf{y}_1,\ldots,\mathbf{y}_M]\in\mathbb{R}^{P\times M}$ the corresponding $P$ outputs for $M$ samples of the input/output pairs. We here note that there are many ML scenarios where the number of input samples and output samples are different (e.g., recurrent NNs have more input samples than output samples).

The use of ML to obtain output $\mathbf{y}$ from features $\mathbf{x}$, as described above, is called {\it supervised learning} (Sec.~\ref{sec:sup_learn}). Often, we wish to discover interesting or useful patterns in the data without explicitly specifying output. This is called {\it unsupervised learning} (Sec.~\ref{sec:unsup}). In unsupervised learning, the goal is to learn interesting or useful patterns in the data. In many cases in unsupervised learning, the input and desired output is the features themselves.

\subsection{Supervised and unsupervised learning}\label{sec:ml_sub}
ML methods generally can be categorized as either supervised or unsupervised learning tasks. In supervised learning, the task is to learn a predictive mapping from inputs to outputs given labeled input and output pairs. Supervised learning is the most widely used ML category and includes familiar methods such as linear regression (a.k.a. ridge regression) and nearest-neighbor classifiers, as well as more sophisticated support vector machine (SVM) and neural network (NN) models--- sometimes referred to as artificial NNs, due to their weak relationship to neural structure in the biological brain. In unsupervised learning, no labels are given, and the task is to discover interesting or useful structure within the data. This has many useful applications, which include data visualization, exploratory data analysis, anomaly detection, and feature learning. Unsupervised methods such as  PCA,  K-means,\cite{macqueen1967some} and Gaussian mixture models (GMMs) have been used for decades. Newer methods include t-SNE,\cite{maaten2008visualizing} dictionary learning,\cite{tosic2011} and deep representations (e.g. autoencoders).\cite{goodfellow2016deep} An important point is that the results of unsupervised methods can be used either directly, such as for discovery of latent factors or data visualization, or as part of a supervised learning framework, where they supply transformed versions of the features to improve supervised learning performance.

\subsection{Generalization: train and test data}

Central to ML is the requirement that learned models must perform well on unobserved data as well as observed data. The ability of the model to predict unseen data well is called {\it generalization}. We first discuss relevant terminology, then discuss how generalization of an ML model can be assessed.

Often the term {\it complexity} is used to denote the level of sophistication of the data relationships or ML task. The ability of a particular ML model to well approximate data relationships (e.g. between features and labels) of a particular complexity is the {\it capacity}. These terms are not strictly defined, but efforts have been made to mathematically formalize these concepts. For example, the Vapnik-Chervonenkis (VC) dimension provides a means of quantifying model capacity in the case of binary classifiers.\cite{hastie2009} Data complexity can be interpreted as the number of dimensions in which useful relationships exist between features. Higher complexity implies higher-dimensional relationships. We note that the capacity of the ML model can be limited by the quantity of training data.

\begin{figure}
\includegraphics[width=\reprintcolumnwidth]{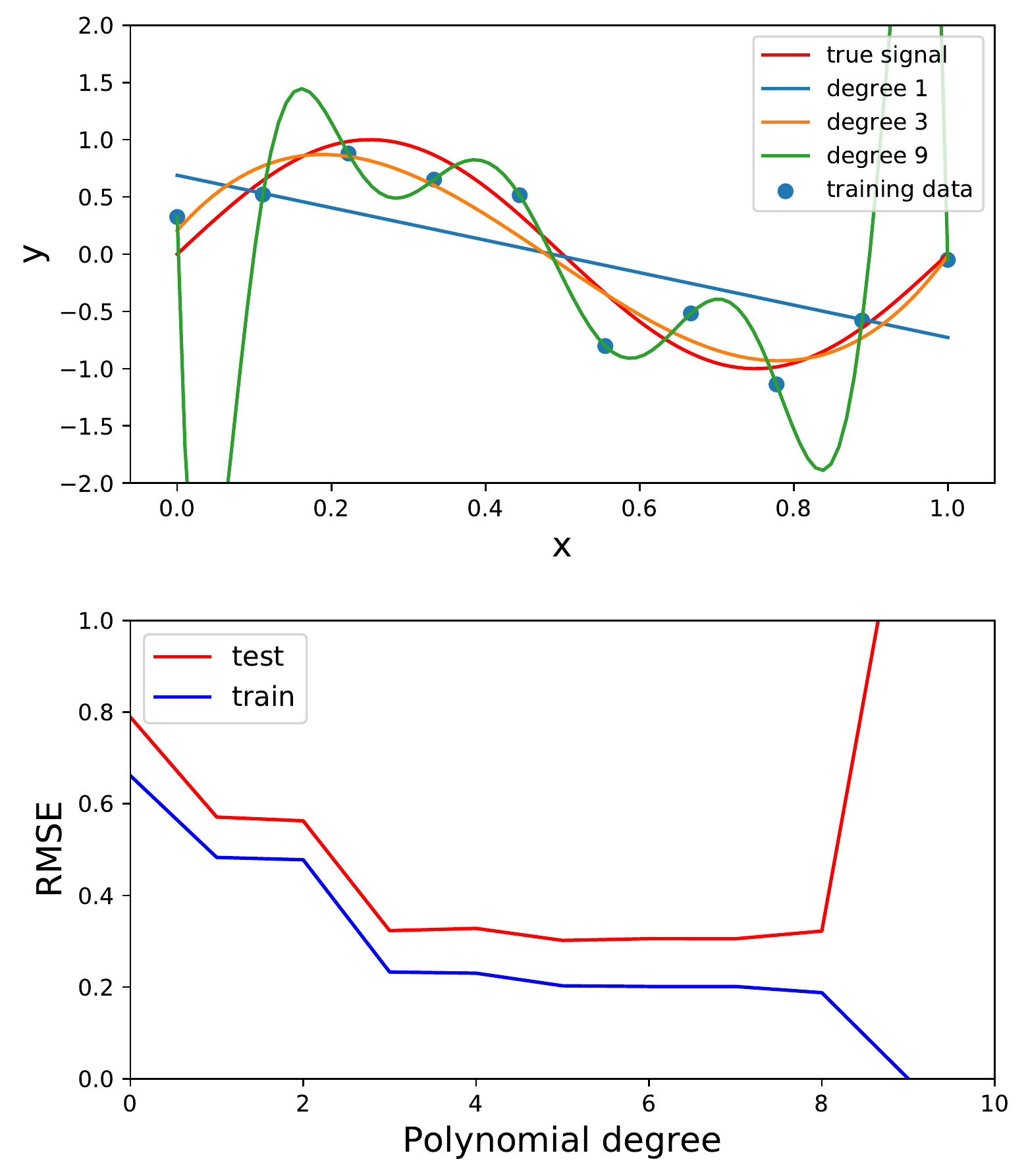}
\caption{\label{fig:train_test}{(Color online) Model generalization with polynomial regression. (Top) The true signal, training data, and three of the polynomial regression results are shown. (Bottom) The root mean square error (RMSE) of the predicted training and test signals were estimated for each polynomial degree.}}
\end{figure}

In general, ML models perform best when their capacity is suited to the complexity of the data provided and the task. For mismatched model-data/task complexities, two situations can arise. If a high-capacity model is used for a low-complexity task, the model will {\it overfit}, or learn the noise or idiosyncrasies of the training set. In the opposite scenario, a low-capacity model trained on a high-complexity task will tend to {\it underfit} the data, or not learn enough details of the underlying physics, for example. Both overfitting and underfitting degrade ML model generalization. The behavior of the ML model on training and test observations relative to the model parameters can be used to determine the appropriate model complexity. We next discuss how this can be done. We note that underfitting and overfitting can be quantified using the {\it bias} and {\it variance} of the ML model. The bias is the difference between the mean of our estimated targets $\widehat{\mathbf{y}}$ and the true mean, and the variance is the expected squared deviation of the estimated targets around the estimated mean value.\cite{hastie2009}

To estimate the performance of ML models on unseen observations, and thereby assess their generalization, a set of {\it test} data drawn from the full training set can be excluded from the model training and used to estimate generalization given the current parameters. In many cases, the data used in developing the ML model are split repeatedly into different sets of training and test data using cross validation techniques (Sec.~\ref{subsec:cross_val}.\cite{kohavi1995} The test data is used to adjust the model hyperparameters (e.g. regularization, priors, number of NN units/layers) to optimize generalization. The hyperparameters are model dependent, but generally govern the model's capacity.

In Fig.~\ref{fig:train_test}, we illustrate the effect of model capacity on train and test error using polynomial regression. Train and test data (10 and 100 points) were generated from a sinusoid ($y=\sin2\pi x$, left) with additive Gaussian noise. Polynomial models of orders 0 to 9 were fit to the training data, and the RMSE of the test and train data predictions are compared. $\text{RMSE}=\sqrt{1/M\sum_m(y_m-\widehat{y}_m)^2}$, with $M$ the number of samples (test or train) and $\widehat{y}_m$ the estimate of $y_m$. Increasing model capacity (complexity) decreases the training error, up to degree 9 where the degree plus intercept matches the number of training points (degrees of freedom). While increasing the complexity initially decreases the RMSE of the test data prediction, errors do not significantly decrease for polynomial degrees greater than 3, and increase for degrees greater than 5. Thus, we would prefer to use a model of degree 3, though the smallest test error was obtained for degree 5. In ML applications on real data, the test/train error curves are generated using cross-validation to improve the robustness of the model selection.

Alternatively, the model can be trained, tuned, and evaluated by dividing the data into three distinct sets: training, validation, and test. In this case the model is fit on the training data, and its performance on the validation data is used to tune the hyperparameters. Only after the hyperparameters are fully tuned on the training and validation data is the model performance evaluated on the test data. Here the test data is kept in a `vault', i.e. it should never influence the model parameters. 

\subsection{Cross-validation} \label{subsec:cross_val}
In many cases, we don't have enough samples to divide the data into 3 fully representative subsets (train, validation, and test). Thus we prefer to use to the tools of cross-validation with only two subsets of data: training and test. Cross-validation evaluates the model generalization by creating multiple training and test sets from the data (without replacement). The model parameters in this case are tuned using the `test' data.

One popular cross-validation technique, called K-fold cross validation,\cite{hastie2009} assesses model generalization by dividing training data into $K$ roughly equal-sized subgroups of the data, called {\it folds}. One fold is excluded from the model training and the error is calculated on the excluded fold. This procedure is executed $K$ times, with the $k$th fold used as the test data and the remaining $K-1$ folds used for model training. With target values divided into folds by $\mathbf{Y}=\mathbf{[Y}_1,\ldots,\mathbf{Y}_K]$ and inputs $\mathbf{X}=\mathbf{[X}_1^\mathrm{T},\ldots,\mathbf{X}_K^\mathrm{T}]^\mathrm{T}$, the cross validation error $\text{CV}_{err}$ is
\begin{equation}
\text{CV}_{err}(f,\mathbf{\theta})=\frac{1}{K}\sum_{i=1}^K L(\mathbf{Y}_i-f^{\sim i}(\mathbf{X}_i^\mathrm{T},\mathbf{\theta})),
\label{eq:kFolds}
\end{equation}
with $f^{\sim i}$ the model learned using all folds except $i$, $\mathbf{\theta}$ the hyperparameters, and $L$ a loss function. $\text{CV}_{err}(f,\mathbf{\theta})$ gives a curve describing the cross-validation (test) error as a function of the hyperparameters.

Some issues arise when using cross-validation. First, it requires as many training runs as subdivisions of the data. Further, tuning multiple hyperparameters with cross-validation can require a number of training runs that is exponential in the number of parameters. Some alternatives to the aforementioned test/train paradigms penalize the model complexity directly in the optimization. Such constraints include the well known Akaike information criterion (AIC) and Bayesian information criterion (BIC). However, AIC and BIC do not account for parameter uncertainty and often favor overly-simple models. In fully Bayesian approaches (as described in Sec.~\ref{sec:Bayes}), parameter uncertainty and model complexity are both well modeled.

\begin{figure}
\includegraphics[width=0.8\reprintcolumnwidth]{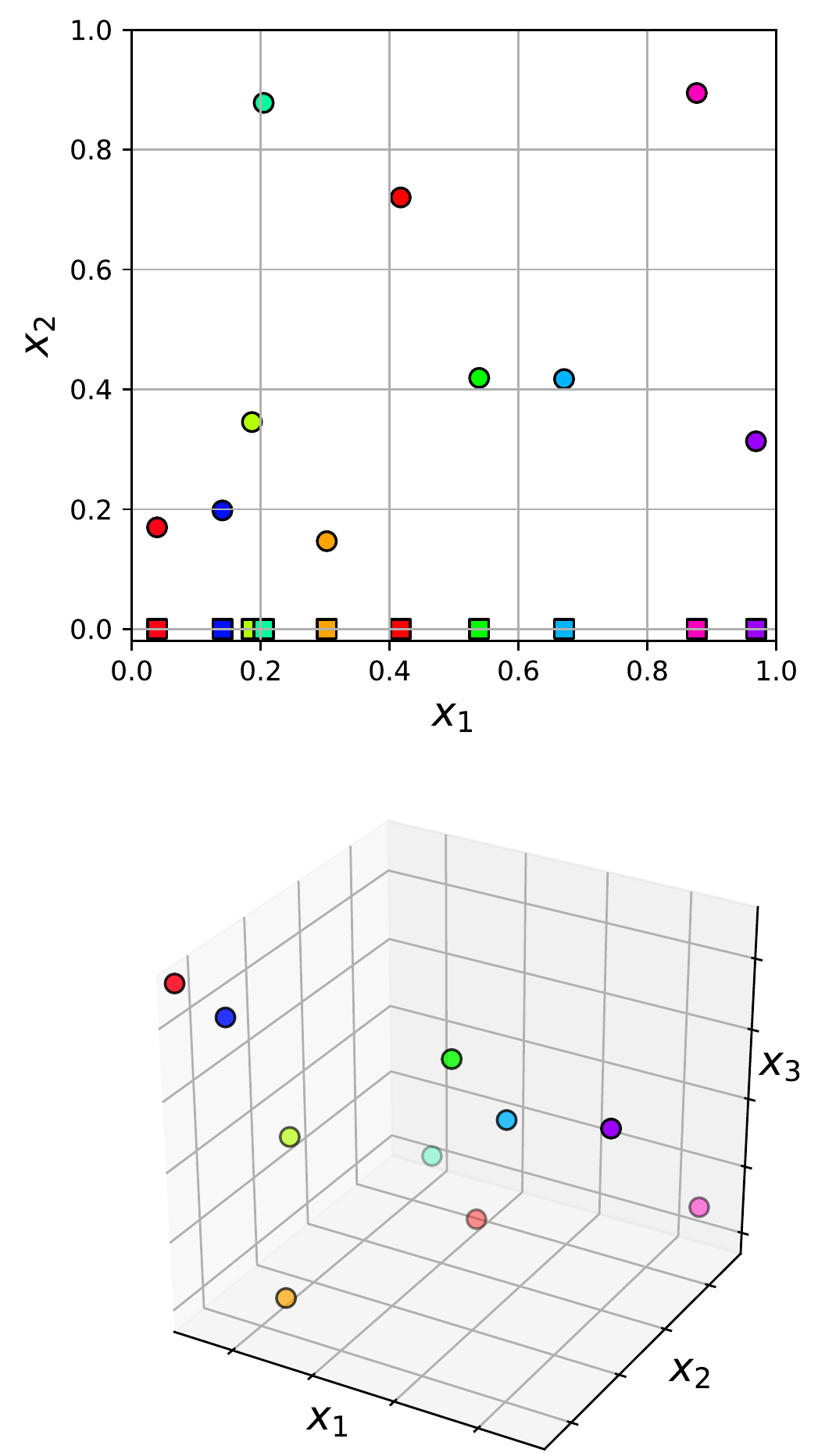}
\caption{\label{fig:curse}{(Color online) Illustration of curse of dimensionality. 10 uniformly distributed data points on the interval (0 1) can be quite close in 1D (top, squares), but as the number of dimensions, $N$, increases, the distance between the points increases rapidly. This is shown for points in 2D (top, circles), and 3D (bottom). The increasing volume $l^N$, with $l$ the normalized feature value scale, presents two issues. (1) local methods (like K-means) break-down with increasing dimension, since small neighborhoods in lower-dimensional space cover an increasingly small volume as the dimension increases. (2) Assuming discrete values, the number of possible data configurations, and thereby the minimum number of training examples, increase with dimension $O(l^d)$.\cite{hastie2009,goodfellow2016deep}}}\
\end{figure}

\subsection{Curse of dimensionality} \label{subsec:curse}
The often high-dimensionality of data also presents a challenge in ML, referred to as the `curse of dimensionality'. Considering features $\mathbf{x}$ are uniformly distributed in $N$ dimensions, (see Fig.~\ref{fig:curse}) with $x_n=l$ the normalized feature value, then $l$ (for example describing a neighborhood as a hypercube) constitutes a decreasing fraction of the features space volume. The fraction of the volume, $l^N$, is given by $\mathrm{f}_v=\mathrm{f}_l^N$, with $\mathrm{f}_v$ and $\mathrm{f}_l$ the volume and length fractions, respectively. Similarly, data tend to become more sparsely distributed in high-dimensional space. The curse of dimensionality most strongly affects methods that depend on distance measures in feature space, such as K-means, since neighborhoods are no longer `local'. Another result of the curse of dimensionality is the increased number of possible configurations, which may lead to ML models requiring increased training data to learn representations.

With prior assumptions on the data, enforced as model constraints (e.g. total variation\cite{chambolle2004} or $\ell_2$ regularization), training with smaller data sets is possible.\cite{goodfellow2016deep} This is related to the concept of learning a {\it manifold}, or a lower-dimensional embedding of the salient features. While the manifold assumption is not always correct, it is at least approximately correct for processes involving images and sound (for more discussion, see Ref.~\onlinecite{goodfellow2016deep}~[pp.~156--159]).

\subsection{Bayesian machine learning}\label{sec:Bayes}

A theoretically principled way to implement ML methods is to use the tools of probability, which have been a critical force in the development of modern science and engineering. Bayesian statistics provide a framework for integrating prior knowledge and uncertainty about physical systems into ML models. It also provides convenient analysis of estimated parameter uncertainty. Naturally, Bayes' rule plays a fundamental rule in many acoustic applications, especially in methods for estimating the parameters of model-based inverse methods. In the wider ML community, there are also attempts to expand ML to be Bayesian model-based, for a review see Ref.~\onlinecite{ghahramani2015}. We here discuss the basic rules of probability, as they relate to Bayesian analysis, and show how Bayes' rule can be used to estimate ML model parameters.

\begin{figure}
\hspace*{-3ex}\includegraphics[width=1.05\linewidth]{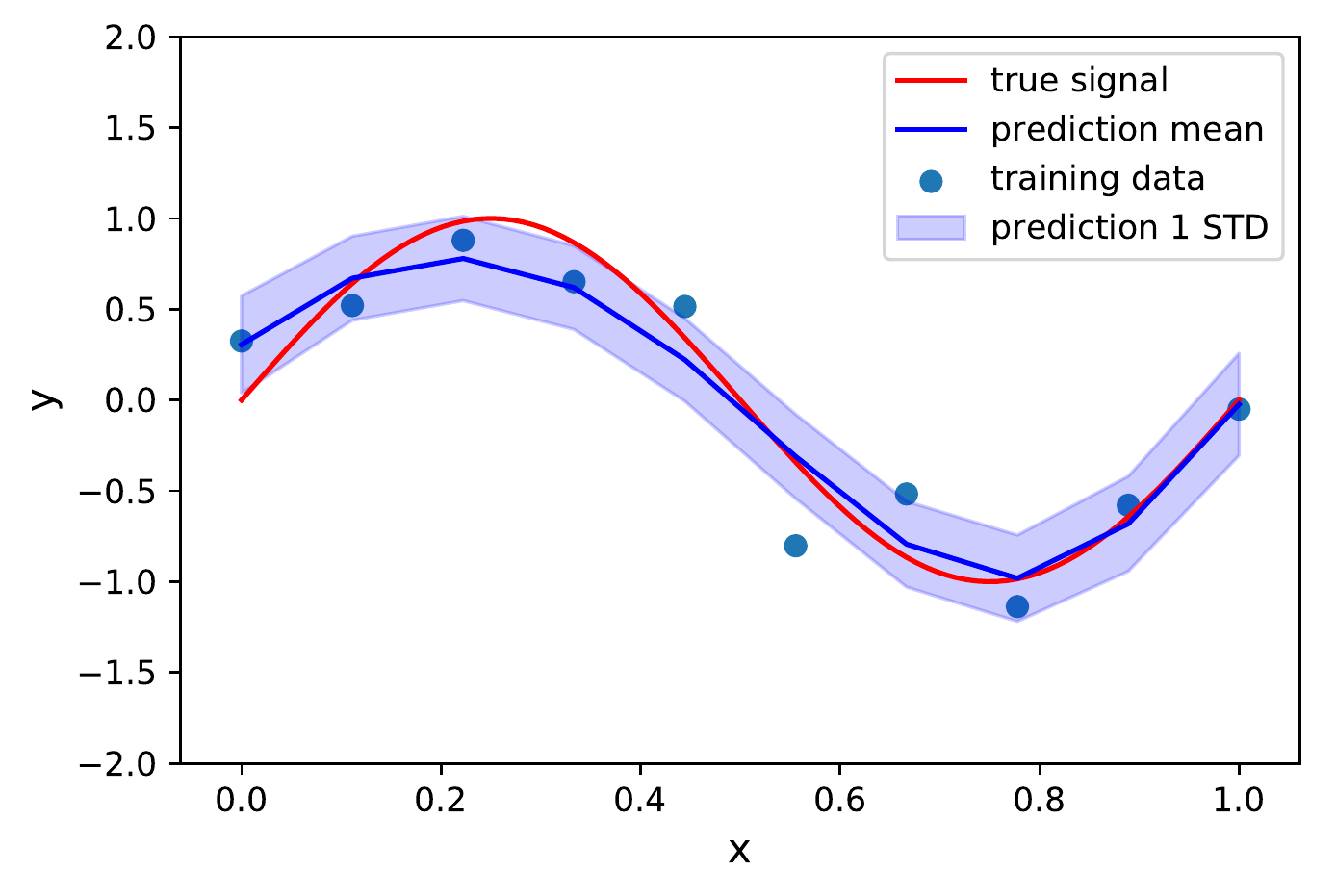}
\caption{\label{fig:bayes_pred}{(Color online) Bayesian estimate of polynomial regression model parameters for sinusoidal data from Fig.~\ref{fig:train_test}. Given prior knowledge and assumptions about the data, Bayesian parameter estimation can help prevent overfitting. It also provides statistics about the predictions. The mean of the prediction (blue line) is compared with the true signal (red) and the training data (blue dots, same as Fig.~\ref{fig:train_test}). The standard deviation of the prediction (STD, light blue) is also given by the Bayesian estimate. The estimate uses prior knowledge about the noise level $\sigma_\epsilon=0.2$ and a Gaussian prior on the model weights $\sigma_\mathbf{w}=10$.}}
\end{figure}

Two simple rules for probability are of fundamental importance for Bayesian ML.\cite{bishop2006} They are the sum rule
\begin{align} \label{eq:rule_sum}
p(\mathbf{x})=\sum_{\mathbf{y}\in Y} p(\mathbf{x,y})~,
\end{align}
and the product rule
\begin{align} \label{eq:rule_prod}
p(\mathbf{x,y})= p(\mathbf{x}|\bf{y})p(\mathbf{y})~.
\end{align}
Here the ML model inputs $\mathbf{x}$ and outputs $\mathbf{y}$ are uncertain quantities. The sum rule \eqref{eq:rule_sum} states that the marginal distribution $p(\mathbf{x})$  is obtained by summing the joint distribution $p(\mathbf{x,y})$ over all values of $\mathbf{y}$. The product rule \eqref{eq:rule_prod} states that $p(\mathbf{x,y})$ is obtained as a product of the conditional distribution, $p(\mathbf{y|x})$, and $p(\mathbf{y})$.

Bayes' rule is obtained from the sum and product rules by
\begin{align}\label{eq:bayesRule}
p({\bf y}|{\bf x})= \frac{p({\bf x},{\bf y})}{\sum_{{\bf y}\in Y} p({\bf x},{\bf y})}= \frac{p({\bf x}|{\bf y})p({\bf y})}{p({\bf x})},
\end{align}
which gives the model output $\mathbf{y}$ conditioned on the input $\mathbf{x}$ as the joint distribution $p(\mathbf{x,y})$ divided by the marginal $p(\mathbf{x})$. 

In ML, we need to choose an appropriate model $f(\mathbf{x})$ \eqref{eq:pred} and estimate the model parameters $\mathbf{\theta}$ to best give the desired output $\mathbf{y}$ from inputs $\mathbf{x}$. This is the inverse problem. The model parameters conditioned on the data is expressed as $p(\mathbf{\theta|x,y})$. From Bayes' rule \eqref{eq:bayesRule} we have
\begin{align}
p(\mathbf{\theta|x,y})&= \frac{p(\mathbf{y|x,\theta})p(\mathbf{\theta|x})}{p(\mathbf{y|x})}\label{eq:bayesRule3}
\\&\propto p(\mathbf{y|x,\theta})p(\mathbf{\theta}).\label{eq:bayesRule4}
\end{align}
$p({\bf \theta})$ is  the prior distribution on the parameters, $p(\mathbf{y|x,\theta})$ called the likelihood, and $p(\mathbf{\theta|x,y})$ the posterior. The quantity $p({\bf y|x})$ is the distribution of the data, also called the evidence or Type II likelihood. Often it can be neglected  (e.g. \eqref{eq:bayesRule4}) as for given data $p({\bf y|x})$ is constant and does not affect the target, ${\bf \theta}$.

A Bayesian estimate of the parameters $\mathbf{\theta}$ is obtained using \eqref{eq:bayesRule3}. Assuming a scalar linear model $y=f(\mathbf{x})+\epsilon$, with $f(\mathbf{x})=\mathbf{x}^\mathrm{T}\mathbf{w}$, where the parameters $\mathbf{\theta=w}\in\mathbb{R}^{N}$ are the weights (see Sec.~\ref{subsec:lr} for more details). A simple solution to the parameter estimate is  obtained if we assume the prior $p({\bf w})$ is Gaussian,
${\cal N} ({\bf \mu}, {\bf \Gamma})$ with  ${\bf \mu}$ mean and covariance ${\bf \Gamma}$.
Often we also assume a Gaussian likelihood $p(\mathbf{x},y|\mathbf{\theta})$, ${\cal N} (\mathbf{x}^\mathrm{T}\mathbf{w}, \sigma_\epsilon)$ with mean $\mathbf{x}^\mathrm{T}\mathbf{w}$ and covariance ${\bf \Sigma}_\epsilon$. We get, see Ref.~\onlinecite{bishop2006}~[p.93], 
\begin{align}
p(\mathbf{w|x},y)=&{\cal N} ({\bf w}_p ,{\bf \Sigma}_p)\\
{\bf w}_p=& {\bf \Sigma}_p ( \frac{1}{\sigma_\epsilon}{\bf x}y + {\bf \Gamma}^{-1}{\bf \mu} )\\
{\bf \Sigma}_p=&(\frac{1}{\sigma_\epsilon}{\bf xx}^\mathrm{T} +{\bf \Gamma}^{-1})^{-1}
\end{align}
The formulas are very efficient for sequential estimation as the prior is conjugated, i.e. it is of the same form as the posterior.
In acoustics this framework has been used for range estimation \cite{michalopoulou2019} and for sparse estimation via the sparse Bayesian learning approach.\cite{Gemba2017SBL,nannuru2019} In the latter, the sparsity is controlled by diagonal prior covariance matrix, where entries with zero prior variance will force the posterior variance and mean to be zero.

With prior knowledge and assumptions about the data, Bayesian approaches to parameter estimation can prevent overfitting. Further, Bayesian approaches provide the probability distribution of target estimates $\widehat{y}$. Fig.~\ref{fig:bayes_pred} shows a Bayesian estimate of polynomial curve-fit developed in Fig.~\ref{fig:train_test}. The mean and standard deviation of the predictions from the model are given. The Bayesian curve fitting is here performed assuming prior knowledge of the noise standard deviation ($\sigma_\epsilon=.2$) and with a Gaussian prior on the weights ($\sigma_\mathbf{w}=10$). The hyperparameters can be estimated from the data using empirical Bayes.\cite{gelman2013bayesian} This is counterpoint to the test-train error analysis (Fig.~\ref{fig:train_test}), where fewer assumptions are made about the data, and the noise is unknown. We note that it is not always practical to formally implement Bayesian parameter estimation due to the increased computational cost of estimating the posterior distribution versus optimization. Where practical, Bayesian models well characterize ML results because they explicitly provide uncertainty in the model parameter estimates with the posterior distribution, and also permit explicit specification of prior knowledge of the parameter distributions (the prior) and data uncertainty.

\section{Supervised learning} \label{sec:sup_learn}

The goal of supervised learning is to learn a mapping from a set of inputs to desired outputs given labeled input and output pairs \eqref{eq:pred}. For discussion, we here focus on real-valued features and labels. The $N$ features in $\mathbf{x}$ can be real, complex, or categorical (binary or integer). Based on the type of desired output $\mathbf{y}$, supervised learning can be divided into two subcategories: regression and classification. When $\mathbf{y}$ is real or complex valued, the task is regression.  When $\mathbf{y}$ is categorical, the task is called classification.

The methods of finding the function $f$ are the core of ML methods and the subject of this section. Generally, we prefer to use the tools of probability to find $f$, if practical. We can state the supervised ML task as the task of maximizing the conditional distribution $p(\mathbf{y}|\mathbf{x})$. One example is the {\it maximum a posteriori} (MAP) estimator
\begin{equation}
    \widehat{\mathbf{y}}=f(\mathbf{x})=\underset{{y}}{\arg\max}~p(\mathbf{y}|\mathbf{x}),
\end{equation}
which gives the most probable value of $\mathbf{y}$, corresponding to the mode of the distribution conditioned on the observed evidence $p(\mathbf{y}|\mathbf{x})$. While the MAP can be considered Bayesian, it is really only a step toward Bayesian treatment (see Sec.~\ref{sec:Bayes}) since MAP returns a point estimate rather than the posterior distribution.

In the following, we further describe regression and classification methods, and give some illustrative applications.

\subsection{ Linear regression, classification} \label{subsec:lr}

We illustrate supervised ML with a simple method: linear regression. We develop a MAP formulation of linear regression in the context of direction-of-arrival (DOA) estimation in beamforming. In seismic and acoustic beamforming, waveforms are recorded on an array of receivers with the goal of finding their DOA. The features are the Fourier-transformed measurements from $M$ receivers, $\mathbf{x}\in\mathbb{C}^M$, and the output $y$ is the DOA azimuth angle (see \eqref{eq:pred}). The relationship between DOA and array power is non-linear, but is expressed as a linear problem by discretizing the array response using basis functions $\mathbf{A}=[\mathbf{a}(\theta_1),\ldots,\mathbf{a}(\theta_N)]\in\mathbb{C}^{M\times N}$, with $\mathbf{a}(\theta_n)$ called steering vectors. The array observations are expressed as $\mathbf{x}=\mathbf{Aw}$. The weights $\mathbf{w}\in\mathbb{C}^N$ relate the steering vectors $\mathbf{A}$ to the observations $\mathbf{x}$. We thus write the linear measurement model as

\begin{equation}\label{eq:doa_lin}
    \mathbf{x}=\mathbf{Aw+\epsilon}.
\end{equation}

In the case of a single source, DOA is $y=\theta_n$ corresponding to $\max\{w_1,\ldots,w_N\}$. $\mathbf{\epsilon}\in\mathbb{C}^M$ is noise (often Gaussian). We seek values of weights $\textbf{w}$ which minimize the difference between the left and right-hand sides of \eqref{eq:doa_lin}. We here consider the case of $L=1$ snapshots.

From Bayes' rule \eqref{eq:bayesRule}, the posterior of the model is
\begin{equation}
p(\mathbf{w}|\mathbf{x})\propto p(\mathbf{x}|\mathbf{w})p(\mathbf{w}),
\end{equation}
with $p(\mathbf{x}|\mathbf{w})$ the likelihood and $p(\mathbf{w})$ the prior. Assuming the noise $\mathbf{\epsilon}$ Gaussian iid with zero-mean,  $p(\mathbf{x}|\mathbf{w})=\mathcal{CN}(\mathbf{x}|\mathbf{Aw},\sigma_\mathbf{\epsilon}^2\mathbf{I})$ with $\mathbf{I}$ the identity,
\begin{equation}
\ln p(\mathbf{w}|\mathbf{x})=
-\frac{1}{\sigma_\mathbf{\epsilon}^2}\|\mathbf{x-Aw}\|_2^2+\ln p(\mathbf{w})+C,
\end{equation}
with $C$ a constant and $\mathcal{CN}$ complex Gaussian. Maximizing the posterior, we obtain
\begin{align}
\max \big\{\ln p(\mathbf{w}|\mathbf{x})\big\}\propto \min \bigg\{\frac{1}{\sigma_\mathbf{\epsilon}^2}\|\mathbf{x-Aw}\|_2^2-\ln p(\mathbf{w})\bigg\}.
\end{align}
Thus, the MAP estimate $\widehat{\mathbf{w}}$, is
\begin{equation}
\widehat{\mathbf{w}}=\underset{\mathbf{w}}{\arg\min}~\frac{1}{\sigma_\mathbf{\epsilon}^2}\|\mathbf{x-Aw}\|_2^2-\ln p(\mathbf{w}).
\end{equation}

Depending on the choice of probability density function for $p(\mathbf{w})$, different solutions are obtained. One popular choice is a Gaussian distribution. For $p(\mathbf{w})$ Gaussian,
\begin{equation}
\widehat{\mathbf{w}}=\underset{\mathbf{w}}{\arg\min}~\|\mathbf{x-Aw}\|_2^2+\lambda_1\|\mathbf{w}\|_2^2,
\label{eq:L2}
\end{equation}
where $\lambda_1=\sigma_\mathbf{\epsilon}^2/\sigma_\mathbf{w}^2$ is a regularization parameter, and $\sigma_\mathbf{w}^2$ the variance of $\mathbf{w}$. This is the classic $\ell_2$-regularized least-squares estimate (a.k.a. damped least squares, or ridge regression). \cite{bishop2006,aster2013} Eq. \eqref{eq:L2} has the analytic solution
\begin{equation}
\widehat{\mathbf{w}}=\big(\mathbf{A}^\mathrm{T}\mathbf{A}+\lambda_1\mathbf{I}\big)^{-1}\mathbf{A}^\mathrm{T}\mathbf{x}.
\label{eq:L2analy}
\end{equation}

\begin{figure}[tb]
\includegraphics[width=\reprintcolumnwidth]{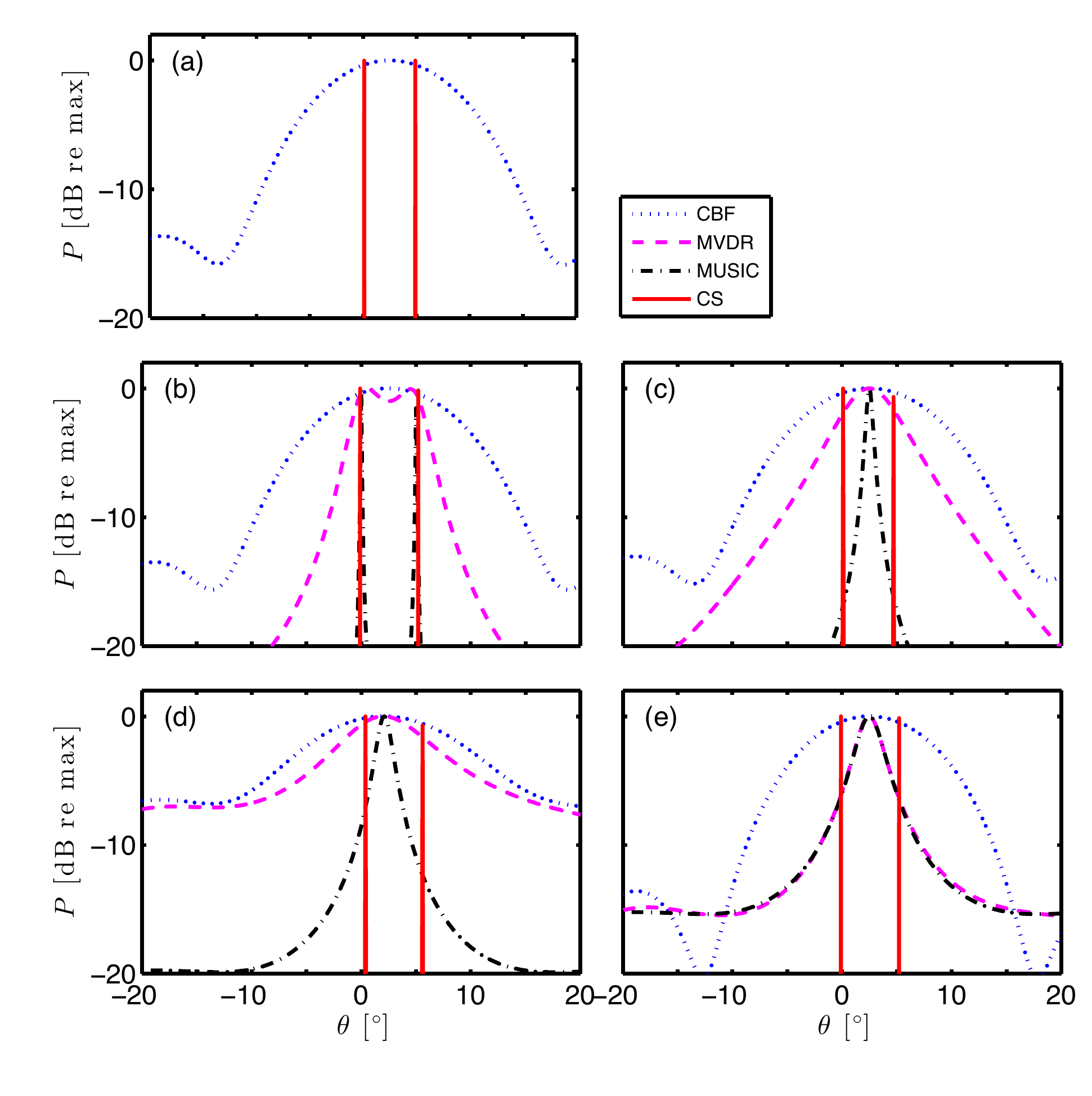}
\caption{\label{fig:cs_doa}{(Color online) DOA estimation from $L$ snapshots for two equal-strength sources at 0$^\circ$ and 5$^\circ$ azimuth with a uniform linear array with $M=8$ sensors and $\lambda/2$ spacing. (a) conventional beamformer (CBF) and compressive sensing (CS) beamforming for uncorrelated sources with 20 dB SNR and one snapshot, $L=1$. CBF, minimum variance distortionless response (MVDR), MUSIC, and CS for uncorrelated sources with (b) $\text{SNR}=20$~dB and $L=50$, (c) $\text{SNR}=20$~dB and $L=4$, (d) $\text{SNR}=0$~dB and $L =50$, and (e) for correlated sources with $\text{SNR}=20$~dB and $L=50$. The array SNR is for one snapshot. From Ref.~\onlinecite{xenaki2014}.}}
\raggedright
\end{figure}

Although the $\ell_2$ regularization in \eqref{eq:L2} is often convenient, it is sensitive to outliers in the data $\mathbf{x}$. In the presence of outliers, or if the true weights $\mathbf{w}$ are sparse (e.g. few non-zero weights), a better prior is the Laplacian, which gives
\begin{equation}
\widehat{\mathbf{w}}=\underset{\mathbf{w}}{\arg\min}~\|\mathbf{x-Aw}\|_2^2+\lambda_2\|\mathbf{w}\|_1,
\label{eq:L1}
\end{equation}
where $\lambda_2=\sigma_\mathbf{\epsilon}/b_\mathbf{w}$ a regularization parameter, and $b_\mathbf{w}$ a scaling parameter for the Laplacian distribution. \cite{murphy2012} Eq.~\eqref{eq:L1} is called the $\ell_1$ regularized least-squares estimator of ${\mathbf{w}}$. While the problem is convex, it is not analytic, though there are many practical algorithms for its solution.\cite{elad2010,mairal2014,gerstoft2015} In sparse modeling, the $\ell_1$-regularization is considered a convex relaxation of $\ell_0$ pseudo-norm, and under certain conditions, provides a good approximation to the $\ell_0$-norm. For a more detailed discussion, please see Refs.~\onlinecite{elad2010,mairal2014}. The solution to \eqref{eq:L1} is also known as the LASSO,\cite{tibshirani1996} and forms the cornerstone of the field of compressive sensing (CS). \cite{candes2006,gerstoft2018}

Whereas in the estimate $\widehat{\mathbf{w}}$ obtained from \eqref{eq:L2} many of the coefficients are small, the estimate from \eqref{eq:L1} has only few non-zero coefficients. Sparsity is a desirable property in many applications, including array processing\cite{haykin2014,gerstoft2018} and image processing.\cite{mairal2014} We give an example of $\ell_1$ (in CS) and $\ell_2$ regularization in the estimation of DOAs on a line array, Fig.~\ref{fig:cs_doa}.

Linear regression can be extended to the binary classification problem. Here for binary classification, we have a single desired output ($N=1$) $y_m$ for each input $\mathbf{x}_m$, and the labels are either 0 or 1. The desired labels for $M$ observations are $\mathbf{y}\in\{0,1\}^{1\times M}$ (row vector),
\begin{equation}
\mathbf{y}=\mathbf{X}\mathbf{w}.
\end{equation}
Here $\mathbf{w}\in\mathbb{R}^N$ is the weights vector. Following the derivation of \eqref{eq:L2}, the MAP estimate of the weights is given by
\begin{equation}
\widehat{\mathbf{w}}=\big(\mathbf{X}^\mathrm{T}\mathbf{X}+\lambda_1\mathbf{I}\big)^{-1}\mathbf{X}^\mathrm{T}\mathbf{y},
\label{eq:L2analy2}
\end{equation}
with $\widehat{\mathbf{w}}$ the ridge regression estimate of the weights.

This ridge regression classifier is demonstrated for binary classification ($C=2$) in Fig.~\ref{fig:linear_vs_svm} (top). The cyan class is $0$ and red is $1$, thus, the decision boundary (black line) is $\mathbf{w}^\mathrm{T}\mathbf{x}_m=0.5$. Points classified as $y_m=1$ are $\{\mathbf{x}_m:\mathbf{w}^\mathrm{T}\mathbf{x}_m>0.5\}$, and points classified as $y_m=0$ are $\{\mathbf{x}_m:\mathbf{w}^\mathrm{T}\mathbf{x}_m\le0.5\}$. In the case where each class is composed of a single Gaussian distribution (as in this example), the linear decision boundary can do well.\cite{hastie2009} However, for more arbitrary distributions, such a linear decision boundary may not suffice, as shown by the poor classification results of the ridge classifier on concentric class distributions in Fig.~\ref{fig:linear_vs_svm} (top-right).

In the case of the concentric distribution, a non-linear decision boundary must be obtained. This can be performed using many classification algorithms, including logistic regression and SVMs.\cite{murphy2012} In the following section we illustrate the non-linear decision boundary estimation using SVMs.

\subsection{Support vector machines} 

Thus far in our discussion of classification and regression, we have calculated the outputs $\mathbf{y}_m$ based on feature vectors $\mathbf{x}_m$ in the raw feature dimension (classification) or on a transformed version of the inputs (beamforming, regression). Often, we can make classification methods more flexible by enlarging the feature space with non-linear transformations of the inputs $\mathbf{\phi(x}_m)$. These transformations can make data, which is not linearly separable, linearly separable in the transformed space (see Fig.~\ref{fig:linear_vs_svm}). However, for large feature expansions, the feature transform calculation can be computationally prohibitive.

Support vector machines (SVMs) can be used to perform classification and regression tasks where the transformed feature space is very large (potentially infinite). SVMs are based on maximum margin classifiers,\cite{murphy2012} and use a concept called the {\it kernel trick} to use potentially infinite-dimensional feature mappings with reasonable computational cost.\cite{bishop2006} This uses kernel functions, relating the transforms of two features as $\kappa(\mathbf{x}_i, \mathbf{x}_j)=\mathbf{\phi(x}_i)^\mathrm{T}\mathbf{\phi(x}_j)\in\mathbb{R}$. They can be interpreted as similarity measures of linear or non-linear transformations of the feature vectors $\mathbf{x}_i,\mathbf{x}_j$. Kernel functions can take many forms (see Ref.~\onlinecite{bishop2006}~[pp.~291--323]), but for this review we illustrate SVMs with the Gaussian radial basis function (RBF) kernel
\begin{equation} \label{RBF}
\kappa(\mathbf{x}_i, \mathbf{x}_j) = \exp(-\gamma ||\mathbf{x}_i - \mathbf{x}_j||^2).
\end{equation}
$\gamma$ controls the length scale of the kernel. RBF can also be used for regression. The RBF is one example of kernelization of an infinite dimensional feature transform.

SVMs can be easily formulated to take advantage of such kernel transformations. Below, we derive the maximum margin classifier of SVM, following the arguments of Ref.~\onlinecite{bishop2006}, and show how kernels can be used to enhance classification.

\begin{figure}[t]
\includegraphics[width = 0.48\textwidth]{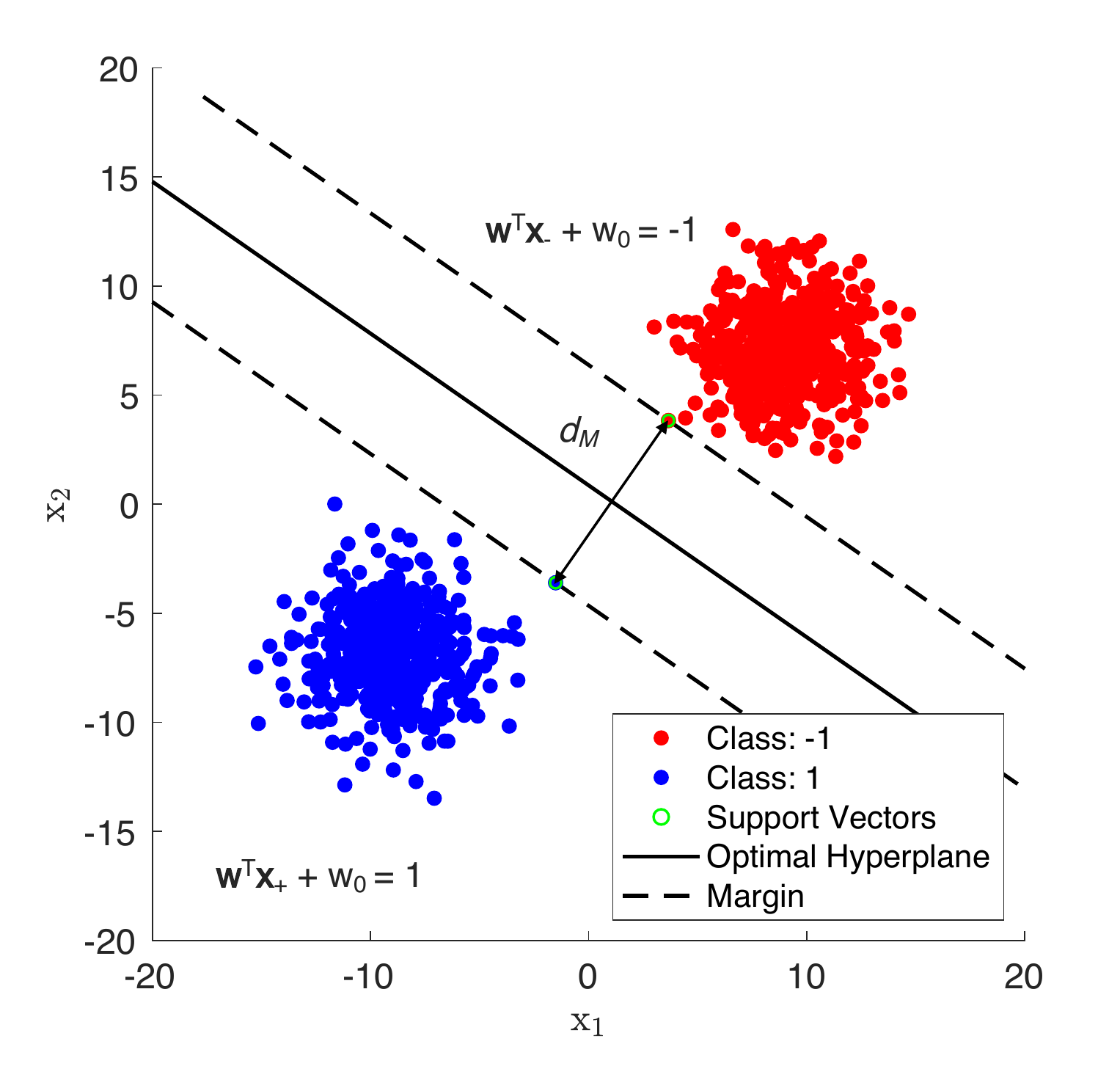}
\caption{\label{fig:FIG2}{(Color online) Support vector machine (SVM) binary classification with separable classes (2D, $N=2$). The hyperplane is estimated by a SVM which maximizes the margin $d_M$ subject to the constraint that none of the data are misclassified (see \eqref{eq:svm_min1}) . When there are only two support vectors (as shown here), the hyperplane is orthogonal to the difference of the support vectors.}}
\end{figure}

\begin{figure}[t]
\includegraphics[width=3.5in]{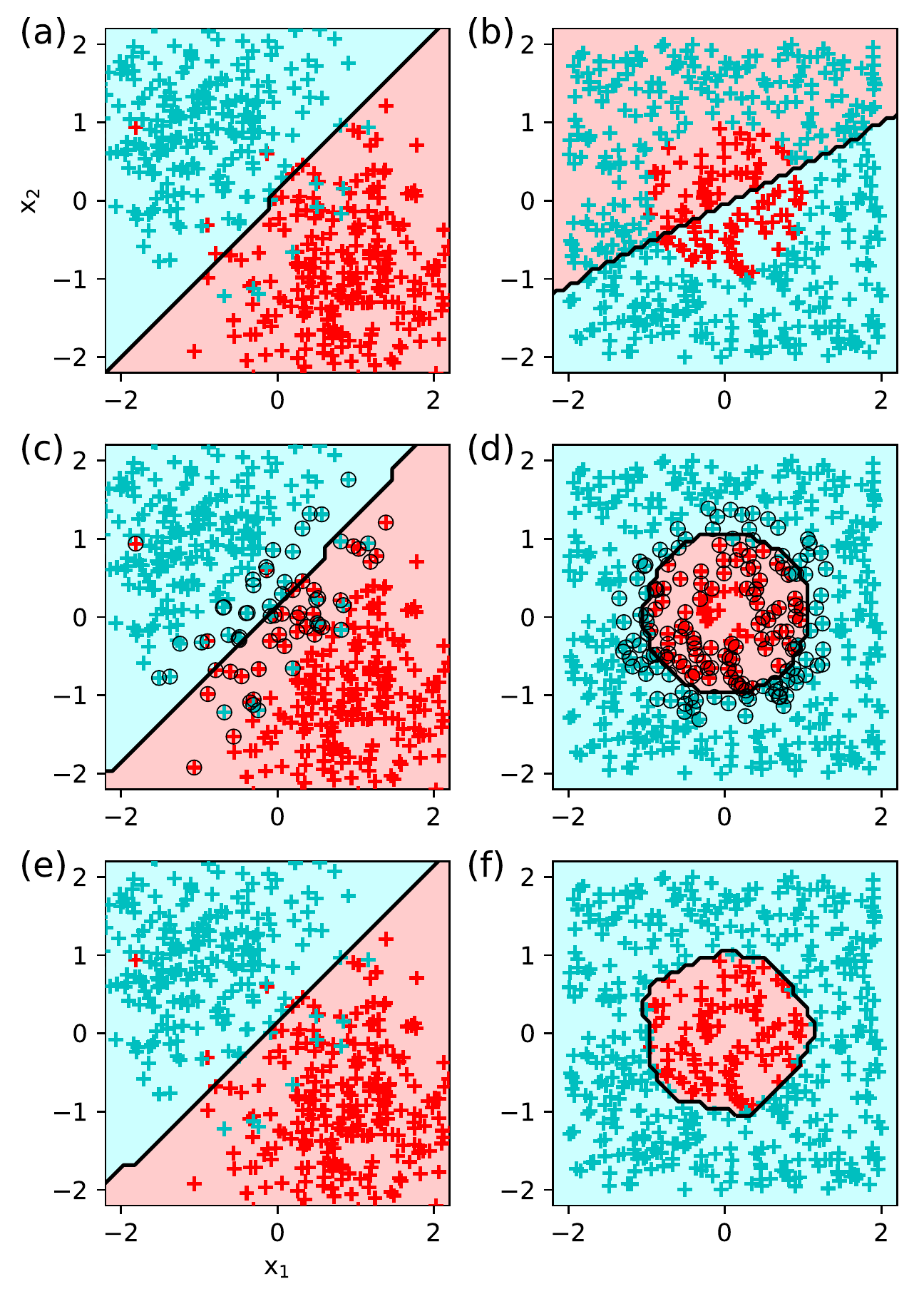}
\caption{(Color online) Binary classification of points with two distributions: (1) two Gaussian distributions (a,c,e) and (2) two uniformly distributions (b,d,f) with a radial boundary (red, cyan) using ridge regression (a,b), SVMs with radial basis functions (RBFs, c,d) with support vectors (black circles), and feed forward NNs (NNs, e,f). SVMs are more flexible than linear regression and can fit more general distributions using the kernel trick with e.g. RBFs. NNs  require fewer data assumptions to separate the classes, instead using non-linear modeling to fit the  distributions.}
\label{fig:linear_vs_svm}
\end{figure}

Initially, we assume linearly separable features $\mathbf{X}$ (see Fig.~\ref{fig:FIG2}) with classes $s_m \in \{1, -1\}$. The class of the objects corresponding to the features is determined by
\begin{equation}
\mathbf{y} = \mathbf{X}\mathbf{w} + \mathbf{w}_0,
\end{equation}
with $\mathbf{w}$ and $\mathbf{w}_0$ the weights and biases. A decision hyperplane satisfying $\mathbf{X}\mathbf{w}+\mathbf{w}_0=\mathbf{0}$ is used to separate the classes. If $y_m$ is above the hyperplane ($y_m>0$), the estimated class label is $\hat{s}_{m}=1$, whereas if $y_m$ is below ($y_m<0$), $\hat{s}_{m}=-1$. This gives the condition $s_my_m>0~\forall~m$. The margin $d_M$ is defined as the distance between the nearest features (Fig.~\ref{fig:FIG2}) with different labels, $\mathbf{x}_{-},~s=-1$ and $\mathbf{x}_{+},~s=+1$. These points correspond to the equations $\mathbf{w}^\mathrm{T}\mathbf{x}_{-}+w_0=-1$ and $\mathbf{w}^\mathrm{T}\mathbf{x}_{+}+w_0=1$. The difference between these equations, normalized by the weights $\|\mathbf{w}\|_2$, yields an expression for the margin
\begin{equation}
\frac{\mathbf{w}^\mathrm{T}}{\|\mathbf{w}\|_2}(\mathbf{x}_{+}-\mathbf{x}_{-})=\frac{2}{\|\mathbf{w}\|_2}.
\end{equation}
The expression says the projection of the difference of $\mathbf{x}_{-}$ and $\mathbf{x}_{+}$ on $\mathbf{w}^\mathrm{T}/\|\mathbf{w}\|_2$ (unit vector perpendicular to the hyperplane) is $2/\|\mathbf{w}\|_2$. Hence, $d_M=2/\|\mathbf{w}\|_2$.

The weights $\mathbf{w}$ and $w_0$ are estimated by maximizing the margin $2/\|\mathbf{w}\|_2$, subject to the constraint that the points $\mathbf{x}_m$ are correctly classified. Observing that $\max~2/\|\mathbf{w}\|_2$ is equivalent to $\min~\frac{1}{2}\|\mathbf{w}\|_2^2$, the optimization is a quadratic program
\begin{align} \label{eq:svm_min1}
&\underset{\mathbf{w}, w_0}{\min}~\frac{1}{2}\|\mathbf{w}\|_2^2, \nonumber\\
&\text{subject to}\ \ s_m (\mathbf{w}^T \mathbf{x}_m + w_0) \geq 1 \quad \forall~m.
\end{align}

If the data are linearly non-separable (class overlapping), slack variables $\xi_m \geq 0$ allows some of the training points to be misclassified.\cite{bishop2006} This gives
\begin{equation} \label{eq:svm_min2}
\begin{aligned}
&\underset{\mathbf{w}, w_0}{\operatorname{\min}} \quad \frac{1}{2} \|\mathbf{w}\|^2 + C \sum_{m=1}^{M}{\xi_m},\\
&\text{subject to  } s_m\mathbf{y}_m  \geq 1-\xi_m~\forall~m.
\end{aligned}
\end{equation}
The parameter $C>0$ controls the trade-off between the slack variable penalty and the margin.

For the non-linear classification problems, the quadratic program \eqref{eq:svm_min2} can be kernelized to make the data linearly separable in a non-linear space defined by feature vectors $\mathbf{\phi}(\mathbf{x}_m)$. The kernel is formed from the feature vectors by $\kappa(\mathbf{x}_m,\mathbf{x}_m')=\mathbf{\phi}(\mathbf{x}_m)^\mathrm{T}\mathbf{\phi}(\mathbf{x}_m')$. Eq.~\eqref{eq:svm_min2} can be rewritten using the Lagrangian dual\cite{bishop2006}
\begin{equation} \label{eq:svm_min3}
\begin{aligned}
&L(\mathbf{a})=\sum_{i=1}^M~a_i-\frac{1}{2}\sum_{i=1}^M\sum_{j=1}^Ma_ia_js_is_j\kappa(\mathbf{x}_i,\mathbf{x}_j),\\
&\text{subject to }~0\le a_i \le C, \\
&\hspace{12ex} \sum_{i=1}^M a_is_i=0
\end{aligned}
\end{equation}
Eq.~\eqref{eq:svm_min3} is solved as a quadratic programming problem. From the Karush-Kuhn-Tucker conditions,\cite{bishop2006} either $a_i=0$ or $s_my_m=1$. Points with $a_i=0$ are not considered in the solution to \eqref{eq:svm_min3}. Thus, only points within the specified slack distance $\xi_m$ from the margin, $s_m\mathbf{y}_m=1-\xi_m$, participate in the prediction. These points are called {\it support vectors}.

In Fig.~\ref{fig:linear_vs_svm} we use SVM with the RBF kernel \eqref{RBF} to classify points where the true decision boundary is either linear or circular. The SVM result is compared with linear regression (Sec.~\ref{subsec:lr}) and NNs (Sec.~\ref{subsec:nn}). Where linear regression fails on the circular decision boundary, SVM with RBF well separates the two classes. The SVM example was implemented in Python using Scikit-learn.\cite{scikit-learn}

We here note that the SVM does not provide probabilistic output, since it gives hard labels of data points and not distributions. Its label uncertainties can be quantified heuristically.\cite{murphy2012}

Because the SVM is a two-class model, multi-class SVM with $K$ classes requires training $K(K-1)/2$ models on all possible pairs of classes. The points that are assigned to the same class most frequently are considered to comprise a single class, and so on until all points are assigned a class from $1$ to $K$. This approach is known as the ``one-versus-rest" scheme, although slight modifications have been introduced to reduce computational complexity.\cite{bishop2006,murphy2012}

SVMs have been used for acoustic target classification,\cite{Cao2003SVMTarget} underwater source localization,\cite{niu2017} and classifying animal calls\cite{acevedo2009automated,fagerlund2007bird} to name a few examples. For large datasets, SVMs suffer from high computational cost. Further, kernel machines with generic kernels do not generalize well. Recent developments in deep learning were designed to overcome these limitations, as evidenced by neural networks (NNs) outperforming RBF kernel SVMs on the MNIST data set.\cite{goodfellow2016deep,hinton2006fast}

\subsection{ Neural networks: multi-layer perceptron} \label{subsec:nn} \label{sec:mlp}

Neural networks (NNs) can overcome the limitations of linear models (linear regression, SVM) by learning a non-linear mapping of the inputs $\phi(\mathbf{x}_m)$ from the data over their network structure. Linear models are appealing because they can be fit efficiently and reliably, with solutions obtained in closed form or with convex optimization. However, they are limited to modeling linear functions. As we saw in previous sections, linear models can use non-linear features by prescribing basis functions (DOA estimation) or by mapping the features into a more useful space using kernels (SVM). Yet these prescribed feature mappings are limited since kernel mappings are generic and based on the principle of local smoothness. Such general functions perform well for many tasks, but better performance can be obtained for specific tasks by training on specific data. NNs (and also dictionary learning, see Sec.~\ref{sec:unsup}) provide the algorithmic machinery to learn representations $\phi(\mathbf{x}_m)$ directly from data. \cite{lecun2015,goodfellow2016deep}

The purpose of feed-forward NNs, also referred to as deep NNs (DNNs) or multi-layer perceptrons (MLPs), is to approximate functions. These models are called {\it feed-forward} because information flows only from the inputs (features) to the outputs (labels), through the intermediate calculations. When feedback connections are included in the network, the network is referred to as a recurrent NN (RNN, for more details see Sec.~\ref{sec:deep_learning}). 

\begin{figure}[t]
\includegraphics[width=3in]{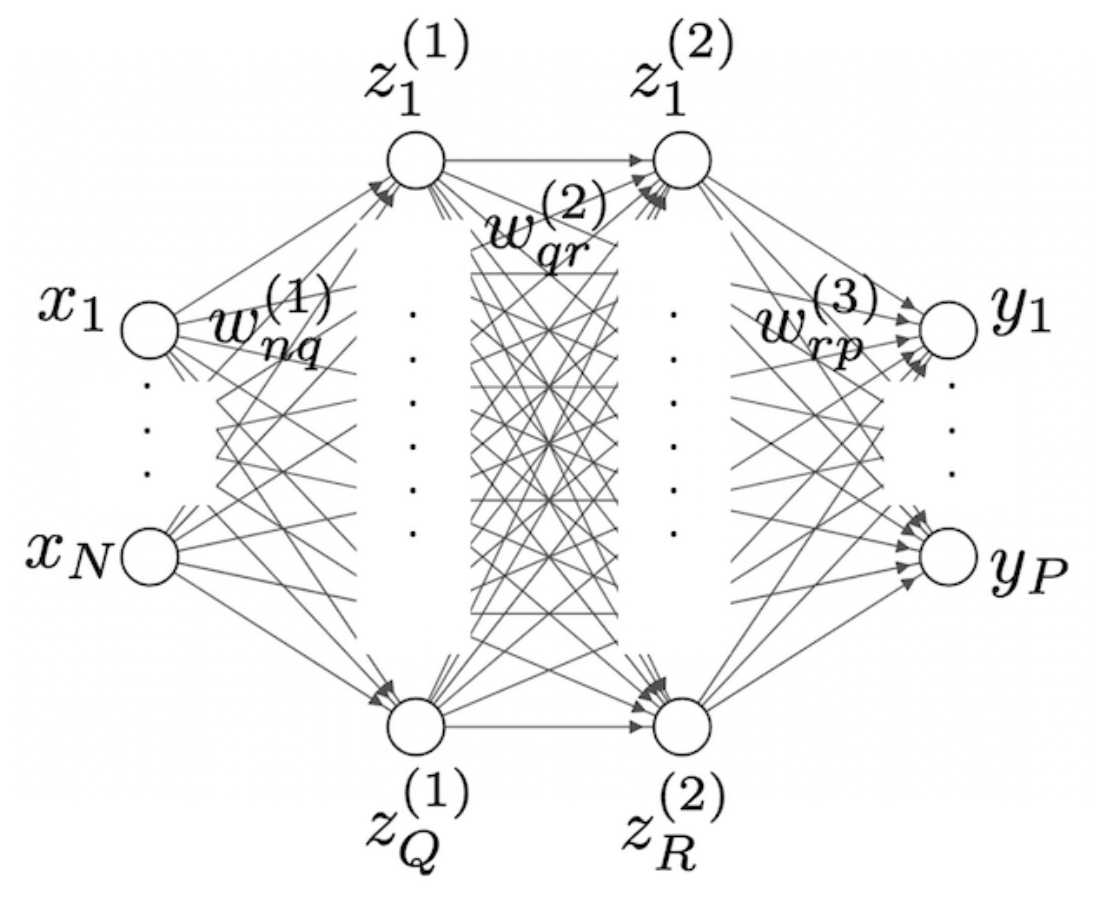}
\caption{\label{fig:fnn_binary}{Feed-forward neural network (NN).}}\
\raggedright
\end{figure}

NNs are called {\it networks} because they are composed of a series of functions associated by a directed graph. Each set of functions in the NN is referred to as a {\it layer}. The number of layers in the network (see Fig.~\ref{fig:fnn_binary}), called the NN {\it depth}, typically is the number of hidden layers plus one (the output layer). The NN depth is one of the parameters that affect the capacity of NNs. The term {\it deep learning} refers to NNs with many layers.\cite{goodfellow2016deep}

In Fig.~\ref{fig:fnn_binary}, an example 3 layer fully-connected NN is illustrated. The first layer, called the {\it input} layer, is the features $\mathbf{x}_m\in\mathbb{R}^N$. The last layer, called the {\it output} layer, is the target values, or labels $\mathbf{y}_m\in\mathbb{R}^P$. The intervening layers of the NN, called {\it hidden} layers since the training data does not explicitly define their output, are $\mathbf{z}^{(1)}\in\mathbb{R}^Q$ and $\mathbf{z}^{(2)}\in\mathbb{R}^R$. The circles in the network (see Fig.~\ref{fig:fnn_binary}) represent network units.

The output of the network units in the hidden and output layers is a non-linear transformation of the inputs, called the {\it activation}. Common activation functions include softmax, sigmoid, hyperbolic tangent, and rectified linear units (ReLU).  Activation functions are further discussed in Sec.~\ref{sec:deep_learning}. Before the activation, a linear transformation is applied to the inputs
\begin{equation} \label{eq:fnn_act1}
a_q=\sum_{n=1}^N w_{nq}^{(1)}x_n+w_{q0}^{(1)},
\end{equation}
with $a_q$ the input to the $q$th unit of the first hidden layer, and $w_{nq}^{(1)}$ and $w_{q0}^{(1)}$ the weights and biases, which are to be learned. The output of the hidden unit $z_q^{(1)}=g_1(a_q)$, with $g_1$ the activation function. Similarly,
\begin{equation} \label{eq:fnn_act2}
\begin{aligned}
&a_r=\sum_{q=1}^Q w_{qr}^{(2)}z_q^{(1)}+w_{r0}^{(2)},\\
&a_p=\sum_{r=1}^R w_{rp}^{(3)}z_r^{(2)}+w_{p0}^{(3)},
\end{aligned}
\end{equation}
and $z_r^{(2)}=g_2(a_r)$, $y_p=g_3(a_p)$.

The NN architecture, combined with the series of small operations by the activation functions, make the NN a {\it general function approximator}. 
In fact, a NN with a single hidden layer can approximate any continuous function arbitrarily well with a sufficient number of hidden units.\cite{hornik1991approximation} We here illustrate a NN with two hidden layers. Deeper NN architectures are discussed in Sec.\ref{sec:deep_learning}.

NN training is analogous to the methods we have previously discussed (e.g. linear regression and SVM models): a loss function is constructed and gradients of the cost function are used to train the model. For NNs, a typical loss function, $L$, for classification is cross-entropy.\cite{goodfellow2016deep} Given the target values (labels) $\mathbf{S=[s}_1,\ldots,\mathbf{s}_m]\in\mathbb{R}^{P\times M}$ and input features $\mathbf{X}$, the average cross-entropy $L$  and weight estimate are given by
\begin{equation} \label{eq:fnn_cost}
\begin{aligned}
L(\mathbf{w})&=-\frac{1}{P}\sum_{p=1}^P\sum_{m=1}^M~s_{pm}\ln y_{pm},\\
\widehat{\mathbf{w}}&=\underset{\mathbf{w}}{\arg\min}~L(\mathbf{w}),
\end{aligned}
\end{equation}
with $\mathbf{w}$ the matrix of the weights and $\widehat{\mathbf{w}}$ its estimate. The gradient of the objective \eqref{eq:fnn_cost}, $\nabla L(\mathbf{w})$, is obtained via backpropagation.\cite{rumelhart1986learning} Backpropagation uses the derivative chain rule to find the gradient of the cost with respect to the weights at each NN layer. With backpropagation, any of the numerous variants of gradient descent can be used to optimize the weights at all layers. 

The gradient information from backpropagation is used to find the optimal weights. The simplest weight update is obtained by taking a small step in the direction of the negative gradient
\begin{equation} \label{eq:fnn_grad_desc}
\mathbf{w}^\text{new}=\mathbf{w}^\text{old}-\eta\nabla L(\mathbf{w}^\text{old}),
\end{equation}
with $\eta$ called the {\it learning rate}, which controls the step size. Popular NN training algorithms are  stochastic gradient descent\cite{goodfellow2016deep} and Adam (adaptive moment estimation).\cite{kingma2014adam}

The choice of activation functions for the hidden and output layers are determined by 4 important NN applications: binary classification, multi-class classification (classes don’t overlap), multi-label classification (classes overlap), regression. For all of these, modern architectures use ReLU for hidden layers (the number and sizes of hidden layers are determined by trials and errors). On a basic level, the architectures only differ in terms of output units (e.g. the final NN layer). These are sigmoid activation for binary classification, softmax for multi-label, multi sigmoid for multi-label, linear for regression. Loss functions should also be adapted accordingly.

NN models have been used extensively in acoustics. Specific applications are discussed in Sec.~\ref{subsec:dl_applications}.

\section{Unsupervised learning} \label{sec:unsup}
Unlike in supervised learning where there are given target values or labels $y_m$, unsupervised learning deals only with modeling the features $\mathbf{x}_m$, with the goal of discovering interesting or useful structures in the data. The structures of the data, represented by the data model parameters $\mathbf{\theta}$, give probabilistic unsupervised learning models of the form $p(\mathbf{X}|\mathbf{\theta})$. This is in contrast to supervised models that predict the probability of labels or regression values given the data and model: $p(\mathbf{Y}|\mathbf{X,\theta})$ (see Sec.~\ref{sec:sup_learn}). We note that the distinction between unsupervised and supervised learning methods is not always clear. Generally, a learning problem can be considered unsupervised if there are no annotated examples or prediction targets provided.

The structures discovered in unsupervised learning serve many purposes. The models learned can, for example, indicate how features are grouped or define latent representations of the data such as the subspace or manifold which the data occupies in higher-dimensional space. Unsupervised learning methods for grouping features include clustering algorithms such as K-means\cite{macqueen1967some} and Gaussian mixture models (GMMs). Unsupervised methods for discovering latent models include principal components analysis (PCA), matrix factorization methods such as non-negative matrix factorization (NMF),\cite{lee2001algorithms}
independent component analysis (ICA),\cite{hyvarinen2001} and dictionary learning.\cite{kreutz2003dictionary,elad2010,tosic2011,mairal2014} Neural network models, called autoencoders, are also used for learning latent models.\cite{goodfellow2016deep} Autoencoders can be understood as a non-linear generalization of PCA and, in the case of sparse regularization (see Sec.~\ref{sec:sup_learn}), dictionary learning.

The aforementioned models of unsupervised learning have many practical uses. Often, they are used to find the `best' representation of the data given a desired task. A special class of K-means based techniques, called vector quantization,\cite{gersho1991} was developed for lossy compression. In sparse modeling, dictionary learning seeks to learn the `best' sparsifying dictionary of basis functions for a given class of data. In ocean acoustics, PCA (a.k.a. empirical orthogonal functions) have been used to constrain estimates of ocean sounds speed profiles (SSPs), though methods based on sparse modeling and dictionary learning have given an alternative representation.\cite{bianco2016,bianco2017a} Recently, dictionary-learning based methods have been developed for travel time tomography.\cite{bianco2018b,bianco2019high} Aside from compression, such methods can be used for data restoration tasks such as denoising and inpainting. Methods developed for denoising and inpainting can also be extended to inverse problems more generally.

In the following, we illustrate unsupervised ML, highlighting PCA, EM with GMMs, K-means, dictionary learning, and autoencoders.

\subsection{ Principal components analysis}

For data visualization and compression, we are often interested in finding a subspace of the feature space which contains the most important feature correlations. This can be a subspace which contains the majority of the feature variance. PCA finds such a subspace by learning an orthogonal, linear transformation of the data. The {\it principal components} of the features are obtained as the right singular vector  of the design matrix $\mathbf{X}$ (or eigenvector of $\mathbf{X}^\mathrm{T}\mathbf{X}$) with
\begin{linenomath*}
\begin{equation}
\mathbf{X}^\mathrm{T}\mathbf{X}=\mathbf{P}\mathbf{\Sigma}^2\mathbf{P}^T.
\label{eq:pcaAnal}
\end{equation}
\end{linenomath*}
$\mathbf{P}=[\mathbf{p}_1,\ldots,\mathbf{p}_N]\in\mathbb{R}^{N\times N}$ are principal components (eigenvectors) and $\mathbf{\Sigma}^2=\rm{diag}([\sigma_1^2,\ldots,\sigma_N^2])\in\mathbb{R}^{N\times N}$ are the total variances of the data along the principal directions defined by principal components $\mathbf{p}_n$, with $\sigma_1^2\ge,\ldots,\ge \sigma_N^2$. This matrix factorization can be obtained using, for example, singular value decomposition.\cite{hastie2009}

In the coordinate system defined by $\mathbf{P}$, with axes $\mathbf{p}_n$, the first coordinate accounts for the highest portion of the overall variance in the data and subsequent axes have equal or smaller contributions.  Thus, truncating the resulting coordinate space results in a lower dimensional representation that often captures a large portion of the data variance.  This has benefits both for visualization of data and modeling as it can reduce the aforementioned curse of dimensionality (see Sec.~\ref{subsec:curse}).  Formally, the projection of the original features $\mathbf{X}$ onto the principal components $P$ is
\begin{linenomath*}
\begin{equation}
\mathbf{B}^\mathrm{T}=\mathbf{XQ},
\label{eq:pcaAnal2}
\end{equation}
\end{linenomath*}
with $\mathbf{Q}\in\mathbb{R}^{N\times P}$ the first $P$ eigenvectors and $\mathbf{B}=[\mathbf{\beta}_1,\ldots,\mathbf{\beta}_M]\in\mathbb{R}^{P\times M}$ the lower-dimensional projection of the data. $\mathbf{X}$ can be approximated by
\begin{linenomath*}
\begin{equation}
\widehat{\mathbf{X}}^\mathrm{T}\approx\mathbf{QB},
\label{eq:pcaAnal3}
\end{equation}
\end{linenomath*}
which give a compressed version data $\widehat{\mathbf{X}}$ with less information than the original data $\mathbf{X}$ (lossy compression).

PCA is a simple example of representation learning that attempts to disentangle the unknown factors generating the data variance. The principal variances quantify the importance of the features, and the principal components are a coordinate system under which the features are uncorrelated. While correlation is an important feature dependency, we often are interested in learning representations that can disentangle more complicated, perhaps correlated, dependencies.

\subsection{ Expectation maximization and Gaussian mixture models}

Often, we would like to model the dependency between observed features. An efficient way of doing this is to assume that the observed variables are correlated because they are generated by a hidden or {\it latent} model. Such models can be challenging to fit but offer advantages, including a compressed representation of the data. A popular latent modeling technique called  Gaussian mixture models (GMMs)\cite{mclachlan2000finite} models arbitrary probability distributions as a linear superposition of $K$ Gaussian densities.

The latent parameters of GMMs (and other mixture models) can be obtained using a non-linear optimization procedure called the expectation-maximization (EM) algorithm.\cite{dempster_maximum_1977} EM is an iterative technique which alternates between (1) finding the expected value of the latent factors given data and initialized parameters, and (2) optimizing parameter updates based on the latent factors from (1). We here derive EM in the context of GMMs and later show how it relates to other popular algorithms, like K-means.\cite{macqueen1967some}

For features $\mathbf{x}_m$, the GMM is
\begin{align} \label{eq:gmm1}
p(\mathbf{x}_m)=\sum_{k=1}^K\pi_{k}\mathcal{N}(\mathbf{x}_m|\mathbf{\mu}_k,\mathbf{\Sigma}_k),
\end{align}
with $\pi_{k}$ the weights of the Gaussians in the mixture, and $\mathbf{\mu}_k$ and $\mathbf{\Sigma}_k$ the mean and covariance of the $k$th Gaussian. The weights $\pi_{k}$ define the marginal distribution of a binary random vector $\mathbf{z}_m\in\{0,1\}^K$, which give membership of data vector $\mathbf{x}_m$ to the $k$th Gaussian ($z_{km}=1$ and $z_{im}=0~\forall~i\ne k$). 

The features $\mathbf{x}_m$ are related to the latent vector $\mathbf{z}_m$ and the parameters $\mathbf{\theta}=\{\mathbf{\pi}_k,\mathbf{\mu}_k,\mathbf{\Sigma}_k\}$ via conditional and joint distributions. The conditional distribution $p(\mathbf{x}_m|\mathbf{\theta})$ is obtained using the sum rule \eqref{eq:rule_sum}),
\begin{align} \label{eq:gmm4}
p(\mathbf{x}_m|\mathbf{\theta})=\sum_{\mathbf{z}_m}p(\mathbf{x}_m|\mathbf{z}_m,\mathbf{\theta})p(\mathbf{z}_m|\mathbf{\theta})=\sum_{\mathbf{z}_m}p(\mathbf{x}_m,\mathbf{z}_m|\mathbf{\theta}).
\end{align}
To find the parameters, the log-likelihood or $p(\mathbf{x}_m|\mathbf{\theta})$ is maximized over observations $\mathbf{X}=[\mathbf{x}_1^\mathrm{T},\ldots,\mathbf{x}_M^\mathrm{T}]$
\begin{align} \label{eq:gmm_loglike2}
\ln p(\mathbf{X}|\mathbf{\theta})=\sum_{m=1}^M \ln \bigg\{ \sum_{\mathbf{z}_m}p(\mathbf{x}_m,\mathbf{z}_m|\mathbf{\theta})\bigg\}.
\end{align}
Eq.~\eqref{eq:gmm_loglike2} is challenging to optimize because the logarithm cannot be pushed inside the summation over $\mathbf{z}_m$.

In EM, a complete data log likelihood
\begin{align} \label{eq:em1}
L(\mathbf{\theta})=\sum_{m=1}^M \ln p(\mathbf{x}_m,\mathbf{z}_m|\mathbf{\theta})
\end{align}
is used to define an auxiliary function, $Q(\mathbf{\theta,\theta}^{\text{old}})=\mathbb{E}[L(\mathbf{\theta})|\mathbf{\theta}^\text{old}]$, which is the expectation of the likelihood evaluated assuming some knowledge of the parameters. The knowledge of the parameters is based on the previous or `old' values, $\mathbf{\theta}^\text{old}$. The EM algorithm is derived using the auxiliary function. For more details, please see Ref.~\onlinecite{murphy2012}~[pp.~350--354]. Helpful discussion is also presented in Ref.~\onlinecite{bishop2006}~[pp.~430--443]

The first step of EM, called the E-step (for expectation), estimates the responsibility $r_{km}$ of the $k$th Gaussian in reconstructing the $m$th data density $p(\mathbf{x}_m)$ given the current parameters $\theta$. From Bayes' rule, the E-step is

\begin{equation} \label{eq:em5}
r_{km}=\frac{\pi_k^\text{old}\mathcal{N}(\mathbf{x}_m|\mathbf{\mu}_k^\text{old},\mathbf{\Sigma}_k^\text{old})}{\sum_{j=1}^K \pi_j^\text{old}\mathcal{N}(\mathbf{x}_m|\mathbf{\mu}_j^\text{old},\mathbf{\Sigma}_j^\text{old})}
\end{equation}

The second step of EM, called the M-step, updates the parameters by maximizing the auxiliary function, $\mathbf{\theta}^{\text{new}}=\underset{\mathbf{\theta}}{\arg\max}~Q(\mathbf{\theta,\theta}^\text{old})$, with the responsibilities $r_{km}$ from the E-step \eqref{eq:em5}.\cite{bishop2006,ng2000cs229} The M-step estimates of $\mathbf{\pi}$ (using also $\sum_{k=1}^K\pi_{k}=1$), $\mathbf{\mu}_k$, and $\mathbf{\Sigma}_k$ are
\begin{align}\label{eq:em6}
\pi_k^\text{new}&=\frac{1}{M}\sum_{m=1}^Mr_{km}=\frac{r_k}{M},\nonumber\\
\mathbf{\mu}_k^\text{new}&=\frac{1}{r_k}\sum_{m=1}^Mr_{km}\mathbf{x}_m,\\
\mathbf{\Sigma}_k^\text{new}&=\frac{1}{r_k}\sum_{m=1}^Mr_{km}(\mathbf{x}_m-\mathbf{\mu}_k^\text{new})(\mathbf{x}_m-\mathbf{\mu}_k^\text{new})^\mathrm{T},\nonumber
\end{align}
with $r_k=\sum_{m=1}^Mr_{km}$ the weighted number of points with membership to centroid $k$. The EM algorithm is run until an acceptable error has been obtained. The error can be obtained for example by evaluating the log likelihood \eqref{eq:gmm_loglike2} with the estimated parameters \eqref{eq:em6}. 

We note that singularities can arise in the maximum likelihood approach to EM, presented here. If only one data point is assigned to a Gaussian (and there is more than one Gaussian), the log likelihood function \eqref{eq:gmm_loglike2} goes to infinity as the variance of the Gaussian component with a single data point goes to zero. This does not occur in a Bayesian formulation.

In EM the objective function is not convex and solutions often can get caught in local minima. These issues can be corrected, in part, using multiple parameter initializations and choosing the results with the smallest residual. In ML, local minima are a common challenge as optimization objectives are rarely convex. This is an especially large issue in DL and has driven significant development in DL algorithms (see Sec.~\ref{sec:deep_learning}).

GMMs (EM) have been used extensively in acoustics. A few of the applications include source localization, separation, and speech enhancement.\cite{vincent2018audio} These applications are further discussed in Sec.~\ref{sec:speech}. GMMs have also been used in animal vocalization classification.\cite{roch2007gaussian}

\subsection{ K-means}

The K-means algorithm\cite{macqueen1967some} is a method for discovering clusters of features in unlabeled data. The goal of doing this can be to estimate the number of clusters or for data compression (e.g. vector quantization\cite{gersho1991}). Like EM, K-means solves \eqref{eq:gmm_loglike2}. Except, unlike EM, $\mathbf{\pi}_k=1/K$ and $\mathbf{\Sigma}_k=\sigma^2 \mathbf{I}$ are fixed. Rather than responsibility $r_{km}$ describing the posterior distribution of $\mathbf{z}_m$ (per \eqref{eq:em5}), in K-means the membership is a `hard' assignment (in the limit $\sigma\to 0$, please see Ref.~\onlinecite{bishop2006} for more details):
\begin{equation}
r_{km}=\bigg\{
\begin{matrix}
\ 1\ \ \text{if}\ \  \widehat{k}=\underset{k}{\argmin}\|\mathbf{x}_m-\mathbf{\mu}_k^\text{old}\|_2 \\
\hspace{-16ex} 0\ \ \text{otherwise}.
\end{matrix}
\label{eq:kmeans1}
\end{equation}
Thus in K-means, each feature vector $\mathbf{x}_m$ is assigned to the nearest centroid $\mathbf{\mu}_k$. The distance measure is the Euclidian distance (defined by the $\ell_2$-norm, \eqref{eq:kmeans1}). Based on the centroid membership of the features, the centroids are updated using the mean of the feature vectors in the cluster
\begin{align} \label{eq:kmeans2}
\mathbf{\mu}_k^{\text{new}}=\frac{1}{r_k}\sum_{i:r_{ki}=1}\mathbf{x}_i.
\end{align}
Sometimes the variances are also calculated. Thus, K-means is a two-step iterative algorithm which alternates between categorizing the features and updating the centroids. Like EM, K-means must be initialized, which can be done with random initial assignments. The number of clusters can be estimated using, for example, the gap statistic.\cite{hastie2009}

\subsection{ Dictionary learning} \label{sec:dictlearn}

In this section we introduce {\it dictionary learning} and discuss one classic dictionary learning method: the K-SVD algorithm.\cite{aharon2006} An important task in sparse modeling (see Sec.~\ref{sec:sup_learn}) is obtaining a dictionary which can well model a given class of signals. There are a number of methods for dictionary design, which can be divided roughly into two classes: analytic and synthetic. Analytic dictionaries have columns, called {\it atoms}, which are derived from analytic functions such as wavelets or the discrete cosine transform (DCT).\cite{mallat1999,elad2010} Such dictionaries have useful properties, which allow them to obtain acceptable sparse representation performance for a broad range of data. However, if enough training examples of a specific class of data are available, a dictionary can be synthesized or learned directly from the data. Learned dictionaries, which are designed from specific instances of data using dictionary learning algorithms, often achieve greater reconstruction accuracy over analytic, generic dictionaries. Many dictionary learning algorithms are available.\cite{mairal2014}

As discussed in Sec.~\ref{sec:sup_learn}, sparse modeling assumes that a few (sparse) atoms from a dictionary $\mathbf{D}\in\mathbb{R}^{N\times K}$ can adequately construct a given feature $\mathbf{x}_m$. With coefficients $\mathbf{\beta}_m\in\mathbb{R}^K$, this is articulated as $\mathbf{x}_m\approx\mathbf{D\beta}_m$. The coefficients can be solved by
\begin{equation}
\begin{aligned}
\widehat{\mathbf{\beta}}_m= \underset{\mathbf{\beta}_m}{\arg\min} \ \|\mathbf{D}\mathbf{\beta}_{m}-\mathbf{x}_m\|_2^2~\text{subject to} \ \|\mathbf{\beta}_{m}\|_0=T,
\label{eq:dlearn1}
\end{aligned}
\end{equation}
with $T$ the number of non-zero coefficients. The penalty $\|\cdot\|_0$ is the $\ell_0$-pseudo-norm, which counts the number of non-zero coefficients. Since least square minimization with an $\ell_0$-norm penalty  is non-convex (combinatorial), solving \eqref{eq:dlearn1} exactly is often impractical. However, many fast-approximate solution methods exist, including orthogonal matching pursuit (OMP)\cite{elad2010} 
and sparse Bayesian learning (SBL).\cite{wipf2004}

Eq.~\eqref{eq:dlearn1} can be modified to also solve for the dictionary\cite{elad2010}
\begin{equation}
\begin{aligned}
\widehat{\mathbf{B}},\widehat{\mathbf{D}}= \underset{\mathbf{D}}{\arg\min}&\big\{\underset{\mathbf{\beta}_m}{\arg\min} \ \|\mathbf{D}\mathbf{\beta}_{m}-\mathbf{x}_m\|_2^2 \\
&\text{subject to} \ \|\mathbf{\beta}_{m}\|_0=T~\forall{m}\big\},
\label{eq:dlearn2}
\end{aligned}
\end{equation}
with $\mathbf{B=[\beta}_1,\ldots,\mathbf{\beta}_M]$ the coefficients for all examples. Eq.~\eqref{eq:dlearn2} is a bi-linear optimization problem for which no general practical algorithm exists.\cite{elad2010} However, it can be solved well using methods related to K-means. Clustering-based dictionary learning methods\cite{mairal2014} are based on the alternating optimization concept introduced in K-means and EM. The operations of a dictionary learning algorithm are (1) sparse coding given dictionary $\mathbf{D}$, and (2) dictionary update based coefficients $\mathbf{B}$.

This assumes an initial dictionary (the columns of which can be Gaussian noise). Sparse coding can be accomplished by OMP or other greedy methods. The dictionary update stage can be approached in a number of ways. We next briefly describe the class K-SVD dictionary learning algorithm\cite{aharon2006,elad2010} to illustrate basic dictionary learning concepts. Like K-means, K-SVD learns $K$ prototypes of the data (in dictionary learning these are called atoms, where in K-means they are called centroids) but, instead of learning them as the means of the data `clusters', they are found using the SVD since there may be more than one atom used per data point.

In the K-SVD algorithm, dictionary atoms are learned based on the SVD of the reconstruction error caused by excluding the atoms from the sparse reconstruction. For more details please see Ref.~\onlinecite{elad2010}.

Expressing the dictionary coefficients as row vectors $\mathbf{\beta}_T^n\in\mathbb{R}^N$ and $\mathbf{\beta}_T^j\in\mathbb{R}^N$, which relate all examples $\mathbf{X}$ to $\mathbf{d}_n$ and $\mathbf{d}_j$, respectively, the $\ell_2$-penalty from \eqref{eq:dlearn2} is rewritten as
%
\begin{align}
\|\mathbf{X}^\mathrm{T}-\mathbf{D}\mathbf{B}\|^2_\mathcal{F} =\left\|\mathbf{X}^\mathrm{T}-\sum_{k=1}^K\mathbf{d}_k\mathbf{\beta}^k_T\right\|^2_\mathcal{F}
 = \|\mathbf{E}_j-\mathbf{d}_j\mathbf{\beta}^j_T\|^2_\mathcal{F},
\label{eq:dlearn3}
\end{align}
%
where
\begin{linenomath*}
\begin{equation}
\mathbf{E}_j = \bigg{(}\mathbf{X}^\mathrm{T}-\sum_{k\ne j}\mathbf{d}_k\mathbf{\beta}^k_T\bigg{)},
\label{eq:dlearn4}
\end{equation}
\end{linenomath*}
and $\|\cdot\|_\mathcal{F}$ is the Frobenius norm.

An update to the dictionary entry $\mathbf{d}_j$ and coefficients $\mathbf{\beta}_T^j$ which minimizes \eqref{eq:dlearn3} is found by taking the SVD of $\mathbf{E}_j$. However, many of the entries in $\mathbf{\beta}_T^j$ are zero (corresponding to examples which do not use $\mathbf{d}_j$). To properly update $\mathbf{d}_j$ and $\mathbf{\beta}_T^j$ with SVD, \eqref{eq:dlearn3} must be restricted to examples $\mathbf{x}_m$ which use $\mathbf{d}_j$
\begin{linenomath*}
\begin{equation}
\|\mathbf{E}_j^R-\mathbf{d}_j\mathbf{\beta}^j_R\|^2_\mathcal{F},
\label{eq:restricted}
\end{equation}
\end{linenomath*}
where $\mathbf{E}_j^R$ and $\mathbf{\beta}_R^j$ are entries in $\mathbf{E}_j$ and $\mathbf{\beta}_T^j$, respectively, corresponding to examples $\mathbf{x}_m$ which use $\mathbf{d}_j$.
Thus for each K-SVD iteration, the dictionary entries and coefficients are sequentially updated as the SVD of $\mathbf{E}^R_j=\mathbf{USV}^{\rm{T}}$. The dictionary entry $\mathbf{d}_j^i$ is updated with the first column in $\mathbf{U}$ and the coefficient vector $\mathbf{\beta}^j_R$ is updated as the product of the first singular value $\mathbf{S}(1,1)$ with the first column of $\mathbf{V}$.

\begin{figure}[t]
  \scriptsize
  \centering
  \includegraphics[width=0.8\linewidth]{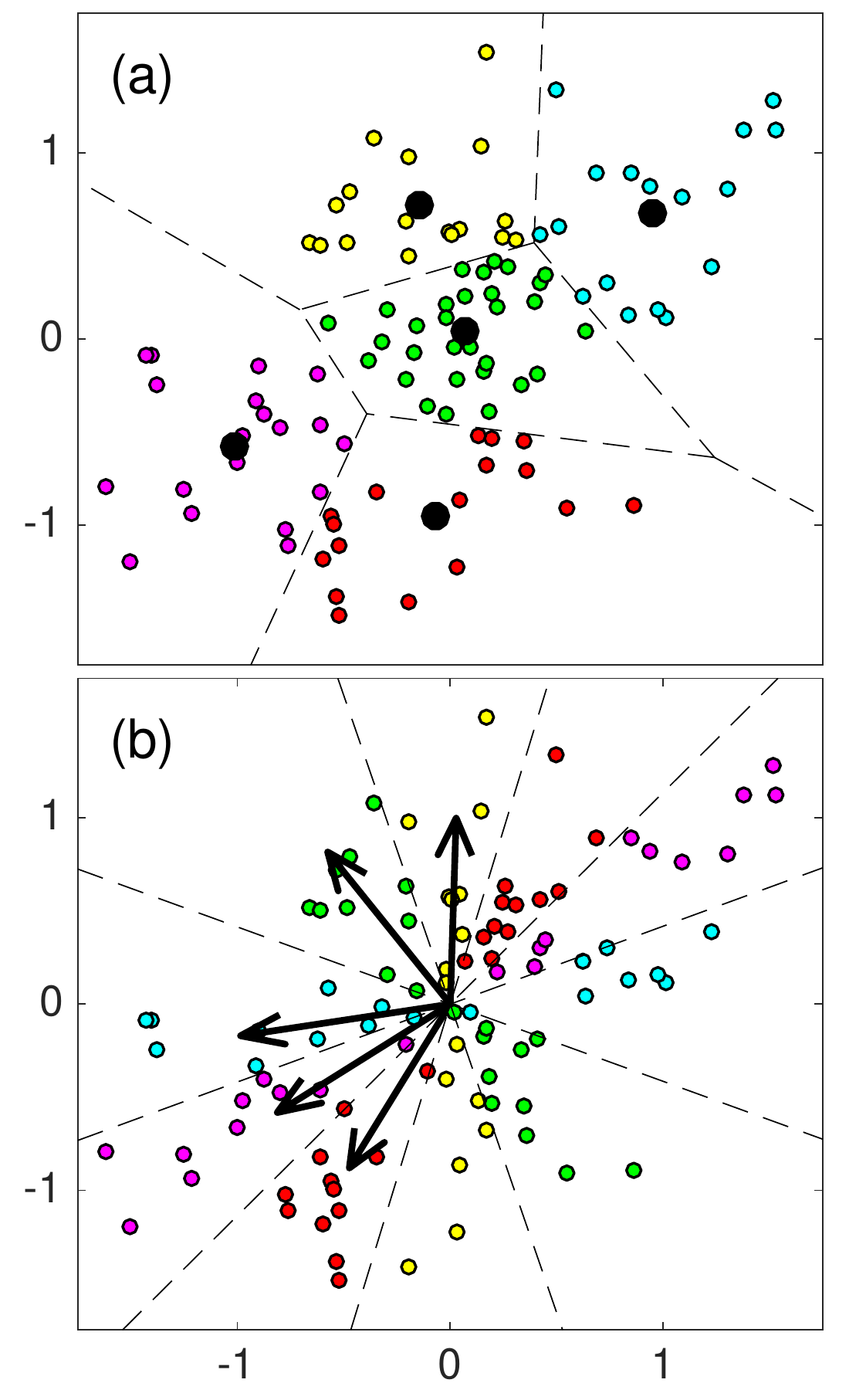}
  \caption{(Color online) Partitioning of Gaussian random distribution using (a) K-means with 5 centroids and (b) K-SVD dictionary learning with $T=1$ and 5 atoms. In K-means, the centroids define Voronoi cells which divide the space based on Euclidian distance. In K-SVD, for $T=1$, the atoms define radial partitions based on the inner product of the data vector with the atoms. Reproduced from \onlinecite{bianco2017a}
  }
  \label{fig:kmeans_dlearn}
\end{figure}

For the case when $T=1$, the results of K-SVD reduces to the K-means based model called gain-shape vector quantization.\cite{gersho1991,elad2010}
When $T=1$, the $\ell_2$-norm in \eqref{eq:dlearn2} is minimized by the dictionary entry $\mathbf{d}_n$ that has the largest inner product with example $\mathbf{x}_m$.\cite{elad2010}  Thus for $T=1$, $[\mathbf{d}_1,\ldots, \mathbf{d}_N]$ define radial partitions of $\mathbb{R}^K$. These partitions are shown in Fig.~\ref{fig:kmeans_dlearn}(b) for a hypothetical 2D ($K=2$) random data set.

Other clustering-based dictionary learning methods are the method of optimal directions\cite{engan2000} and the iterative thresholding and signed K-means algorithm.\cite{schnass2015} Alternative methods include online dictionary learning.\cite{mairal2009}

Dictionary learning has been applied in a number of acoustics problems. The applications include acoustic signal denoising\cite{taroudakis2015noising}, geophysical parameter compression (ocean acoustics)\cite{bianco2017a}, seimic tomography \cite{zhu2015seismic,bianco2018b}, and damage detection. \cite{alguri2018baseline}

\subsection{Autoencoder networks} \label{sec:autoencoders}
Autoencoder networks are a special case of NNs (Sec.~\ref{sec:sup_learn}), in which the desired output is an approximation of the input. Because they are designed to only approximate their input, autoencoders prioritize which aspects of the input should be copied. This allows them to learn useful properties of the data. Autoencoder NNs are used for dimensionality reduction and feature learning, and they are a critical component of modern generative modeling.\cite{goodfellow2016deep} They can also be used as a pretraining step for DNNs (see Sec.~\ref{subsec:unsup_pretr}). They can be viewed as a non-linear generalization of PCA and dictionary learning. Because of the non-linear encoder and decoder functions, autoencoders potentially learn more powerful feature representations than PCA or dictionary learning.

Like feed-forward NNs (Sec.~\ref{subsec:nn}), activation functions are used on the output of the hidden layers (Fig.~\ref{fig:fnn_binary}). In the case of an autoencoder with a single hidden layer, the input to the hidden layer is $\mathbf{z}_1=g_1(a_q(\mathbf{x}))$ and the output is $\widehat{\mathbf{x}}=g_2(a_p(\mathbf{z}_1))$, with $P=M$ (see Fig.~\ref{fig:fnn_binary}). The first half of the NN, which maps the inputs to the hidden units is called the {\it encoder}. The second half, which maps the output of the hidden units to the output layer (with same dimension $N$ of input features) is called the {\it decoder}. The features learned in this single layer network are the weights of the first layer.

If the code dimension is less than the input dimension, the autoencoder is called {\it undercomplete}. In having the code dimension less than the input, undercomplete networks are well suited to extract salient features since the representation of the inputs is `compressed', like in PCA. However, if too much capacity is permitted in the encoder or decoder, undercomplete autoencoders will still fail to learn useful features.\cite{goodfellow2016deep}

Depending on the task, code dimension equal to or greater than the inputs is desireable. Autoencoders with code dimension greater than the input dimension are called {\it overcomplete} and these codes exhibit redundancy similar to overcomplete dictionaries and CNNs. This can be useful for learning shift invariant features. However, without regularization, such autoencoder architectures will fail to learn useful features. Sparsity regularization, similar to dictionary learning, can be used to train overcomplete autoencoder networks.\cite{goodfellow2016deep} For more details and discussion, please see Sec.~\ref{sec:deep_learning}.

Like other unsupervised methods, autoencoders can be used to find transformations of the parameters for data interpretation and visualization. They can also be used for feature extraction in conjunction with other ML methods. Applications of autoencoders in acoustics include speech enhancement\cite{araki2015exploring} and acoustic novelty detection.\cite{marchi2017deep}

\section{Deep learning}
\label{sec:deep_learning}

Deep learning (DL) refers to ML techniques that are based on a cascade of non-linear feature transforms
trained during a learning step.\cite{deng2014deep} In several scientific fields,
decades of research and engineering have led to elegant ways to model data. Nevertheless,
the DL community argues that these models often do not have enough capacity to capture the subtleties
of the phenomena underlying data and are perhaps too specialized. And often it is beneficial to learn the representation directly from a large collection of examples using high-capacity ML models.
DL leverages a fundamental concept shared by many successful handcrafted features: all analyze the data by applying filter banks at different scales. These multi-scale representattions include Mel frequency cepstrum
used in speech processing, multi-scale wavelets,\cite{mallat1989theory}
and scale invariant feature transform (SIFT)\cite{lowe1999object} used in image processing. DL
mimics these processes by learning a cascade of features capturing information at different levels of abstraction. Non-linearities between these features allow deep NNs (DNNs) to learn complicated manifolds.
Findings in neuroscience also suggest that mammal brains process information in a similar way.

In short, a NN-based ML pipeline is considered DL if it satisfies:\cite{deng2014deep}
(i) features are not handcrafted but learned,
(ii) features are organized in a hierarchical manner from low-~to high-level abstraction,
(iii) there are at least two layers of non-linear feature transformations.
As an example, applying DL on a large corpus of conversational text must uncover meanings behind words, sentences and paragraphs (low-level) to further extract
concepts such as lexical field, genre, and writing style (high-level).

To comprehend DL, it is useful to  look at what it is not.
MLPs with one hidden layer (aka, shallow NNs)
are not deep as they only learn one level of feature extraction. Similarly, non-linear SVMs are analogous to shallow NNs.
Multi-scale wavelet representations\cite{mallat2016understanding} are a hierarchy of features (sub-bands) but
the relationships between features are linear.
When a NN classifier is trained on (hand-engineerd) transformed data,
the architecture can be deep, but it is not DL as the first transformation
is not learned.
 
Most DL architectures are based on DNNs, such as MLPs, and their early development can be traced to the 1970-80s. 
Three decades after this early development, only a few deep architectures emerged. And these architectures were limited to process data of no more than a few hundred dimensions. Successful examples developed over this intervening period are the two
handwritten digit classifiers: Neocognitron\cite{fukushima1980neocognitron} and LeNet5.\cite{lecun1998}
Yet the success of DL  started
at the end of the 2000s on what is called the third wave of artificial NNs.
This  success is attributed to the large increase in  available data
and computation power, including parallel architectures and GPUs.
Nevertheless,  several open-source DL toolboxes\citep{Torch,tensorflow2015,chollet2015keras,vedaldi2015matconvnet}
 have helped the community in introducing
a multitude of new strategies. These aim at fighting the limitations of back-propagation:
its slowness and tendency to get trapped in poor stationary points
(local optima or saddle points). The following subsections describe some of these strategies, see
Ref.~\onlinecite{goodfellow2016deep}
for an exhaustive review.

\begin{figure}[t]
  \scriptsize
  \centering
  \includegraphics[width=\linewidth]{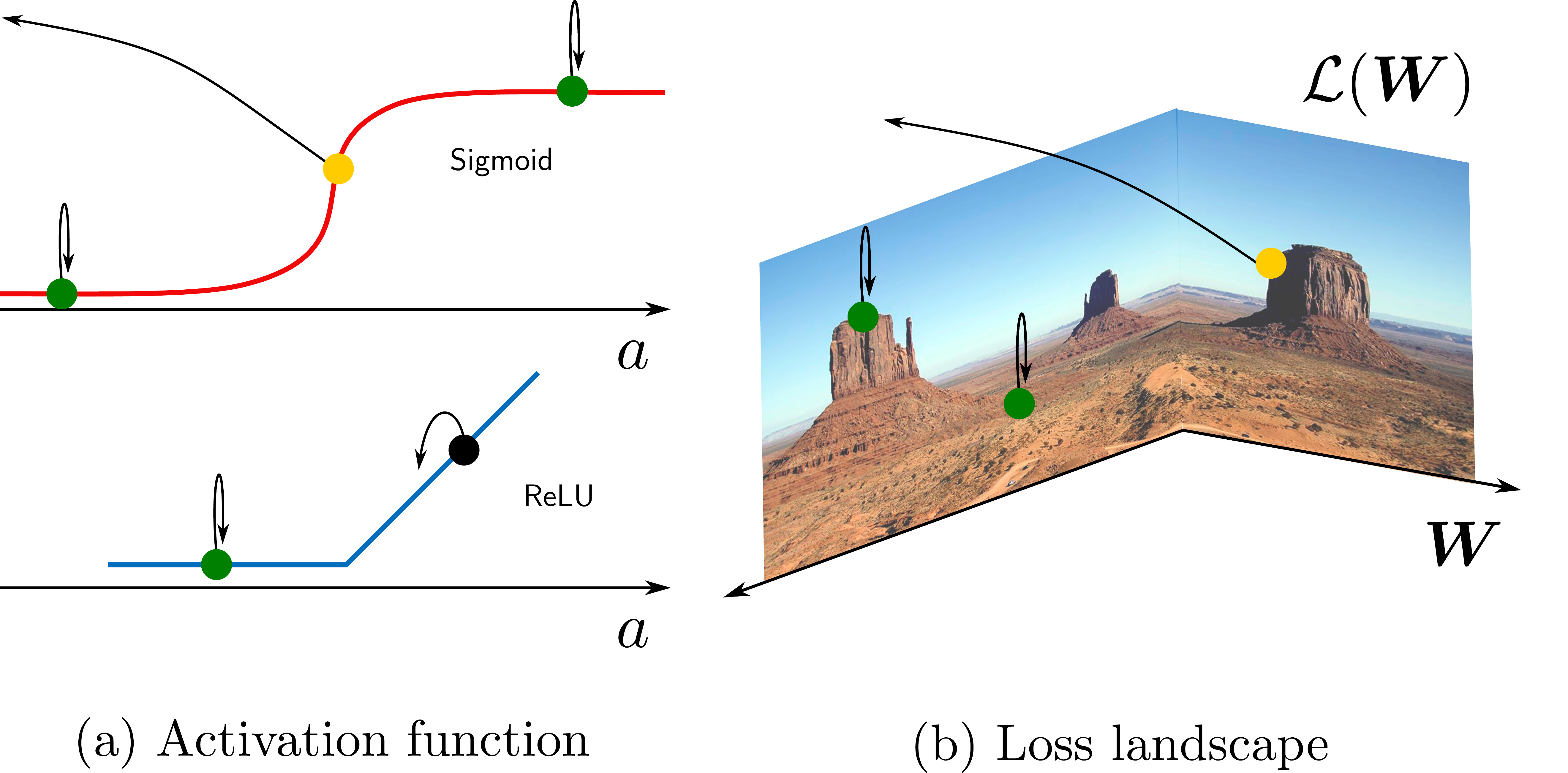}
  \caption{(Color Online) Illustration of the vanishing and exploding gradient problems. (a) The sigmoid and ReLU activation functions. (b) The loss $L$ as a function of the network weights $\bf W$ when
  using sigmoid activation functions is shown as a `landscape'.
    Such landscapes are hilly with large plateaus delimited by cliffs.
    Gradient-based updates (arrows)
    vanish on plateaus (green dots) and explode
    on cliffs (yellow dots). On the other hand, by using ReLU, backpropagation is less subject to the exploding gradient problem as there are fewer plateaus and cliffs in the associated cost landscape.
  }
  \label{fig:gradient_problems}
\end{figure}

\begin{figure*}
  \scriptsize
  \centering
  \includegraphics[width=5in]{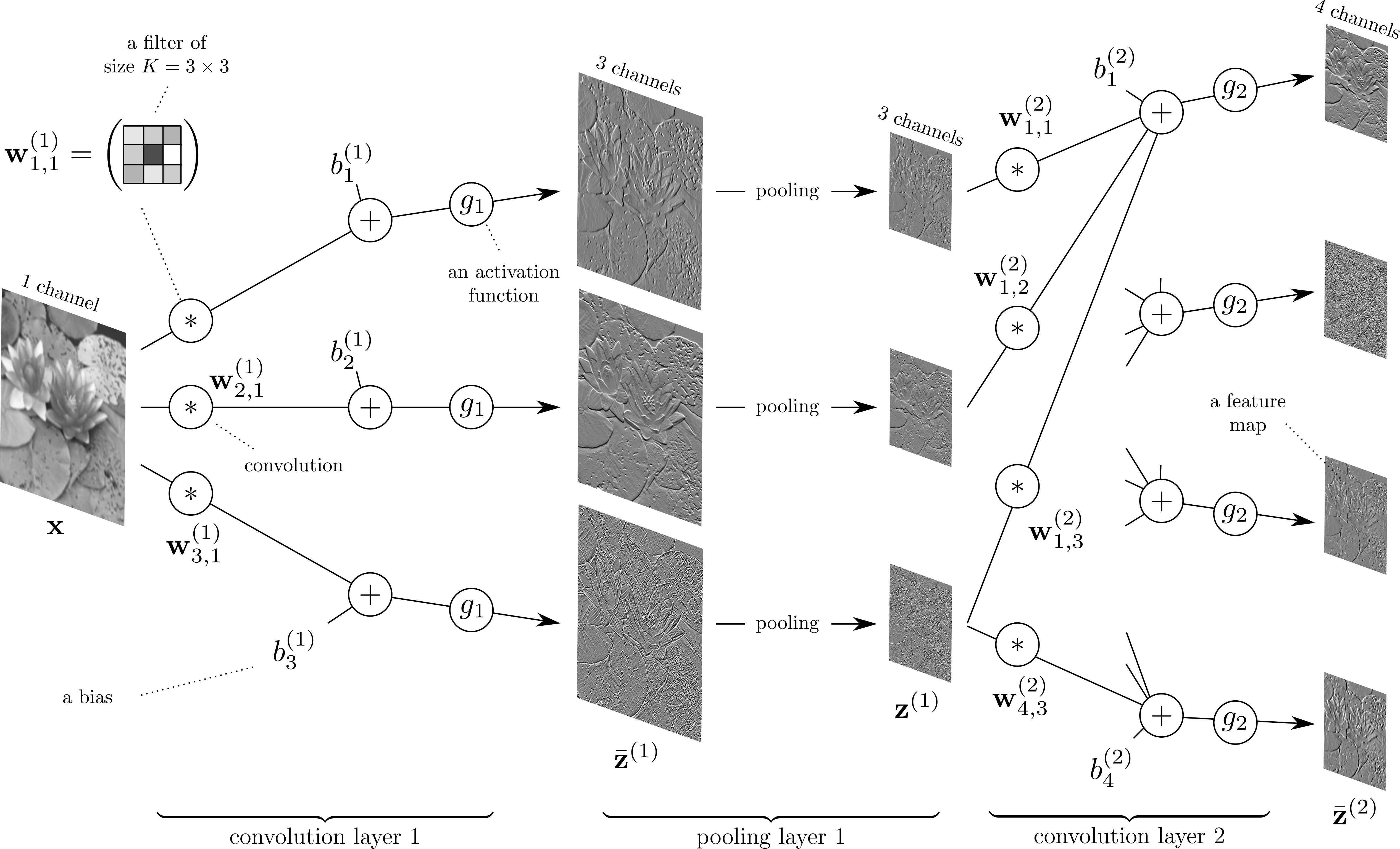}
  \caption{
  The first layer of a traditional CNN.
  For this illustration we chose a first hidden layer extracting 3
  feature maps.
  The filters have the size $K = 3 \times 3$.
  }
  \label{fig:CNN_two_first_layers}
\end{figure*}

\subsection{Activation Functions and Rectifiers}

The earliest multi-layer NN used logistic sigmoids
(Sec.~\hyperlink{sec:mlp}{\ref*{sec:sup_learn}-c}) or hyperbolic tangent for the non-linear activation function $g$:
\begin{align}
  {\bf z}^l_i = g({\bf a}^l_i)
  \quad\text{where}\quad
  {\bf a}^l = {\bf W}^l {\bf z}^{l-1} + {\bf b}^l~,
\end{align}
where ${\bf z}^l$ is the
vector of features at layer $l$ and ${\bf a}^l$ are the vector of potentials
(the affine combination of the features from the previous layer).
For the sigmoid  activation function in Fig.~\ref{fig:gradient_problems}(a), the derivative  is significantly
non-zero for only $a$ near $0$. With such
functions, in a randomly initialized NN, half of the hidden units are expected to activate ($f(a)>0$) for a given
training example, but only a few will influence the gradient, as $a\gg 0$.
In fact, many hidden units will have  near-zero gradient for all training samples,
and the parameters responsible for those units will be slowly updated.
This is called the {\it vanishing gradient problem}.
A na\"{i}ve repair to the problem is to increase the learning rate.
However, parameter updates will become too large for small $a$.  
Due to this, the overall training procedure might be unstable: this is the {\it gradient exploding problem}.
Fig.~\ref{fig:gradient_problems}(b) indicates of these two problems.
Shallow NNs are not necessarily susceptible to these
these problems, but they become harmful in DNNs.
Back-propagation with the aforementioned activation functions in DNNs
is slow, unstable, and leads to poor solutions.

Alternative activations have been developed to address these issues. One important class is rectifier units. Rectifiers are activation functions that are zero-valued for negative-valued inputs and linear for positive-valued inputs. Currently, the most popular  is the
Rectifier Linear Unit (ReLU),\cite{nair2010rectified} defined as (see Fig.~\ref{fig:gradient_problems}):
\begin{align}
  g(a) = {\rm ReLU}(a) \triangleq \max(a, 0)~.
\end{align}
While the derivative is zero for negative potentials $a$, the derivative
is one for $a>0$ (though non-differentiable at 0,
ReLU is continuous and then back-propagation is a sub-gradient descent). Thus,
in a randomly initialized NN, half of the hidden units fire and 
influence the gradient, and half do not fire (and do not influence the gradient). If the weights are randomly initialized with zero-mean and variance that preserves the range of variations of all potentials
across all NN layers, most units get significant gradients from at least
half of the training samples, and all parameters in the NN are expected
to be equally updated at each epoch.\cite{glorot2010understanding,he2015delving}
In practice, the use of rectifiers leads to tremendous improvement  in convergence.
Regarding exploding gradients, an efficient solution called gradient clipping\cite{pascanu2012understanding}
simply consists in thresholding the  gradient.

\subsection{End-to-End Training}
 \label{subsec:unsup_pretr}

While important for successful DL models, only addressing vanishing or exploding gradient problems is not alone enough for back-propagation. It is also important to avoid poor stationary points in DNNs. Pioneering methods for avoiding these stationary points included
training DNNs by successively training
shallow architectures in an unsupervised way.\cite{hinton2006fast,bengio2007greedy} Because the individual layers in this case are initially trained sequentially, using the output of preceding layers without optimizing jointly the weights of the preceding layer, these approaches are termed as {\it greedy layer-wise unsupervised pretraining}.

However, the benefits of unsupervised pretraining are not always clear. Many modern DL approaches prefer to train networks end-to-end, training all the network layers jointly from initialization instead of first training the individual layers.\cite{goodfellow2016deep} They rely on variants of gradient descent that aim at fighting
poor stationary solutions. These approaches include
stochastic gradient descent,
adaptive learning rates\cite{duchi2011adaptive},
and momentum techniques.\cite{sutskever2013importance}
Among these concepts, two main notions emerged:
(i) {\it annealing} by randomly exploring configurations first and exploiting them next,
(ii) {\it momentum} which forms a moving average of the negative gradient called velocity.
This tends to give faster learning, especially for noisy gradients or high-curvature loss functions.

Adam\cite{kingma2014adam} is based on adaptive learning rate and moment estimation. It is currently the most popular optimization approach for DNNs. Adam updates each weight $w_{i,j}$ at each step $t$ as follows:
\begin{equation} {w}_{i,j}^{(t+1)}={w}_{i,j}^{(t)}-\frac{\eta}{\sqrt{\hat{v}_{i,j}^{(t)}} + \epsilon} \hat{m}_{i,j}^{(t)},
\end{equation}
with $\eta > 0$ the learning rate, $\epsilon>0$ a smoothing term, and $\hat{m}_{i,j}^t$ and $\hat{v}_{i,j}^t$ the first and second moment of the velocity estimated,
for $0 < \beta_1 < 1$
and $0 < \beta_2 < 1$, as:
\begin{align}
    \hat{m}_{i,j}^{(t)} &= \frac{m_{i,j}^{(t)}}{1 - \beta_1^t},
    \quad
    \hat{v}_{i,j}^{(t)} = \frac{v_{i,j}^{(t)}}{1 - \beta_2^t},
    \\
    m_{i,j}^{(t)} &= \beta_1 m_{i,j}^{(t-1)} + (1 - \beta_1) \frac{\partial L(\mathbf{W}^\text{(t)})}{\partial w_{i,j}^{(t)}},
    \\
    v_{i,j}^{(t)} &= \beta_2 v_{i,j}^{(t-1)} + (1 - \beta_2)
    \left(\frac{\partial L(\mathbf{W}^\text{(t)})}{\partial w_{i,j}^{(t)}}\right)^2.
    \end{align}

\begin{figure*}[t]
  \scriptsize
  \centering
  \includegraphics[width=1.\linewidth]{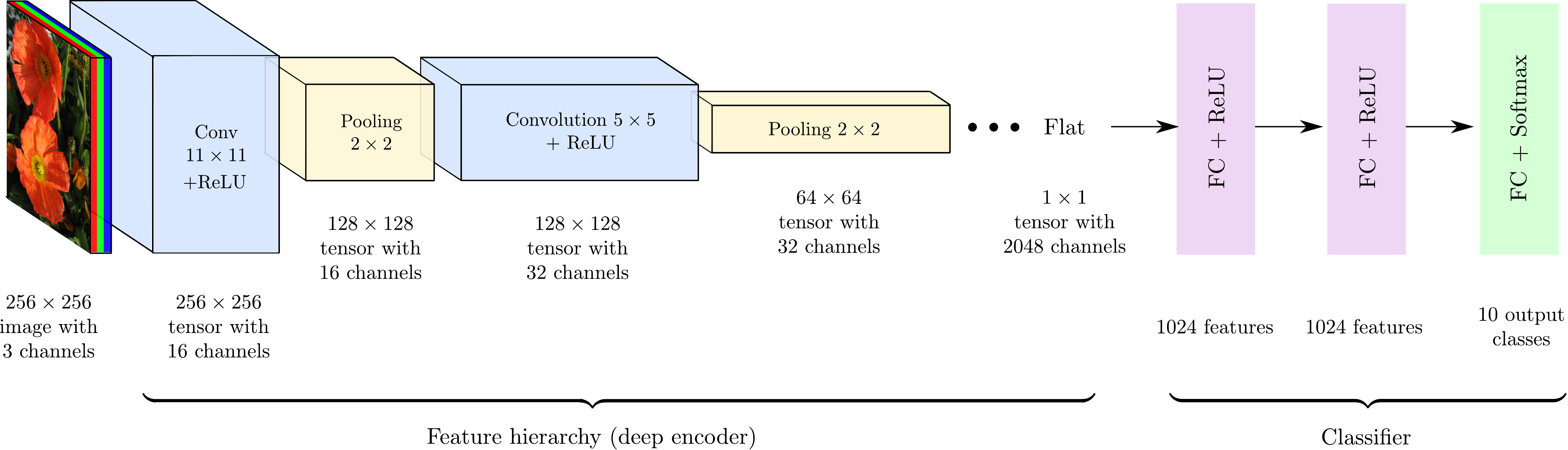}
  \caption{(Color Online) Deep CNN architecture for
    classifying one image into one of ten possible classes.
    Convolutional layers create redundant information by increasing
    the number of channels in the tensors.
    ReLU is used to capture non-linearity in the data.
    Max-pooling operations reduce spatial dimension to get abstraction
    and robustness relative to the exact location of objects.
    When the tensor becomes flat ({\it i.e.}, the spatial dimension is
    reduced to $1 \times 1$), each coefficient serves
    as input to a fully connected NN classifier. The feature dimensions, filter
    sizes, and  number of output classes are only for illustration.}
  \label{fig:deep_CNN_classifier}
\end{figure*}

Gradient descent methods can fall into the local minima near the parameter initialization, which
leads to underfitting. On the contrary, stochastic gradient descent and variants
are expected to find solutions with  lower loss and are more
prone to overfitting.
Overfitting occurs when learning a model
with many degrees of freedom compared to the number
of training samples. The curse of dimensionality (Sec.~\ref{subsec:curse}) claims that, without assumptions on the data,
the number of training data should grow exponentially with the number of
free parameters.
In classical NNs,
an output feature is influenced by all input features,
a layer is fully-connected (FC).
Given an input of size $N$ and a feature vector of size $P$,
a FC layer is then composed of $N \times (P+1)$ weights (including a bias term, see Sec.~\hyperlink{sec:mlp}{\ref*{sec:sup_learn}-c}).
Given that the signal size $N$ can be large, FC NNs are prone to overfitting.
Thus, special care should be taken for initializing the weights,\cite{glorot2010understanding,he2015delving} and
specific strategies must be employed to have some regularization,
such as dropout\cite{srivastava2014dropout}
and batch-normalization\cite{ioffe2015batch}.

With dropout, at each epoch during training, different units for each sample are dropped randomly with probability $1-p$, $0 < p \leq 1$.
This encourages NN units to specialize in detecting particular patterns, and subsequently features to be sparse. In practice, this also makes the optimization faster. During testing, all units are used and the predictions are multiplied by $p$ (such that all units behave as if trained without dropout).

With batch-normalization, the outputs of units are normalized for the given mini-batch. After normalization into standardized features (zero mean with unit variance), the features are shifted and rescaled to a range of variation that is learned by backpropagation. This prevents units having to constantly adapt to large changes in the distribution of their inputs (a problem know as internal covariate shift). Batch-normalization has a slight regularization effect, allowing for a higher learning rate and faster optimization.

\subsection{Convolutional Neural Networks}

Convolutional NNs (CNNs)\cite{fukushima1980neocognitron,lecun1998} are an alternative
to conventional, fully-connected NNs  for temporally or spatially correlated signals.
They limit dramatically the number of parameters of the model and memory requirements by relying on two main concepts:
{\it local receptive fields} and {\it shared weights}. In fully-connected NNs, for each layer, every output interacts with every input. This results in an excessive number of weights for large input dimension (number of weights is $O(N\times P)$). In CNNs, each output unit is connected only with subsets of inputs corresponding to given filter (and filter position). These subsets constitute the local receptive field. This significantly reduces the number of NN multiplication operations on the forward pass of a convolutional layer for a single filter to $O(N\times K)$, with $K$, typically a factor 100 smaller than $N$ and $P$. Further, for a given filter, the same $K$ weights are used for all receptive fields. Thus the number of parameters for each layer and weight is reduced from $O(N\times P)$ to  $O(K)$.

Weight sharing in CNNs gives another important property called {\it shift invariance}. Since for a given filter, the weights are the same for all receptive fields, the filter must model well signal content that is shifted in space or time. The response to the same stimuli is unchanged whenever the stimuli occurs within overlapping receptive fields.
Experiments in neuroscience reveal the existence of such a behavior
(denoted {\it self-similar receptive fields}) in simple cells of
the mammal visual cortex \cite{hubel1962receptive}.
This principle leads CNNs to consider convolution layers with
linear filter banks on their inputs.

Fig.~\ref{fig:CNN_two_first_layers} provides an illustration of
one convolution layer. The convolution layer applies three filters
to an input signal $\bf x$ to produce three  feature maps.
Denoting the $q$th input feature map at layer $l$ as ${\bf z}_q^{(l-1)}$ and
the $p$th output feature map at layer $l$ as $\bar {\bf z}_p^{(l)}$,
a convolution layer at layer $l$ produces $C_{\rm out}$ new feature maps from $C_{\rm in}$ input feature maps as follows
\begin{align}
  \bar {\bf z}_{p}^{(l)}
  =
  g\left(
  \sum_{p = 1}^{C_{\rm in}} {\bf w}^{(l)}_{pq} * {\bf z}^{(l-1)}_{q} + b^{(l)}_p
  \right)
  \;\;\text{for}\;\;
  p=1,\ldots, C_{\rm out}
  \label{eq:cnn}
\end{align}
where $*$ is the discrete convolution, ${\bf w}_{pq}^{(l)}$ are $C_{\rm out} \times C_{\rm in}$ learned linear filters,
$b_p^{(l)}$ are $C_{\rm out}$ learned scalar bias, $p$ is an output channel index and $q$ an input channel index.
Stacking all  feature maps ${\bf z}^{(l)}_p$ together, the set of hidden features is represented as a tensor ${\bf z}^{(l)}$ where each channel corresponds to a given feature map.

For example, a spectrogram is represented
by a $N \times C$ tensor  where $N$ is the signal length and the number of channels $C$ is the number of frequency sub-bands. Convolution layers preserve the spatial or temporal resolution of the input tensor, but  usually increase the number of channels:
 $C_{\rm out} \ge C_{\rm in}$.
This produces a redundant representation which  allows for sparsity in the feature tensor. Only a few units should fire for a given stimuli: a concept that has also been influenced by vision research experiments.\cite{olshausen1997}
Using tensors is a common practice allowing us to represent CNN architectures in a condensed way, see Fig.~\ref{fig:deep_CNN_classifier}.

Local receptive fields impose that an output feature is influenced by
only a small temporal or spatial region of the input feature tensor.
This implies that each convolution is restricted to a small sliding
centered kernel window of odd size $K$, for example, $K=3\times 3=9$ is
a common practice for images.
The number of parameters to learn for that layer is then
$C_{\rm out} \times (C_{\rm in} \times K+1)$
and is independent on the input signal size $N$.
In practice $C_{\rm in}$, $C_{\rm out}$ and $K$ are chosen so small that it is  robust against overfitting.
Typically, $C_{\rm in}$ and $C_{\rm out}$ are less than a few hundreds. A byproduct is that processing becomes much faster for both learning and testing.

Applying $D$ convolution layers of support size $K$
increases the region of influence
(called {\it effective receptive field}) to a $D(K-1)+1$ window.
With only convolution layers,
such an architecture must be very deep
to capture long-range dependencies.
For instance, using filters of size $K=3$, a $10$ deep architecture will
process inputs in sliding windows of only size $21$.

To capture larger-scale dependencies,
CNNs introduce a third concept: {\it pooling}. While convolution layers preserve the
spatial or temporal resolution,
pooling preserves the number of channels but reduces the signal resolution. Pooling is applied independently on each feature map  as
\begin{align}
    {\bf z}_p^{(l)} = {\rm pooling}(\bar{\bf z}_p^{(l)}), \quad\text{for } p = 1, \ldots, C_{\rm out}
\end{align}
and such that ${\bf z}_p^{(l)}$ has a smaller resolution than $\bar {\bf z}_p^{(l)}$.
Max-pooling of size 2 is commonly employed by replacing in all directions two successive values by their maximum. By alternating $D$ convolution and pooling layers, the effective receptive field becomes of size $2^{D-1} (K+1) - 1$.
Using filters of size $K=3$, a $10$ deep architecture will have
an effective receptive field of size $2047$ and can thus capture long-range dependencies.

Pooling is also grounded on neuroscientific findings about the mammalian visual cortex.\cite{hubel1962receptive} Neural cells in the visual cortex condense the information to gain invariance and robustness against small distortions of the same stimuli. Deeper tensors become more elongated with more channels and smaller signal resolution. Hence, the deeper the CNN architecture, the more robust the CNN becomes relative to the exact locations of stimuli in the receptive field.
Eventually the tensor becomes flat, meaning that it is reduced to a vector. Features in that tensor
are no longer temporally or spatially related and they can
serve as input feature vectors for a classifier. The output tensor is not
always  exactly flat, but then the  tensor is mapped into a vector.
In general, a MLP with two hidden FC layers is employed and the architecture
is trained end-to-end by backpropagation or variants, see Fig.~\ref{fig:deep_CNN_classifier}.

This type of architecture is typical of modern image classification NNs
such as AlexNet\cite{krizhevsky2012imagenet} and ZFnet\cite{zeiler2014visualizing},
but was already employed in Neocognitron\cite{fukushima1980neocognitron} and LeNet5.\cite{lecun1998}
The main difference is that modern architectures can deal with data of much higher dimensions as they employ
the aforementioned strategies (such as
rectifiers, Adam, dropout, batch-normalization).
A trend in DL is to make such CNNs as deep as possible with
the least number of parameters by employing specific architectures
such as inception modules, 
depth-wise separable convolutions, 
skip connections, 
and dense architectures.\cite{goodfellow2016deep} 

Since 2012, such architectures have led to state of the art classification
in computer vision,\cite{krizhevsky2012imagenet} even rivaling human performances
on the ImageNet challenge.\cite{he2015delving}
Regarding acoustic applications, this architecture has been
employed for broadband DOA estimation\cite{chakrabarty2017broadband}
where each class corresponds to a given time frame.

\begin{figure*}[t]
  \scriptsize
  \centering
  \includegraphics[width=1.5\reprintcolumnwidth]{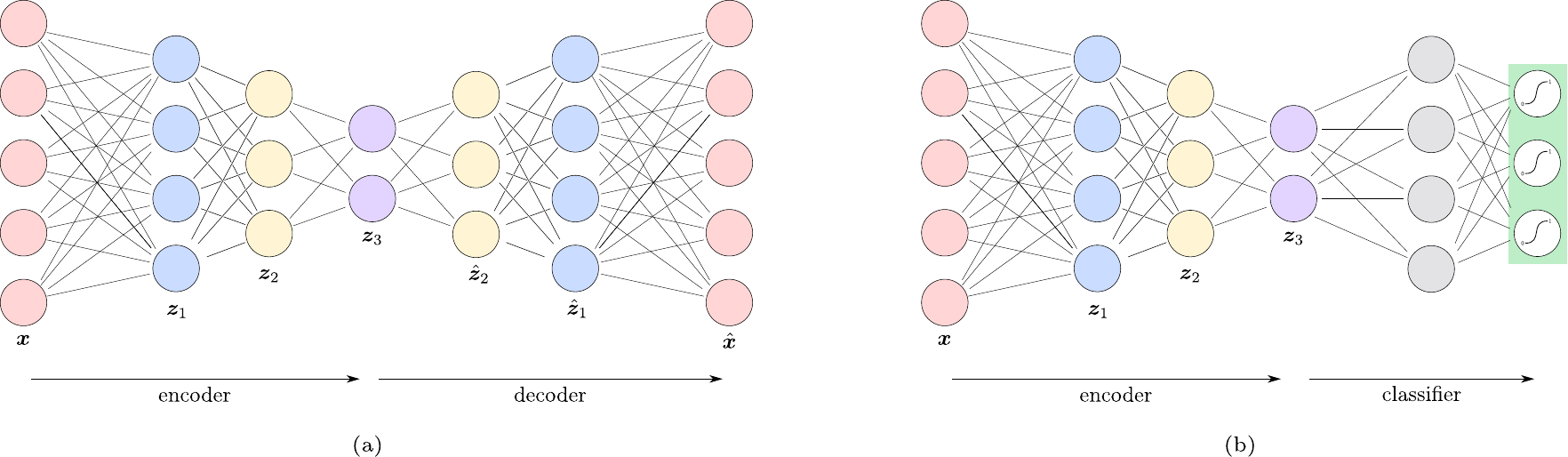}
  \caption{(Color Online)
  Transfer of (a) autoencoders trained in an unsupervised way to (b) a network for supervised classification problem. This illustrates autoencoder architectures, as well as unsupervised pretraining, an early method for initializing NN optimization.
  }
  \label{fig:auto_encoders}
\end{figure*}

\subsection{Transfer learning}

Training deep classifiers from scratch requires using large
labeled datasets. In many applications, these
are not available. An alternative
is using  {\it transfer learning}.\cite{pratt1993discriminability}
Transfer learning reuses parts of a network that were trained on a
large and potentially unrelated dataset for a given ML task. The key idea in transfer learning is that early stages of a deep network learn generic features that may be applicable to other tasks.  Once a network has learned such a task, it is often possible to remove the feed forward layers at the end of the network that are tailored exclusively to the trained task.  These are then replaced with new classification or regression layers, and the learning process finds the appropriate weights of these final layers on the new task.  If the previous representation captured information relevant to the new task, they can be learned with a much smaller data set.
In this vein, deep autoencoders (see Sec.~\ref{sec:autoencoders}) can be used to learn features from a large unlabeled dataset. The learned encoder is next used as a feature extractor after which a classifier can be trained on a small labeled dataset (see Fig.~\ref{fig:auto_encoders}).
Eventually, after the classifier has been trained,
all the layers will be slightly adjusted by performing a few  backpropagation steps end-to-end (referred to as {\it fine tuning}).
Many modern DL techniques rely on this principle.

\subsection{Specialized architectures}

Beyond classification, there exists myriad NN and CNN
architectures. Fully convolutional and U-net architectures, which are enhanced CNNs, are widely used for for regression problems such as signal enhancement,\cite{zhang2017beyond}
segmentation\cite{ronneberger2015u} or object localization.\cite{dai2016r}
Recurrent NNs 
(RNNs) are an alternative to
classical feed-forward NNs to process or produce sequences of variable length.
In particular, long short term memory networks
\cite{goodfellow2016deep}
(called LSTMs) are a specific type of RNN that have
produced remarkable results in several applications where temporal correlations in the data is significant. Applications include speech processing and natural language processing.
Recently, NNs have gained much attention
in unsupervised learning tasks. One key example is data generation with  variational autoencoders and
generative adversarial networks\cite{goodfellow2014generative} (GANs).
The later relies on an original idea grounded on game theory.
It performs a two player game between a generative network and a discriminative one. The generator learns the distribution of the data such that it can produce fake data from random seeds. Concurrently, the discriminator learns the boundary between real and fake data such that it can distinguish the fake data from the ones of the training set. Both NNs compete against each other. The generator tries to fool the discriminator such that the fake data cannot be distinguished from the ones of the training set.

\subsection{Applications in Acoustics} \label{subsec:dl_applications}
DL has yielded promising advances in acoustics. The data-driven DL approaches provide good results relative to conventional or hand-engineered signal processing methods in their respective fields. Aside from improvements in performance, DL (and also ML generally) can provide a general framework for performing acoustics tasks. This is an alternative paradigm to developing highly specialized algorithms in the individual subfields. However, an important challenge across all fields is obtaining sufficient training data. To properly train DNNs in audio processing tasks, hours of representative audio data may be required.\cite{vincent2018audio,purwins_deep_2019} Since large amounts of training data might not be available, DL is not always practical. Though scarcity of training data can be addressed partly by using synthetic training data or data augmentation.\cite{mesaros2017dcase,niu2019deep} In the following we highlight recent advances in the application of DL in acoustics.\cite{cakir2017convolutional,mesaros2017dcase,dibias,adavanne2019sound,trees2002optimum,He2016ResNet,hershey2016deep,ernst2018speech,parviainen2018time,diment2017transfer,shen2018natural,nugraha2016multichannel,Perotin2019CRNN}

\begin{figure}[tb]
\includegraphics[width=\reprintcolumnwidth]{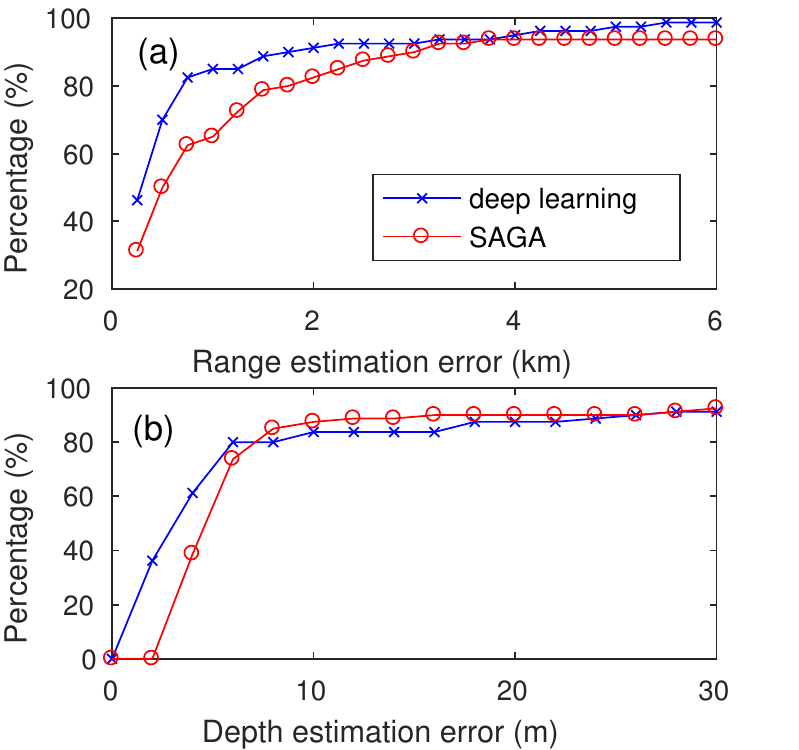}
\caption{\label{fig:niu2019deep}(Color Online) Prediction error of (a) source range and (b) depth from deep learning (DL)-based and conventional underwater source localization from acoustic data. Source locations are obtained using DL-based method (ResNet)\cite{niu2019deep} and Seismo-Acoustic inversion using Genetic Algorithms (SAGA).\cite{gerstoft1994inversion} The prediction error is the percentage of total predictions with maximum error below a given value, where the maximum error value is shown on the x-axis).}
\end{figure}

Two tasks in acoustics and audio signal processing that have benefited from DL are sound event detection and source localization. These methods replace physics-based acoustic propagation models or hand-engineered detectors with deep-learning architectures. In Ref.~\onlinecite{cakir2017convolutional}, convolutional recurrent NNs achieve state-of-the art results in the sound event detection task in the 2017 Detection and Classification of Acoustic Scenes and Events (DCASE) challenge.\cite{mesaros2017dcase} In Ref.~\onlinecite{chakrabarty2017broadband}, CNNs are developed for broadband DOA estimation which use only the phase component of the STFT. The CNNs obtain competitive results with steered response power phase transform (SRP-PHAT) beamforming.\cite{dibias} The CNN was trained using synthetically generated noise and it generalized well to speech signals. In Ref.~\onlinecite{adavanne2019sound} the event detection and DOA estimation tasks are combined into a signal DNN architecture based on convolutional RNNs. The proposed system is used with synthetic and real-world, reverberant and anechoic data, and the DOA performance is competitive with MUSIC.\cite{trees2002optimum} In Ref.~\onlinecite{niu2019deep}, DL is used to localize ocean sources in a shallow ocean waveguide using a single hydrophone, as shown in Fig.~\ref{fig:niu2019deep}. Two deep residual NNs (50-layers each, ResNet50\cite{He2016ResNet}) are trained to localize the range and depth of a source using millions of synthetic acoustic fields.  The ResNet50 DL model achieves competitive source range and depth prediction error when compared to popular genetic algorithm-based inversion methods Fig.~\ref{fig:niu2019deep}.\cite{gerstoft1994inversion} The source (range or depth) prediction error defined here is the percentage of predictions with maximum error below a given value, with given values for range and depth defined along the x-axis in Fig.~\ref{fig:niu2019deep}.

DL has also been applied in speech modeling, source separation, and enhancement. In Ref.~\onlinecite{hershey2016deep} a deep clustering approach is proposed, based on spectral clustering, which uses a DNN to find embedding features for each time-frequency region of a spectrogram. This is applied to the problem of separating two speakers of the same gender, but can be applied to problems where multiple sources of the same class are active.
In Ref.~\onlinecite{ernst2018speech} DNNs are used to remove reverberation from speech recordings using a single microphone. The system works with the STFT of the speech signals. Two different U-net architectures, as well as adversarial training with GAN are implemented. The dereverberation performance of the proposed DL architectures outperform competing methods in most cases.

Much like in acoustics, seismic exploration research has traditionally focused on advanced signal processing algorithms, with only occasional applications of pattern recognition techniques. ML and especially DL methods, have recently seen significant increases in seismic exploration applications. One area of the field has obtained many benefits from DL models is the interpretation of geological structure elements. Classification and interpretation of these structures, such as salt domes, channels, faults and folds, from seismic images faces several challenges, including handling extremely large volumes of 3D seismic data and sparse and uncertainty manual image annotations from geologists.\cite{chopra_seismic_2005,qi_image_2019} Many benefits are achieved by automating these procedures. Several recently-developed ML techniques construct attributes adapted to specific data via ML algorithms\cite{tosic2011,goodfellow2016deep} instead of hand-engineering them. 

DL has been applied to the seismic interpretation of faults\cite{huang2017,wu2019a,wu2019b}, channels\cite{pham2019}, salt domes, as well as seismic facies classification using 3D CNNs and GANs \cite{liu2019}. In Ref.~\onlinecite{wu2019a}, a 3D U-net was applied to detect or segment faults from a 3D seismic images. In Ref.~\onlinecite{liu2019}, a semi-supervised facies classifier based on 3D CNNs with GANs was developed to handle large volumes of data from new exploration fields which might have few labels. There have also been interesting developments in seismic data post-processing, including automated sesimic facies classification.\cite{li_classifying_2018}

\section{Speaker localization in reverberant environments} \label{sec:speech}

Speech enhancement is a core problem in audio signal processing, with commercial applications in devices as diverse as mobile phones, hands-free systems, human-car communication, smart homes or hearing aids. An essential component in the design of speech enhancement algorithms is acoustic source localization. Speaker localization is also directly applicable to many other audio related tasks, e.g. automated camera steering, teleconferencing systems and robot audition.

Driven by its large number of applications, the localization problem has attracted significant research attention, resulting in a plethora of localization methods proposed during the last two decades.\cite{pertila2018multichannel}
Nevertheless, robust localization in adverse conditions, namely in the presence of background noise and reverberation, still remains a major challenge.

A recent challenge on acoustic source LOCalization And TrAcking (LOCATA), endorsed by the IEEE Audio and Acoustic Signal Processing technical committee, has established a database to encourage research teams to test their algorithms.\cite{LOCATA2018a} The challenge dataset consists of acoustic recordings from real-life scenarios. With this data, the performance of source localization algorithms in real-life scenarios can be assessed.

There is a growing interest in supervised-learning for audio source localization using NNs. In the recent issue on ``Acoustic Source Localization and Tracking in Dynamic Real-Life Scenes'' in the IEEE Journal on Selected topics in Signal Processing, three papers used variants of NNs for source localization.\cite{Perotin2019CRNN,Chakrabarty2019Multi,adavanne2019sound}
We expect this trend to continue, with an emphasis on methods that do not require a large set of labeled data. Such labeled data is very difficult to obtain in the localization problem. For example, in Ref.~\onlinecite{Opoc1910:Deep}, a weakly-labeled ML paradigm is presented. The approach used few labeled samples with known positions along with larger set of unlabeled samples, for which only their \textit{relative} physical ordering is known.

In this short survey, we explore two families of learning-based approaches. The first is an unsupervised method based on GMM classification. The second is a semi-supervised method based on manifold learning.

Despite the progress that has been made in the recent years in the manifold-learning approach for localization, some major challenges remain to be solved, e.g. robustness to changes in array constellation and the acoustic environment, and the multiple concurrent speakers case.

\subsection{Localization and tracking based on the expectation-maximization procedure}
In this section we review an unsupervised learning methodology for speaker localization and tracking of an unknown number of concurrent speakers in noisy and reverberant enclosures, using a spatially distributed microphone array. We cast the localization problem as a classification problem in which the measurements (or features extracted thereof) can be associated with a grid of candidate positions\cite{mandel2010model} $\mathcal{P} = \{ \mathbf{p}_1,\ldots,\mathbf{p}_M\}$, where $M = |\mathcal{P}|$ is the number of candidates. The actual number of speakers is always significantly lower than $M$.

The speech signals, together with an additive noise, are captured by an array of microphones ($N>1$). The binaural case ($N=2$) was presented in\cite{mandel2010model}. We assume a simple sound propagation model with a dominant direct-path and potentially a spatially-diffuse reverberation tail. The $n$th microphone signal in the {STFT} domain is given by:
\begin{equation}
z_n(t,k)=\sum_{m=1}^M d_m(t,k)  g_{m,n}(k) s_m(t,k) +v_{n}(t,k)
\label{eq:signal_model}
\end{equation}
where $t=0,\ldots,T-1$ is the time index, $k=0,\ldots,K-1$ is the frequency index, $g_{m,n}(k)$ is the direct-path transfer function from the speaker at the $m$-th position to the $n$-th microphone:
\begin{equation}
g_{m,n}(k) =  \frac{1}{\|\mathbf{p}_m-\mathbf{p}_n\|}\exp\left(-j\frac{2\pi k}{K} \frac{\tau_{m,n}}{T_s} \right)
\end{equation}
where $T_s$ is the sampling period, and $\tau_{m,n}=\frac{\|\mathbf{p}_m-\mathbf{p}_n\|}{c}$ is the {TDOA} between candidate position $\mathbf{p}_m$ and microphone position $\mathbf{p}_m$ and $c$ the sound velocity.
This {TDOA} can be calculated in advance from the predefined grid points and the array geometry, which is assumed to be known.

$s_m(t,k)$ is the speech signal uttered by a speaker at grid point $m$ and $v_{n}(t,k)$ is either an ambient noise or a spatially-diffused  reverberation tail. The indicator signal $d_m(t,k)$ indicates whether speaker $m$ is active in the $(t,k)$-th STFT bin:
\begin{equation} \label{d_definition}
d_m(t,k) =
\begin{cases}
      1, & \text{if speaker $m$ is active in STFT bin } (t,k) \\
      0, & \text{otherwise}
\end{cases}.
\end{equation}
Note that, according to the sparsity assumption\cite{yilmaz2004blind} the vector $\mathbf{d}(t,k) =  \mathrm{vec}_{m} \{ d_m(t,k) \}\in \{\mathbf{e}_1,\ldots,\mathbf{e}_M\}$,  where
$\mathrm{vec}_{m} \{\cdot\}$ is a concatenation of the elements along the $m$th index
and $\mathbf{e}_m$ is a ``one-hot" vector (equals `1' in its $m$th entry, and zero elsewhere).
The $N$ microphone signals are concatenated in a vector form:
\begin{equation} \label{eq:signal_model_vec}
 \mathbf{z}(t,k) = \sum_{m=1}^{M} d_m(t,k)\mathbf{g}_{m}(k) s_m(t,k) +  \mathbf{v}(t,k),
\end{equation}
where $\mathbf{z}(t,k)$, $\mathbf{g}_{m}(k)$ and $\mathbf{v}(t,k)$ are the respective concatenated vectors.

We will discuss several alternative feature vector selections from the raw data. Based on the W-disjoint orthogonality property of the speech signal, \cite{rickard2002approximate,yilmaz2004blind} these features can be attributed a GMM\eqref{eq:gmm1}, with each Gaussians associated with a candidate position in the enclosure on the predefined grid.

An alternative is to organize the microphones in  dual-microphone nodes and to extract the pair-wise relative phase ratio (PRP)
\begin{equation}
\phi_n(t,k) \triangleq \frac{z_{n}^1(t,k)}{z_{n}^2(t,k)} / \frac{|z_{n}^1(t,k)|}{|z_{n}^2(t,k)|},
\label{eq:prp}
\end{equation}
with $n$ the node index (number of microphones in this case is $2N$), the superscript is the microphone-pair index (either 1 or 2) within the pair $n$. Under the assumptions that 1) the inter-microphone distance is small compared with the distance of grid points from the node center, and 2) the reverberation level is low, the PRP of a signal impinging the microphones located at $\mathbf{p}_{n}^1$ and $\mathbf{p}_{n}^2$ from a grid point $\mathbf{p}_m$ can be approximated by
\begin{equation}
\tilde{\phi}^k_{n}(\mathbf{p}_m) \triangleq \exp\left(-j\frac{2\pi
k}{K}\frac{\cdot
(||\mathbf{p}_m-\mathbf{p}_{n}^2||-||\mathbf{p}_m-\mathbf{p}_{n}^1||) }{c \cdot
T_s}\right).
\label{eq:prp_cent}
\end{equation}
Since this approximation is often violated, we use $\tilde{\phi}^k_{n}(\mathbf{p}_m)$ as the centroid of a Gaussian that describes the PRP. For multiple speakers in unknown positions we can use the W-disjoint orthogonality to express the distribution of the PRP as a GMM:
\begin{equation}
f(\boldsymbol{\phi}) =\prod_{t,k} \sum_{m=1}^M \pi_m \prod_{n} \mathcal{CN}(\phi_n(t,k);\tilde{\phi}^k_n(\mathbf{p}_m),\sigma^2).
\label{eq:stat_model_phi}
\end{equation}
We will also assume for simplicity that $\sigma^2$ is set in advance.

\begin{figure}[t]
\subfigure[steered response power phase transform (SRP-PHAT) beamformer. \cite{dibias}]
{\includegraphics[width=0.49\columnwidth, trim = 0mm 0mm 0mm 0mm, clip=true]{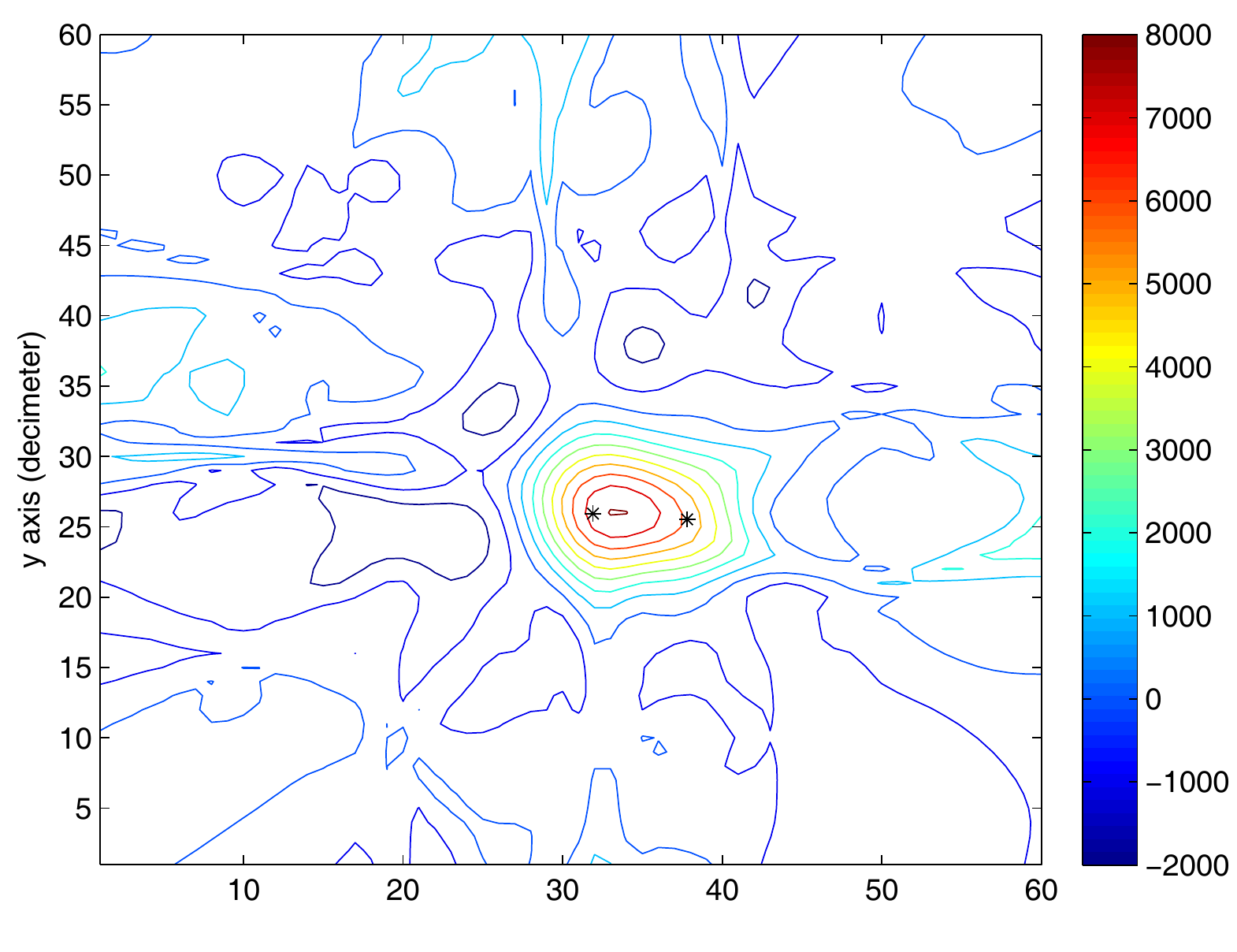}}
\subfigure[Batch expectation maximization (EM)\cite{schwartz2014speaker,dorfan2015tree}]
{\includegraphics[width=0.49\columnwidth, trim = 0mm 0mm 0mm 0mm, clip=true]{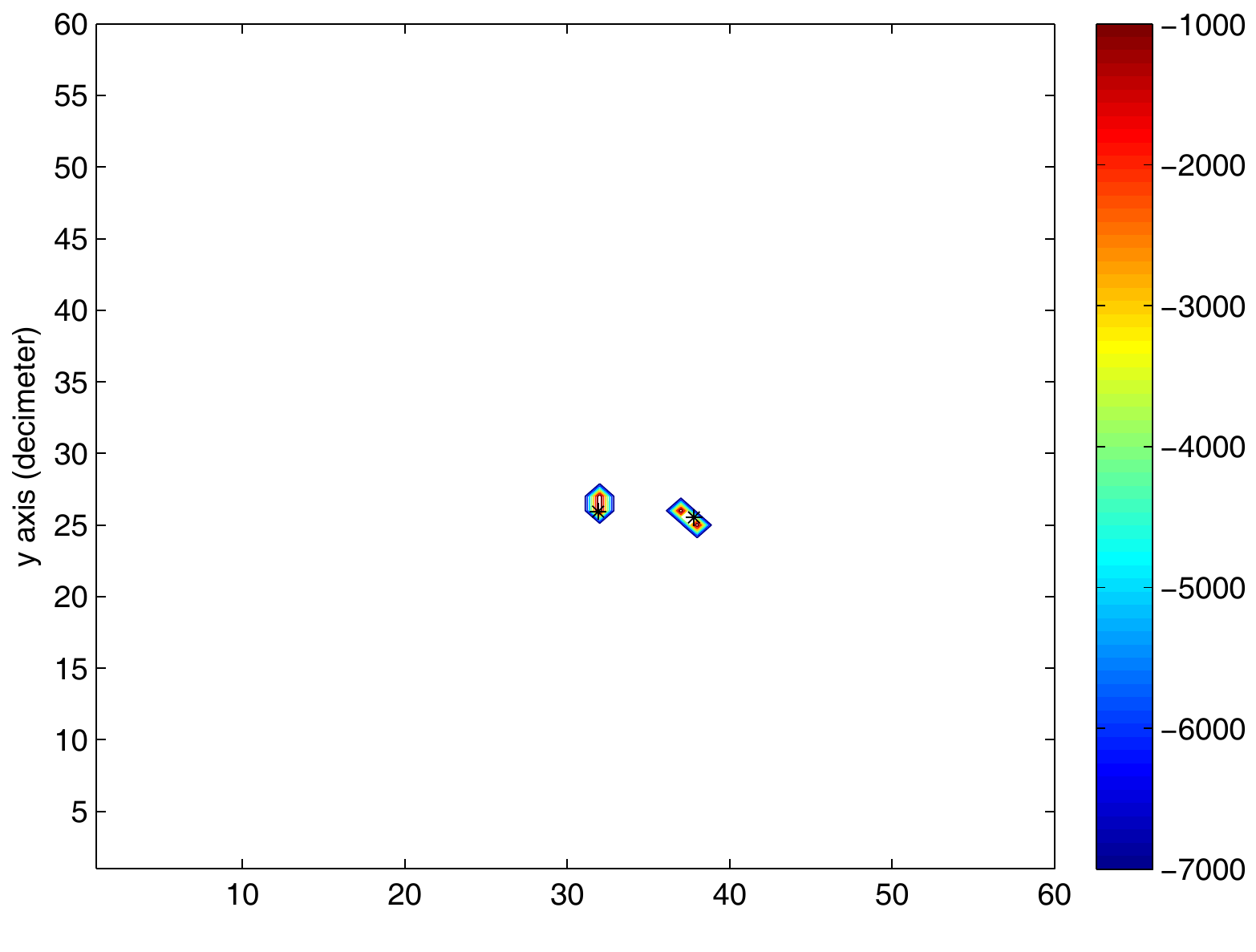}}
\caption{(Color Online) Two targets localization. $10\times 10$~cm grid, 12 nodes, inter-microphone distance per node 50~cm, $T_{60}=300$~ms.}
\label{fig:srp_em}
\end{figure}

Using the GMM, the localization task can be formulated as a maximum likelihood parameter estimation problem. The number of active speakers in the scene and their position will be indirectly determined by examining the GMM weights, $\pi_m,\,m=1,\ldots,M$, and selecting their peak values. As explained above, the ML parameter estimation problem cannot be solved in closed-form. Instead, we will resort to the expectation-maximization (EM) procedure.\cite{dempster_maximum_1977}
The E-step results in the estimate of the indicator signal (here the hidden data):
\begin{multline}
\hat{d}^{(\ell-1)}(t,k,m)\triangleq E\left\{d(t,k,m)|\boldsymbol{\phi}(t,k);\hat{\boldsymbol{\pi}}^{(\ell-1)}\right\}\\
=\frac{\hat{\pi}^{(\ell-1)}_m\prod_n\mathcal{CN}
		\left(\phi_n(t,k);\tilde{\phi}_n^{k}(\mathbf{p}_m),\sigma^2\right)} {\sum_{m=1}^M\hat{\pi}^{(\ell-1)}_m\prod_n\mathcal{CN}
		\left(\phi_n(t,k);\tilde{\phi}_n^{k}(\mathbf{p}_m),\sigma^2\right)}.
\end{multline}
In the M-step the GMM weights are estimated:
\begin{equation}
 	\hat{\pi}_m^{(\ell)}= \frac{\sum_{t,k}\hat{d}^{(\ell-1)}(t,k,m)}{T \cdot K}.
\end{equation}
The procedure is repeated until a number of predefined iterations $\ell=L$ is reached. We refer to this procedure as \emph{batch} EM, as opposed to the \emph{recursive} and \emph{distributed} variants that will be later introduced.
In Fig.~\ref{fig:srp_em} a comparison between the classical SRP-PHAT and the batch EM is depicted. It is evident that the EM algorithm (which maximizes the ML criterion) achieves much higher resolution.

In Ref.~\onlinecite{dorfan2015tree}, a distributed version of this algorithm was presented, suitable for wireless acoustic sensor networks (WASNs) with known microphone positions.
WASNs are characterized by low computational resources in each node and by a limited connectivity between the nodes. A bi-directional tree-based distributed EM (DEM) algorithm that circumvents the inherent network limitations was proposed, by substituting the standard EM iterations by iterations across nodes. Furthermore, a recursive distributed EM (RDEM) variant, which is better suited for online applications, is proposed.

In Ref.~\onlinecite{dorfan2018distributed}, an improved, bio-inspired, acoustic front-end that enhances the direct-path, and consequently increasing the robustness of the proposed schemes to high reverberation, is presented. An alternative method for enhancing the direct-path is presented in \onlinecite{li2017multiple}, where the multi-path propagation model of the sound is taken into account, by the so-called convolutive transfer function (CTF) model.\cite{talmon2009relative}

In another variant of the classification paradigm, the GMM is substituted by a mixture of von Mises,\cite{Brendel2018Localization} which is a suitable distribution for the periodic phase of the microphone signals.

Here we will elaborate on another alternative feature, namely the raw microphone signals (in the STFT domain). According to our measurement model (\ref{eq:signal_model},\ref{eq:signal_model_vec}) the raw data can also be described by a GMM:\cite{dorfan2016multiple,schwartz2016multi,schwartz2017doa}
\begin{equation} \label{eq:stat_model_z}
f_z(\mathbf{z}) = \prod_{t,k}\sum_{m=1}^M \pi_m\mathcal{CN}(\mathbf{z};\mathbf{0},\boldsymbol{\Phi}_{\mathbf{z},m}(t,k))
\end{equation}
where the covariance matrix of each Gaussian is given by:
\begin{equation}
\boldsymbol{\Phi}_{\mathbf{z},m}(t,k)  = \mathbf{g}_{m}(k) \mathbf{g}^H_{m}(k) \phi_{s,m}(t,k) + \boldsymbol{\Phi}_{\mathbf{v}}(k).
\end{equation}
Here we assumed that the noise is stationary and its PSD known.
In this case (frequency index $k$ is omitted for brevity), the E-step simplifies to:\cite{Weisberg2019online}
\begin{equation}\label{eq:E-step2}
  \hat{d}_m^{(\ell-1)}(t) = \frac{ \hat\pi_m^{(\ell-1)} T_m(t) }{\sum_m  \hat\pi_m^{(\ell-1)}T_m(t)}.
\end{equation}	
with the likelihood ratio test (LRT):
\begin{equation}\label{eq:simplified_indicator}
  \textrm{LRT}_m(t) =  \frac{1}{\mathrm{SNR}_m^{\textrm{post}}(t)} \exp \left( \mathrm{SNR}_m^{\textrm{post}}(t)-1\right)
\end{equation}
where $\mathrm{SNR}_m^{\textrm{post}}(t)=\frac{\left|\hat{s}_{m,\mathrm{MVDR}}(t) \right|^2}{\phi_{v,m}}$ is the posterior SNR of a signal from the $m$th candidate position.  $\hat{s}_{m,\mathrm{MVDR}}(t) \equiv \mathbf{w}^H_m \mathbf{z}(t)$ is an estimate of the speech using the minimum variance distortionless response beamforming (MVDR-BF), $\mathbf{w}_m = \frac{ \Phi^{-1}_{\mathbf{v}} \mathbf{g}_{m} }{\mathbf{g}^H_{m} \Phi^{-1}_{\mathbf{v}} \mathbf{g}_{m}}$, which constitutes a sufficient statistic for estimating the speech {PSD} $\phi_{s,m}(t)$ given the observations $\mathbf{z}(t)$, and $\phi_{v,m} \equiv \frac{1}{\mathbf{g}^H_{m} \Phi^{-1}_{\mathbf{v}} \mathbf{g}_{m}}$ is the PSD of the residual noise at the output of the MVDR-BF, directed towards the $m$th position candidate.

\begin{figure}[!t]
  \centering
  \includegraphics[width=1.0\reprintcolumnwidth]{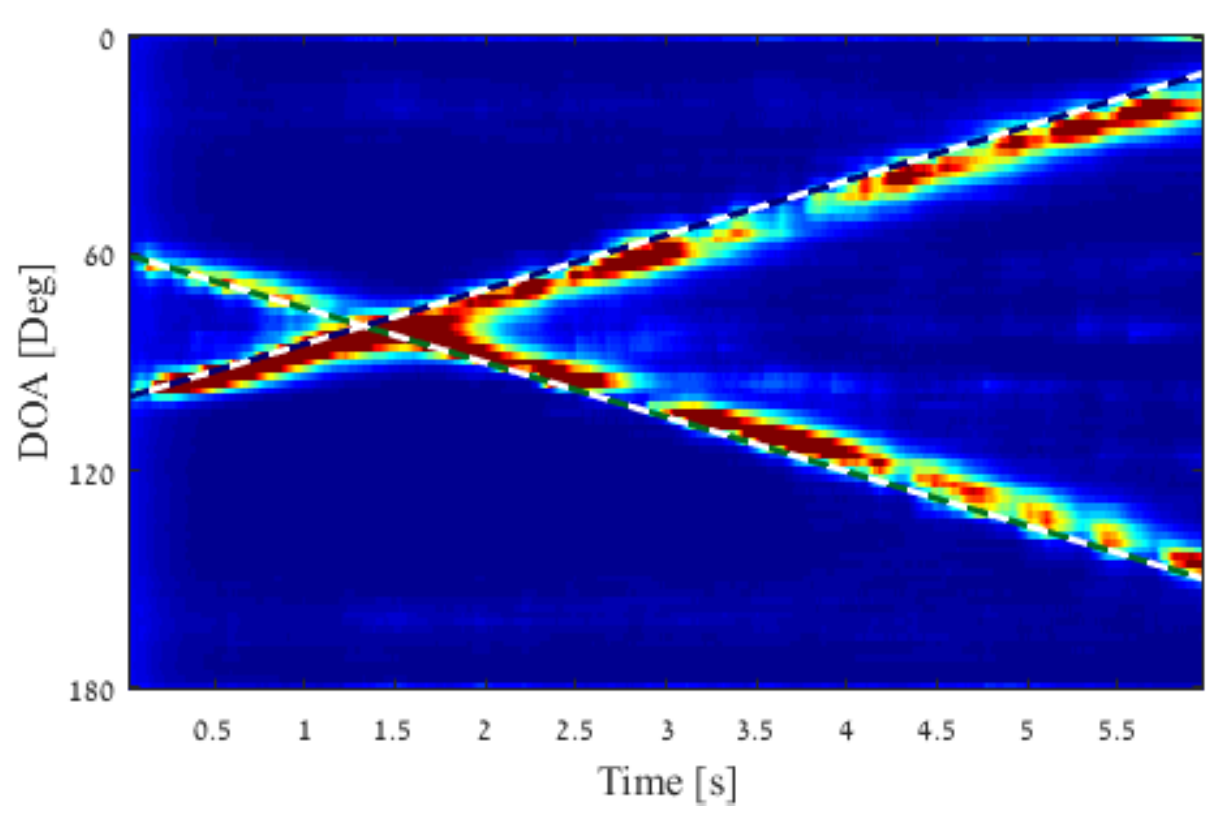}
  \caption{(Color Online) Two speaker tracking using recursive expectation-maximization\cite{Weisberg2019online}. True DOAs are indicated with dashed lines.}
  \label{fig:two_track}
\end{figure}

Two recursive EM (REM) variants can be found in literature,  see Capp\'{e} and Moulines\cite{cappe_-line_2009} and Titterington.\cite{titterington_recursive_1984,wang_almost_2006} The former is based on recursive calculation of the auxiliary function, and the latter utilizes a Newton-based recursion for the maximization, with the Hessian substituted by the Fisher information matrix (FIM). Recursive EM algorithms for source localization were analyzed and developed in Refs.~\onlinecite{chung2001comparative,chung2005tracking}. Titterington's method was extended to deal with constrained maximization, encountered in the problem at hand in Ref.~\onlinecite{schwartz2014speaker}.
Applying these procedures to both data models in (\ref{eq:stat_model_phi}) and (\ref{eq:stat_model_z}) results in:\cite{schwartz2014speaker}
\begin{equation}
\label{eq:REM}
\hat{\pi}_m^R(t)= \hat{\pi}_m^R(t-1)+\gamma_{t}(\hat{\pi}_m(t) -\hat{\pi}_m^R(t-1))
\end{equation}
where $ \hat{\pi}_m(t)= \frac{\sum_{k}\hat{d}(t,k,m)}{K}$ is the instantaneous estimate of the indicator and $\hat{\pi}_m^R(t)$ is the recursive estimator.
The tracking capabilities of the algorithm in Ref.~\onlinecite{Weisberg2019online} in simulated data with low noise level and two speakers in reverberation time $T_{60}\approx300$~ms is depicted in Fig.~\ref{fig:two_track}.
For this experiment we used an 8-microphone linear array, hence only DOA estimation capabilities were examined. For the DOA candidates we used a grid of possible azimuth angles between $-90^\circ$ and $90^\circ$, with a resolution of $2^\circ$. The proposed algorithm provides speaker DOA probability distributions as a function of time, as depicted in Fig.~\ref{fig:two_track}, and not directly the DOA estimates. To estimate the actual trajectory of the speakers, the speaker locations at each time are found by the probability maxima. In another variant that uses the same features \eqref{eq:simplified_indicator}, the tracking problem is recast as an hidden Markov model (HMM) and the time-varying DOAs are inferred by the forward-backward algorithm\cite{weisberg2019HMM} rather than the recursive EM in \eqref{eq:REM}.

\subsection{Speaker localization and tracking using manifold learning}

 Until recently, a main paradigm in localization research was based on certain statistical and physical assumptions regarding the propagation of sound sources,\cite{Image79,polack1993playing} and mainly focused on robust methods to extract the direct-path. However, valuable information on the source location can also be extracted from the reflection pattern in the enclosure.

The main claim here is that the intricate reflection patterns of the sound source on the room facets and the objects in the enclosure define a \emph{fingerprint}, uniquely characterizing the source location, and that meaningful location information can be inferred from the data by harnessing the principles of \emph{manifold learning}.\cite{talmon2011supervised,laufer2013relative}
Yet, the intrinsic degrees of freedom in the acoustic responses have a limited number. Hence, we can conclude that the variability of the acoustic response in specific enclosures depends only on a small number of parameters. This calls upon manifold learning approaches to improve localization abilities.
We first consider recordings from two microphones
\begin{align}
y_1(n)&=a_1(n)*s(n)+u_1(n) \nonumber \\
y_2(n)&=a_2(n)*s(n)+u_2(n) \
\end{align}
with $s(n)$  the source signal,
$a_i(n),~i=\{1,2\}$ the acoustic impulse responses (AIRs) relating the source and each of the microphones,
and $v_i(n)$ noise signals which are independent of the source.
Define the acoustic transfer functions (ATFs) $A_i(k)$ as the Fourier transform of the AIRs $a_i(n)$, respectively. Then, the relative transfer function (RTF) is defined as\cite{gannot2001signal}
\begin{equation}
    H(k) = \frac{A_2(k)}{A_1(k)}.
\end{equation}
The RTF represents the acoustic path, encompassing all sound reflection paths. As such, it can be viewed as a generalization of the PRP centroid~\eqref{eq:prp_cent}.  A plethora of blind RTF estimation procedure exists.\cite{Markovich-Golan2018Performance}
Finally, we define the RTF vector by concatenating several values of $H(k)$ in the relevant frequency band (where the speech power is significant):
\begin{equation}
    \mathbf{h}=[\begin{array}{cccc}
    H(k_1)& H(k_2)&\cdots& H(k_D)
    \end{array}]^T,
    \label{eq:RTF_vec}
\end{equation}
with $k_1$ and $k_D$ are the lower and upper frequencies of the significant frequency band.
Note that the RTF is independent of the source signal, hence can serve as an acoustic feature, as required in the following method.

Our goal is to find a representation of RTF vectors, as defined in \eqref{eq:RTF_vec}. This representation should reflect the intrinsic degrees-of-freedom that control the variability of a set of RTF. To this end, we collect a set of $N$ RTF vectors from the examined environment: $\mathbf{h}_i,\,i=1,2,\ldots,N$. We then construct a graph that empirically represents the acoustic manifold. The RTFs are used as the graph nodes (not to be confused with the microphone constellation as defined above), and the edges are defined using an RBF kernel
$k(\mathbf{h}_i,\mathbf{h}_j)=\exp\left\{-\frac{\Vert\mathbf{h}_i-\mathbf{h}_j\Vert^2} {\varepsilon} \right\}$ between two RTF vectors, $\mathbf{h}_i,\mathbf{h}_j$.
Define the $N\times N$ kernel matrix $\mathbf{K}_{ij}=k(\mathbf{h}_i,\mathbf{h}_j)$. Let $\mathbf{D}$ be a diagonal matrix whose diagonal elements are the sums of rows of $\mathbf{K}$.
Define the row stochastic $\mathbf{P}=\mathbf{D}^{-1}\mathbf{K}$, a non-symmetric transition matrix with elements defining a Markov process on the graph $\mathbf{P}_{ij}=p(\mathbf{h}_i,\mathbf{h}_j)$, which is a discretization of a diffusion process on the manifold.\cite{Coifman2006}
Since $\mathbf{P}$ is non-symmetric but similar to a symmetric matrix, we can define the left- and right-eigenvectors of the matrix with shared non-negative eigenvalues: $\mathbf{P}\boldsymbol{\phi}^{(i)}=\lambda_i\boldsymbol{\phi}^{(i)}$ and $\boldsymbol{\psi}^{(i)}\mathbf{P}=\lambda_i\boldsymbol{\psi}^{(i)}$. In these definitions,  $\boldsymbol{\phi}^{(i)}$ is the right (column) eigenvector, $\boldsymbol{\psi}^{(i)}$ is the left (row) eigenvector and $\lambda_i$ is the corresponding eigenvalue.

This decomposition induces a nonlinear mapping of the RTF into a low-dimensional Euclidean space:
\begin{equation}
\boldsymbol{\Phi}_d: \mathbf{h}_i \mapsto \left[ \lambda_1\boldsymbol\phi_1^{(i)}, \ldots, \lambda_d\boldsymbol\phi_d^{(i)}\right]^T.
\label{eq:map}
\end{equation}
The nonlinear operator, defined in \eqref{eq:map}, is referred to as the diffusion mapping.\cite{Coifman2006} It maps the $D$-dimensional RTF vector $\mathbf{h}_i$ in the original space to a lower $d$-dimensional Euclidian space, constructed as the $i$th component of the most significant $d$ eigenvectors (multiplied by the corresponding eigenvalue). Note, that the first eigenvector $\boldsymbol\phi_0$ is an all-ones trivial vector since the rows of $\mathbf{P}$ sum to 1.

The diffusion distance reflects the flow between two RTFs on the manifold, which is related to the geodesic distance on the manifold, namely, two RTFs are close to each other if their associated nodes of the graph are well-connected. It can be proven that the diffusion distance:
\begin{equation}
D_\mathrm{Diff}^2(\mathbf{h}_i,\mathbf{h}_j)
=\sum_{r=1}^N\left(p\left(\mathbf{h}_i,\mathbf{h}_r\right)-p\left(\mathbf{h}_j,\mathbf{h}_r\right)\right)^2/\boldsymbol{\psi}_0^{(r)} 
\label{eq:diff_dist}
\end{equation}
is equal to the Euclidean distance in the diffusion maps space when using all $N$ eigenvectors, and that it can be well-approximated by using only the first few $d$ eigenvectors:
\begin{equation}
D_\mathrm{Diff}(\mathbf{h}_i,\mathbf{h}_j)\cong\|\mathbf{\Phi}_d(\mathbf{h}_i)-\mathbf{\Phi}_d(\mathbf{h}_j)\|. 
\end{equation}
This constitutes the basis of the embedding from the high-dimension RTFs with their intricate geodesic distance, to the simple Euclidean distance in the low-dimension space. Thus, distances and ordering in the low-dimensional space can be easily measured. As we will next demonstrate, the low-dimensional representation inferred from this mapping has a one-to-one correspondence with physical quantities, here the location of the source.

\begin{figure}[!t]
\centering
\includegraphics[width=\reprintcolumnwidth]{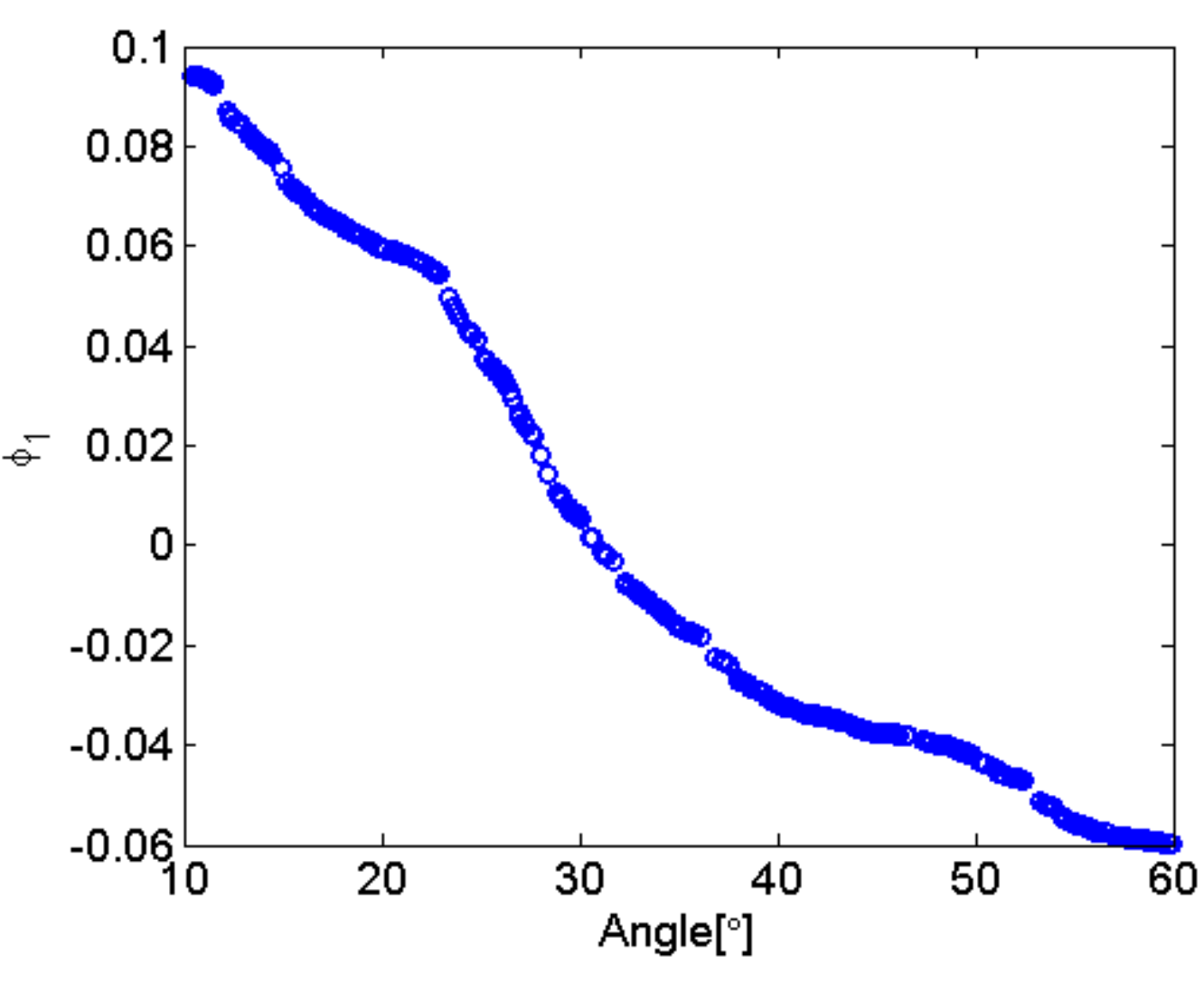}
\caption{(Color Online) Diffusion Mapping. A set of $N$ RTF vectors $\mathbf{h}_i,\,i=1,\ldots,N$ is considered. Using diffusion mapping each RTF $D$-dimensional vector $\mathbf{h}_i$ is mapped into the $i$th component the first non-trivial eigenvector $\boldsymbol{\phi}_1^{(i)}$ (the eigenvalue is shared by all components and hence ignored here). By mapping the entire set we get $N$ embedded values. These values constitute the y-axis of the graph. In the x-axis we draw the known angle of arrival of associated with the RTF vector $\mathbf{h}_i$. A clear correspondence is demonstrated, proving that the diffusion mapping indeed blindly extracts the intrinsic degree-of-freedom of the RTF set, and hence can be utilized for data-driven localization.\cite{laufer2015manifolds}}
\label{fig:diffusion}
\end{figure}

To demonstrate the ability of this nonlinear diffusion mapping to capture the controlling parameters of the acoustic manifold, the following scenario was simulated.\cite{laufer2015manifolds} Two microphones were positioned at $[3,3,1]$~m and $[3.2,3,1]$~m in $6\times 6.2\times 3$~m room with reverberation time of $T_{60}=500$~ms and SNR$=20$~dB. The source position was confined to a circle around the microphone pair with 2~m radius.
It is evident from Fig.~\ref{fig:diffusion} that the dominant eigenvector indeed corresponds to the angle of arrival of the source signal in the range $10^{\circ}-60^{\circ}$.
This forms the basis for a semi-supervised localization method.\cite{Laufer2016semi}

It is acknowledged that collecting labeled data in reverberant environment is a cumbersome task, however measuring RTFs in the enclosure is relatively easy. It is therefore proposed to collect a large number of RTFs in the room where localization is required. These unlabeled RTFs can be collected whenever a speaker is active in the environment. These RTFs will be used to infer the structure of the manifold. A small number of labelled RTFs, i.e. RTFs with an associated accurate position label, will also be collected.  These points will be used to anchor the inferred manifold to the physical world, thereby facilitating the position estimation of an unknown RTF at test time.
In this method, a mapping from an RTF to a position $p=g(\mathbf{h})$ (here we define a mapping from the RTF to each coordinate, hence a vector to scalar mapping) is inferred by the following optimization problem
\begin{equation}
\widehat{g}=\argmin_{g\in \mathcal{H}_k}\frac{1}{n_L}\sum_{i=1}^{n_L} (\bar{p}_i-g(\mathbf{h}_i))^2+\gamma_k\|g\|_{\mathcal{H}_k}^2+\gamma_M{\|g\|_\mathcal{M}^2}, \label{eq:man_reg}
\end{equation}
with $n_L$ labeled pairs  $\{\mathbf{h}_i,\bar{p}_i\}$ and $n_U \gg n_L$ unlabeled RTFs. The optimization has two regularization terms, a Tikhonov regularizer $\|g\|_{\mathcal{H}_k}^2$ that controls the smoothness of the mapping and a manifold-regularization $\|g\|_\mathcal{M}^2$ that controls the smoothness along the inferred manifold.

The minimizer of the regularized optimization problem can be found by optimization in a reproducing kernel Hilbert space (RKHS)\cite{Belkin2003}
\begin{equation}
g(\mathbf{h})=\sum_{i=1}^{n_D} a_i k(\mathbf{h}_i,\mathbf{h})
\label{eq:luexpen}
\end{equation}
where $k:\mathcal{M}\times\mathcal{M}\rightarrow\mathbb{R}$ is the reproducing kernel of $\mathcal{H}_k$, 
 with $k(\mathbf{h}_i,\mathbf{h}_j)$ as defined above, 
and $n_D=n_L+n_U$ the total number of training points.
Using this semi-supervised method may improve significantly the localization accuracy, e.g. the RMSE of this method in reverberation level of $T_{60}=600$~ms and SNR=5~dB is $\approx3^\circ$, while the classical generalized cross-correlation (GCC) method\cite{knapp1976generalized} achieves an RMSE of $18^\circ$ at the same acoustic conditions.

\begin{figure}[!t]
    \centering
\includegraphics[width=\reprintcolumnwidth]{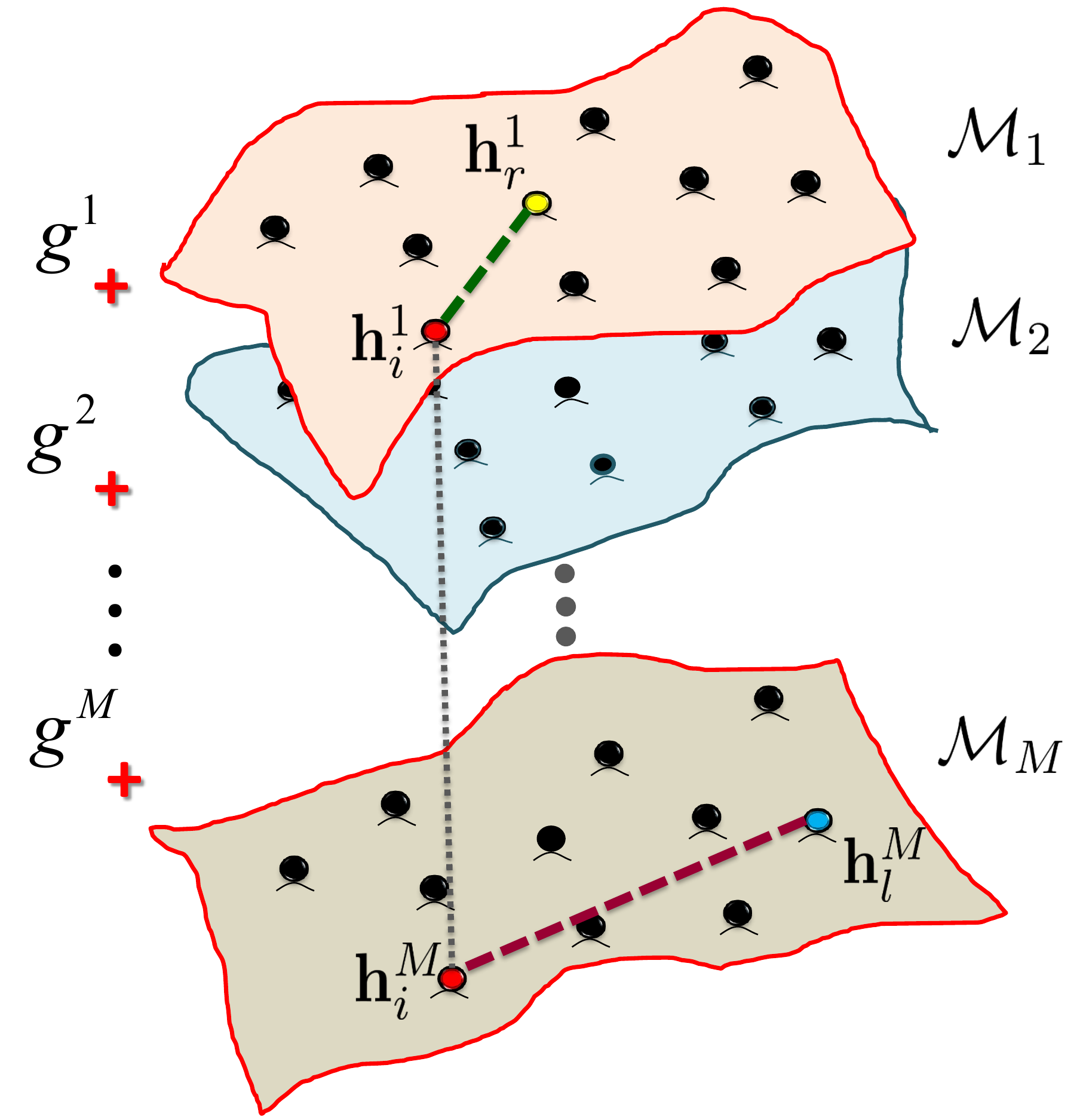}
    \caption{(Color Online) A multi-view perspective of the acoustic scene with each manifold defining a mapping from the RTFs to position estimates.\cite{laufer2017semi}}
    \label{fig:multi_manifold}
\end{figure}

An extension to the multiple microphone case is presented in Ref.~\onlinecite{laufer2017semi}. In this case, it is necessary to fuse the viewpoints of all nodes into one mapping from a set of RTFs to a single position estimate $g: \cup_{m=1}^M\mathcal{M}_m \mapsto \mathbb{R}$.

Using a Bayesian perspective of the RKHS optimization,\cite{sindhwani2007semi} in which the mapping $p=g(\mathbf{h})$ is modelled as a Gaussian process, it is easy to extend the single-node problem to a multiple-node problem, by using an average Gaussian process $g=\frac{1}{M}(g^1+g^2+\ldots+g^M) \sim\mathcal{GP}(0,\tilde{k})$. The covariance of this Gaussian process can be calculated from the training data
\begin{multline}
\textrm{cov}(g(\mathbf{h}_r),g(\mathbf{h}_l))\equiv\tilde{k}(\mathbf{h}_r,\mathbf{h}_l)=\\\frac{1}{M^2}\sum_{q,w=1}^M \sum_{i=1}^{n_D}{k_q(\mathbf{h}^q_r,\mathbf{h}^q_i)}{k_w(\mathbf{h}^w_l,\mathbf{h}^w_i)}.
\end{multline}
See Fig.~\ref{fig:multi_manifold} for a schematic depiction of the multi-manifold fusion paradigm.

\begin{figure}[t]
\centering
\includegraphics[width=\reprintcolumnwidth]{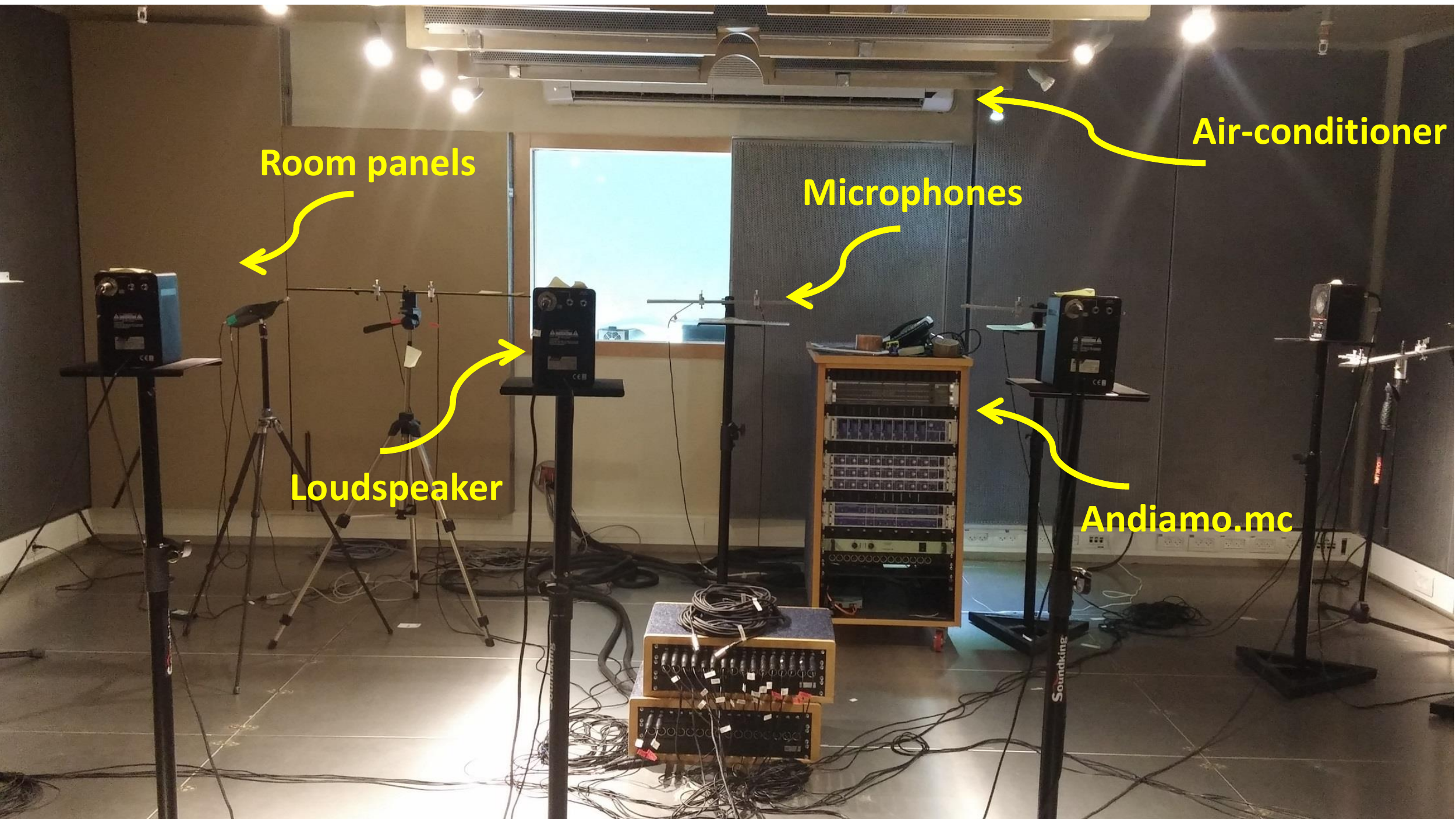}
\centering
\caption{(Color Online) The acoustic lab at Bar-Ilan university with controllable reverberation levels.}
\label{fig:lab}
\end{figure}

\begin{figure}[t]
\centering
\includegraphics[width=\reprintcolumnwidth]{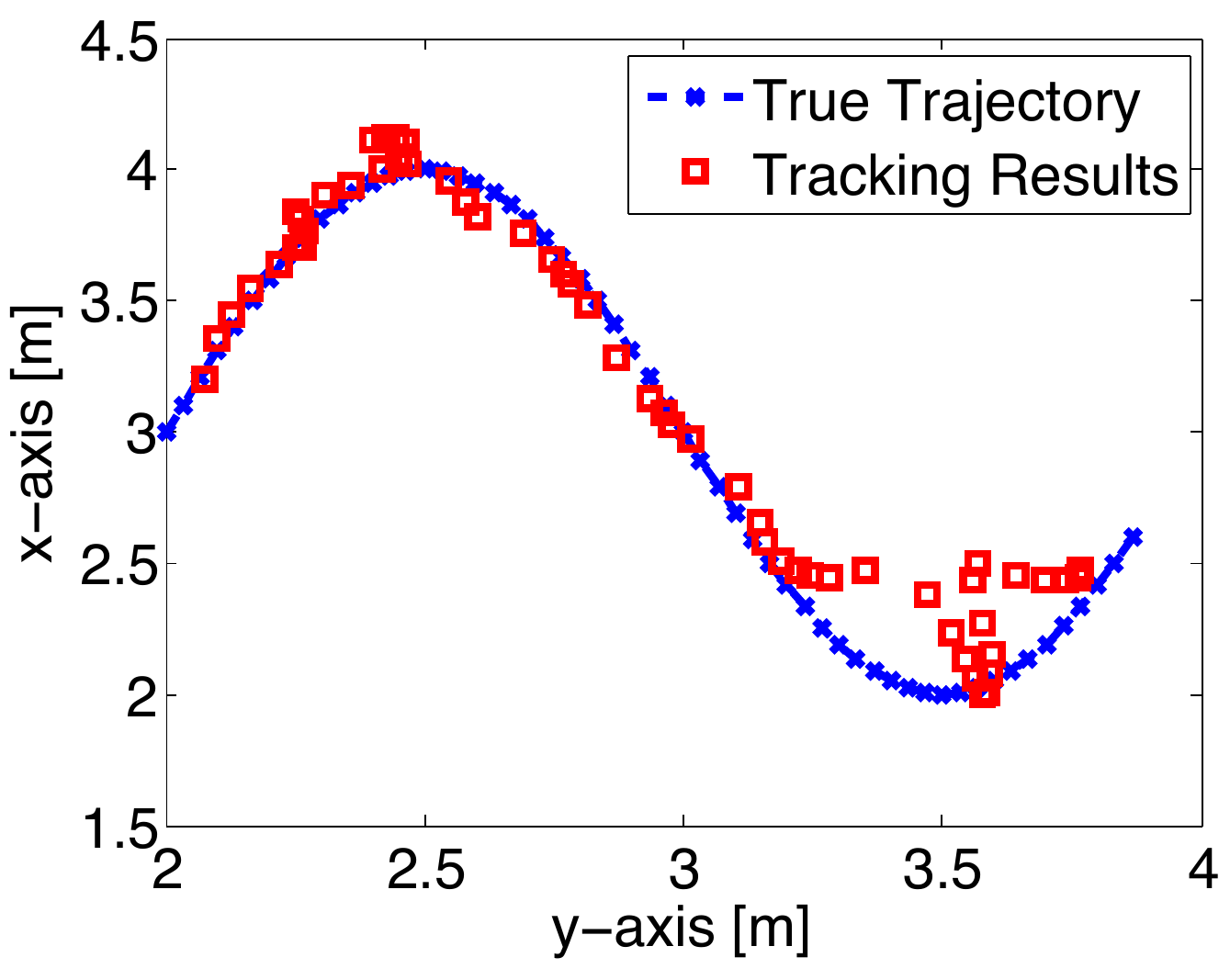}
\caption{(Color Online) Manifold-based tracking algorithm.\cite{Laufer-Goldshtein2017track}}
\label{fig:manifold_tracking}
\end{figure}

The multi-manifold localization scheme\cite{laufer2017semi} was evaluated using real signals recorded at the Bar-Ilan acoustic lab, see Fig.~\ref{fig:lab}. This $6\times6\times2.4$~m room is covered by two-sided panels allowing to control the reverberation level. In the reported experiment reverberation time was set to $T_{60}=620$~ms.
The source position was confined to a $2.8\times2.1$~m area. 3 microphone pairs with inter-distance of 0.2~m were used.
For the algorithm training 20 labelled samples  with 0.7~m resolution and 50 unlabeled samples were used. The algorithm was tested with two noise types: air-conditioner noise and babble noise. As an example, for SNR=15~dB the SRP-PHAT\cite{dibias} achieves an RMSE of 58~cm (averaged over 25 samples in the designated area) while the multi-manifold algorithm achieves 47~cm.
Finally, in dynamic scenarios, recursive versions using the Kalman filter and its extensions, with the covariance matrices of the propagation and measurement processes inferred from the manifold structure.\cite{Laufer-Goldshtein2017track,laufer2018hybrid}
Simulation results for $T_{60}=300$~ms and a sinusoidal movement, 5~s long and approximate velocity of 1~m/s is depicted in Fig.~\ref{fig:manifold_tracking}. Very good tracking capabilities are demonstrated.
In the simulations, the room size was set to $5.2\times 6.2 \times3$~m, the number of microphone pairs was $M = 4$ with 0.2~m inter-distance between microphone pairs.
The training comprised 36 samples with 0.4~m resolution.

\begin{figure}[tb]
\includegraphics[width=\reprintcolumnwidth]{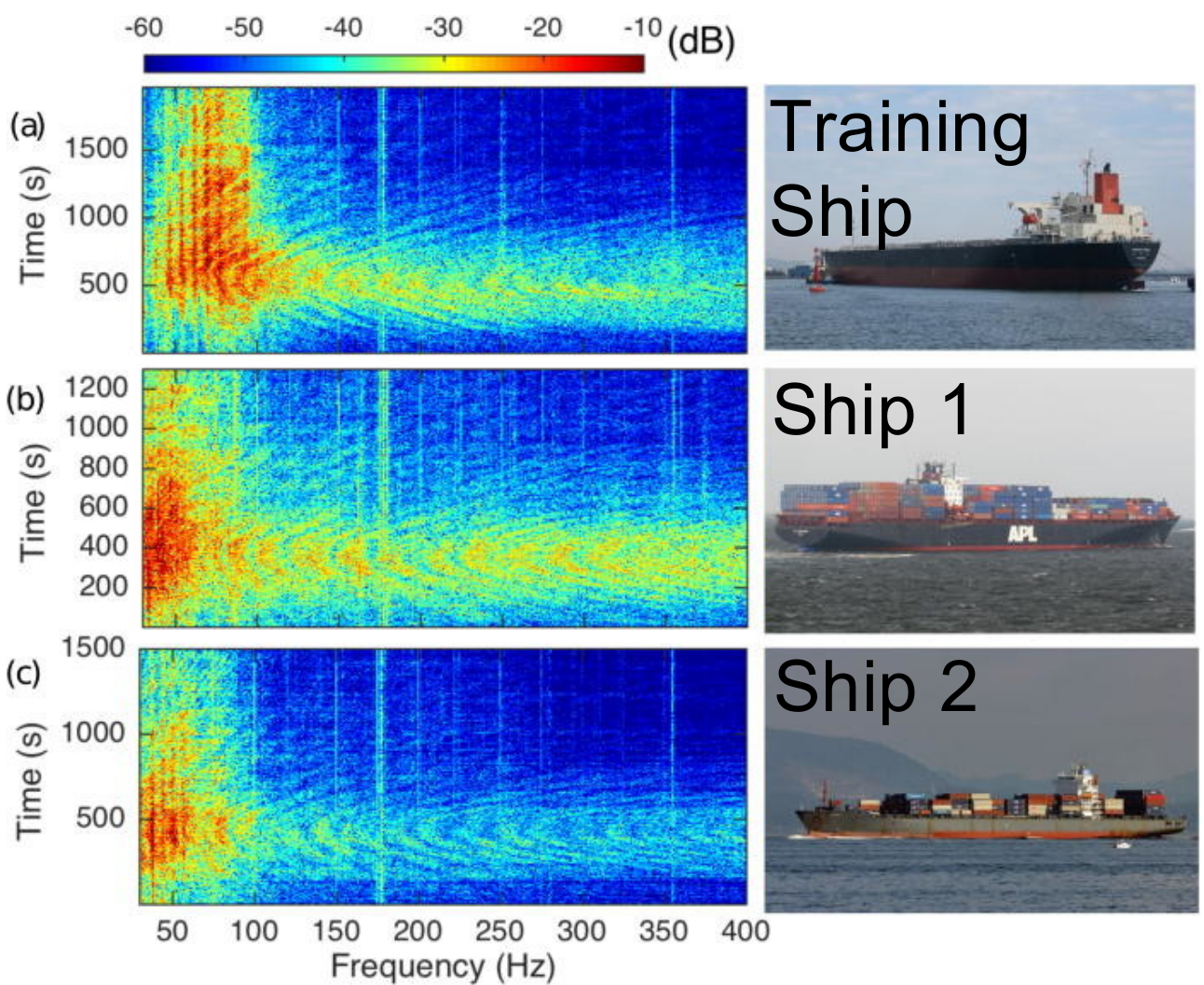}
\caption{\label{fig:shipnoise} (Color Online) Spectrograms of shipping noise in the Santa Barbara Channel during 2016, (a) September 15, 13:00-13:33, (b) September 16, 19:11-19:33, and (c) September 17, 19:29-19:54.\cite{niu2017ship}}
\end{figure}

\begin{figure}[tb]
\includegraphics[width=\reprintcolumnwidth]{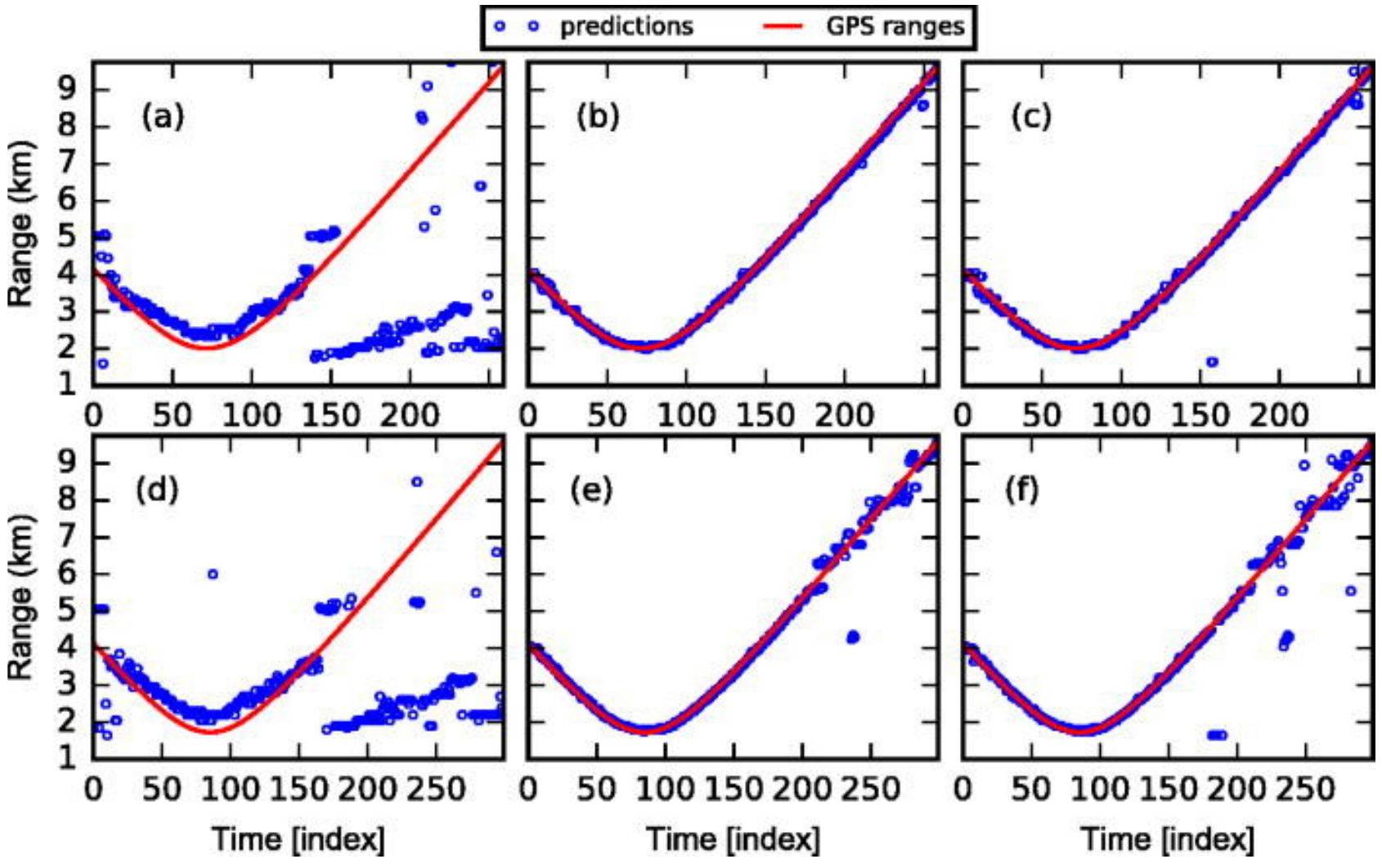}
\caption{\label{fig:ML_ShipLocalization} (Color Online) Ship range localization in the Santa Barbara Channel, 53-200 Hz, using (a,d) MFP, (b,e) Support Vector Classifiers, and (c,f) feed-forward neural network classifier, tested on (a--c) Track 1 and (d--f) Track 2. The time index is 5 s.\cite{niu2017ship}}
\end{figure}

\section{Source localization in ocean acoustics} \label{sec:underwater_acoustics}

Underwater source localization methodologies in ocean acoustics have conventionally relied on physics-based propagation models of the known environment. Unlike conventional methods, ML methods may be considered ``model-free" as they do not rely on physics-based forward-modeling to predict source location. ML instead infers patterns from acoustic data which allows for a purely data-driven approach to source localization. However, in lieu of sufficient data, model simulations can also be incorporated with experimental data for training, in which case ML may not be fully model-free.  The application of ML to underwater source localization\cite{lefort2017direct, niu2017} is a relatively new research area with the potential to leverage recent advancements in computing for accurate, real-time prediction.

Matched-field processing\cite{bag93} (MFP) has been applied to ocean source localization for decades with reasonable success.\cite{Richardson1991APosteriori, Porter1994MFPbenchmark} Recent MFP modifications incorporate compressive sensing since there are only a few source locations.\cite{forero2014shallow,Gemba2017SBL,Gemba2017MFP,gemba2019robust} 
However, MFP is prone to model mismatch.\cite{Tolstoy1989Mismatch, Hamson1989MismatchShallow} Model mismatch has been alleviated by data-replica MFP where closely matched data is available.\cite{Kuperman1993Replicas, Hursky2001MatchedModes}

The earliest ML-based approach to underwater source localization was implemented by a NN trained on modeled data to learn a forward model consisting of weighted transformations.\cite{Ozard1991ann, steinberg1991localization} Early ML methods were also applied to seabed inversion with limited success\cite{Benson2000inversion, Caiti1996RBF} and to seafloor classification using both supervised and unsupervised learning\cite{michalopoulou1995}. The models were linear in the weight space due to lack of widespread knowledge about efficient nonlinear inference algorithms. Also at this time, NN performance was hindered by computational limitations.

Due to these early computational limitations, alternative methods replaced NNs in the state-of-the-art. These methods included Bayesian inference with physical forward models\cite{Michalopoulou2006Gibbs, Dosso2007Bayes} and model-free localization methods, including the waveguide and array invariant methods, which are effective in well-studied waveguide environments.\cite{LeeMakris2006AI, Thode2000RangingWI, Song2015AIWI}
The field of ML once-again gained momentum with the growth of computational efficiency, the advent of open-source software\cite{Torch, tensorflow2015, Theano} and, notably, improved learning algorithms for deep, nonlinear inference.

More recent developments in ML for underwater acoustics concerned target classification. \cite{fischell2015classification, Cao2003SVMTarget} Studies of ocean source localization using ML appeared soon thereafter,\cite{lefort2017direct, niu2017} and include applications to experimental data for broadband ship localization,\cite{niu2017ship}
target characterization,\cite{fischell2017supervised} and post-processing of time difference of arrival estimates.\cite{rauchenstein2018tdoa} Recently, studies have examined underwater source localization with CNNs \cite{ferguson2018sound} and DL,\cite{WangPeng2018, Huang2018,chi_sound_2019} taking advantage of 2D data structure, shared weighting, and huge model-generated datasets. Other recent applications of ML in ocean acoustics include geoacoustic inversion.\cite{piccolo2019}

\begin{figure}[tb]
\includegraphics[width=\reprintcolumnwidth]{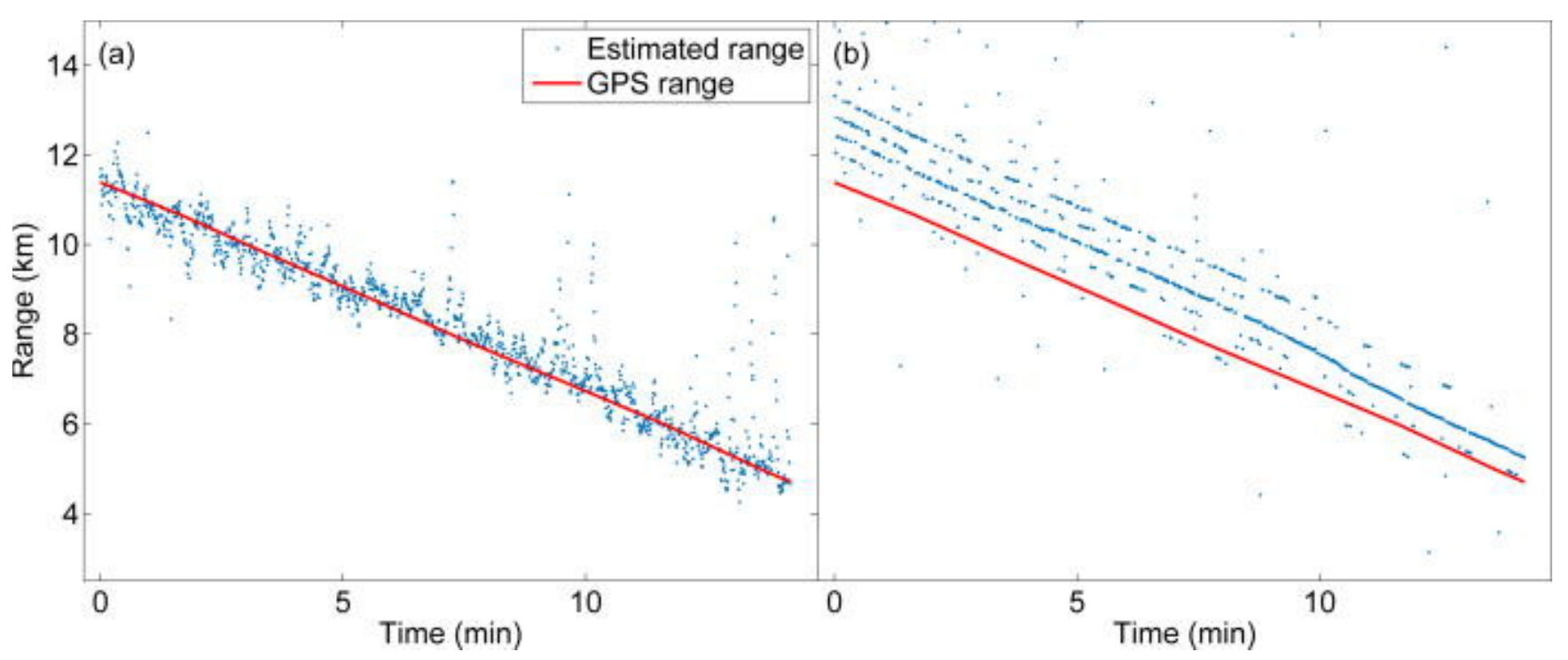}
\caption{\label{fig:DeepML_ShipLocalization}(Color Online)  Ship range localization in the Yellow Sea, 100-150 Hz, using (a) time-delay neural network and (b) MFP with depth mismatch (model: 36 m, ocean: 35.5 m). The neural network was trained on a large, simulated dataset with various environments.\cite{Huang2018}}
\end{figure}

In Niu et al. 2017,\cite{niu2017} the sample covariance matrix was used in a feed-forward neural network (NN) classifier to predict source range. The NN performed well on simulated data and localized cargo ships from Noise09 and Santa Barbara Channel experiments\cite{niu2017ship} (Fig.~\ref{fig:shipnoise}). While NNs achieved high accuracy, MFP was challenged by solution ambiguity  (Fig.~\ref{fig:ML_ShipLocalization}). Huang et al.\cite{Huang2018} used the eigenvalues of the sample covariance matrix in a deep time-delay neural network (TDNN) regression, which they trained on simulated data from many environments. For a shallow, sloping ocean environment, the TDNN was trained at multiple ocean depths to avoid model mismatch. It tracked the ships location accurately, whereas MFP always overestimated the ship range (Fig.~\ref{fig:DeepML_ShipLocalization}). Recently, Niu et al. 2019\cite{niu2019deep} input the acoustic amplitude on a single hydrophone into a deep residual CNN (Res-Net)\cite{He2016ResNet} to predict source range and depth (Fig.~\ref{fig:niu2019deep}). The deep model was trained with tens of millions of samples from numerous environmental configurations. The deep Res-Net had lower range prediction error and competitive depth error compared to the Seismo-Acoustic Genetic Algorithm (SAGA) inversion method.

Future source localization research will benefit from combining the developments in propagation modeling, parallel and cloud computation tools, big data storage for long-term or large-scale acoustic recordings. Powerful new ML methods uilizing these techniques will achieve real-time, accurate ocean source localization.

\section{Bioacoustics} \label{sec:bioacoustics}
Bioacoustics is the study of sound production and perception, including the role of sound in communication and the effects of natural and anthropogenic sounds on living organisms.  ML has the potential to address many questions in this field.  In some cases, ML is directly applied to answer specific questions:  When are animals present and vocalizing? \cite{mellinger1997methods,mellinger2004comparison}  Which animal is vocalizing?\cite{clemins2005automatic,bermant2019deep}  What species produced a vocalization?\cite{steiner1981species} What call or song was produced and how do these sounds relate to one another?\cite{kershenbaum2016acoustic,ten2013analyzing} Among these questions, species detection and identification is a primary driver of many bioacoustics studies due to the reasonably direct implications for conservation and mitigation.

Information mined from these direct acoustic measurements and  can be used to answer specific biological, ecological, and management questions. Examples of this include:  What is the density of animals in an area\cite{marques2009estimating}, and how is the density changing over time?\cite{hildebrand2019assessing} How do lunar patterns affect foraging behavior?\cite{simonis2017lunar}  Many of the issues presented throughout this section are also relevant to soundscape ecology, which is the study of all sounds within an environment.\cite{pijanowski2011soundscape}

\begin{figure}[t]
\includegraphics[width=\reprintcolumnwidth]{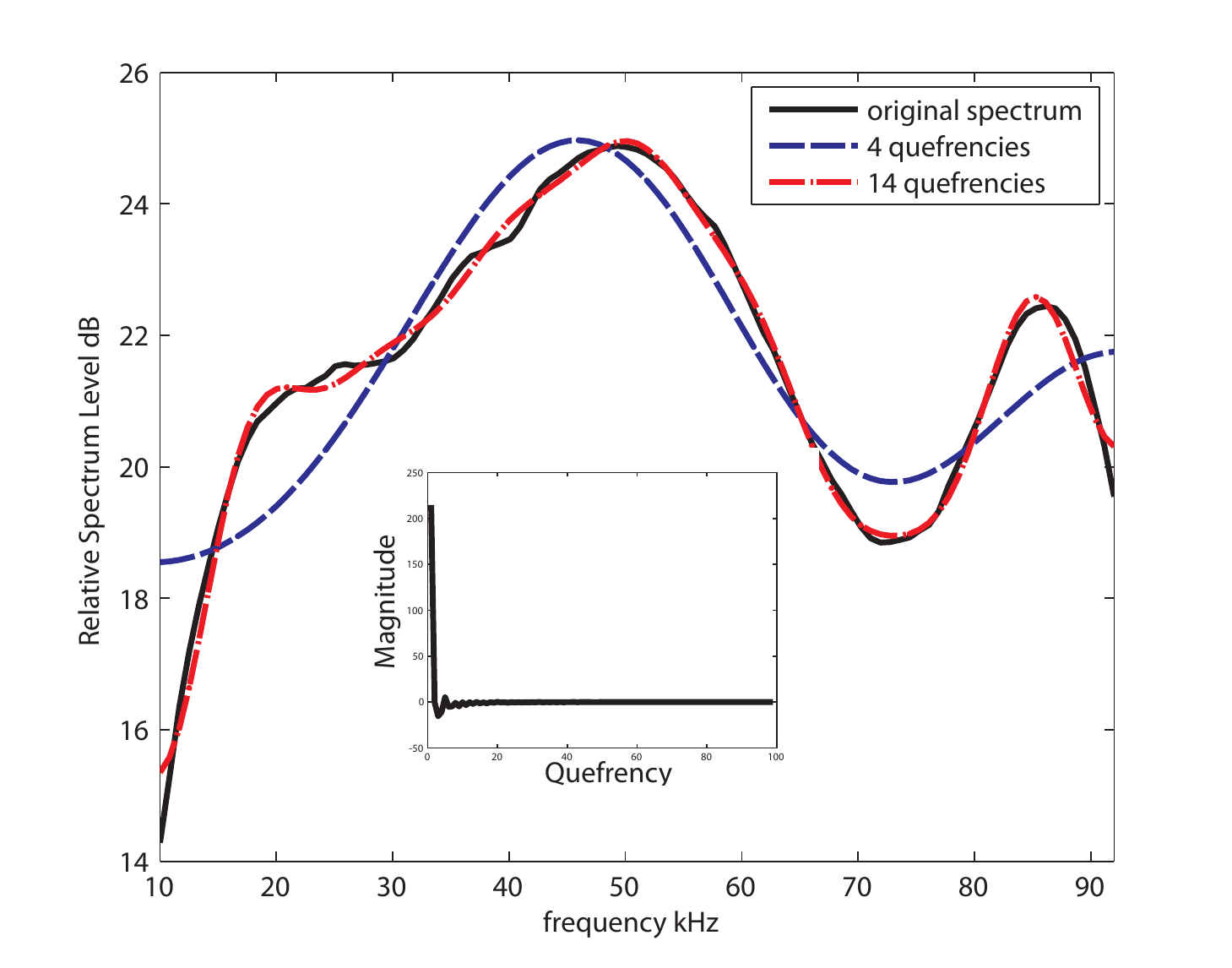}
\caption{(Color Online) Spectrum of a common dolphin echolocation click with inlay of the cepstrum of the spectrum.  Dashed lines show reconstruction of spectrum from truncated cepstral series, showing that gross characteristics of the spectrum can be captured with a low number of coefficients.  Adding coefficients increases amount of detail captured.  From, \cite{roch2011classification} used with permission.}  \label{fig:roch1}
\end{figure}

Although many recent works are starting to use learned features, such as those produced by autoencoder NNs or other dimensionality reduction techniques mentioned in the introduction, much of the bioacoustics literature uses hand-selected features.  These are either applied across the spectrum, such as cepstral representation of spectra \cite{oppenheim2004frequency} which capture the shape of the spectral envelope of a short segment of the signal \cite{kogan1998automated,roch2011classification} in a low number of dimensions (Fig.~\ref{fig:roch1}) or engineered towards specific calls.  Many of the features designed for specific calls tend to concentrate on statistics of acoustic parameters such as mean or center frequency, bandwidth, time-bandwidth products, number of inflections in tonal calls, etc.  
It is fairly common to use psychoacoustic scales such as the Melodic (Mel) scale
which recognizes that humans (and most other animals) have an acoustic fovea, a frequency range where they can most accurately perceive frequency differences. However, it is important to remember that this varies between species and the standard Mel scale is weighted towards humans whose hearing characteristics may vary from the target species.

Learned features attempt to determine the feature set from the data and include any type of manifold learner such as principal component analysis or autoencoders.  In most cases, the feature learners are given standard features such as cepstral coefficients or statistics of a call (in which case they are simply learning a manifold of the features) or attempt to learn from relatively unprocessed data such as time-frequency representations.  Stowell and Plumbley (2014)\cite{stowell2014automatic} provide an example of using a spherical K-means learner to construct features from Mel-filtered spectra.  Spherical K-means normalizes the input vectors and uses a cosine distance as its distortion metric.  Other feature learners that have been used in bioacoustics include sparse autoencoders\cite{halkias2013classification}, and CNNs that learn weights associated with features of interest\cite{smirnov2013north}.

There are many examples of template-based methods that can work well when calls are highly stereotyped.  The simplest type of template method is the time-domain matched filter, but in bioacoustics matched filters are typically implemented in time-frequency space\cite{mellinger1997methods}.  More complex matched filters permit non-linear compression or elongation of the filter with dynamic time warping\cite{sakoe1978dynamic}, which has been used for both delphinid whistles\cite{buck1993quantitative} and bird calls\cite{kogan1998automated}.  However, even these so-called “stereotyped” calls have, in many species, been shown to drift over time. It has been shown that the tonal frequency of blue whale calls have decreased \cite{mcdonald2009worldwide}.  These types of changes can cause matched-template methods to require recalibration.

Supervised learning is the primary learning paradigm that has been used in ML bioacoustics research and can be traced back to the use of linear discriminant analysis.  An early example of this was the work of Steiner (1981)\cite{steiner1981species} that examined classifying delphinid whistles by species.  GMMs have been used to capture statistical variation of spectral parameters of the calls of toothed whales\cite{roch2007gaussian} and sequence information has been exploited with hidden Markov models for classifying bird song by species.\cite{kogan1998automated,somervuo2006parametric}  Multi-level perceptron NNs also have a rich history of being applied in bioacoustics, with varied uses such as bat species identification, bowhead whale ({\em Balaena mysticetus}) call detection, and recognizing killer whale ({\em Orcinus orca}) dialects.\cite{deecke2006automated,parsons2000acoustic,potter1994marine} Decision tree methods have been used, with early approaches using classification and regression trees for species identification.\cite{oswald2003acoustic}  SVM-based methods have also had considerable success, examples include classifying the calls of birds and anurans to species.\cite{acevedo2009automated,fagerlund2007bird}

Ensemble learning is a well-known method of combining classifiers to improve results by reducing the variance of the classification decision through the use of multiple classifiers that learn from different data, with well-known examples such as random forest\cite{breiman2001random} and adaptive boosting.\cite{schapire1998boosting}  These techniques have been leveraged by the bioacoustics community, such as the work by Gradišek et al. (2016)\cite{gradivsek2017predicting} that used random forests to distinguish bumble bee species based on characteristics of their buzz.

One of the most recent trends in bioacoustic pattern recognizers is the use of DNNs that have reduced classification error rates in many fields\cite{lecun2015} and have reduced many of the issues of overfitting NNs seen in earlier artificial NNs through a variety of methods such as increased training data, architectural changes, and improved regularization techniques. An early use of this in bioacoustics can be seen in the work of Halkias et al.\cite{halkias2013classification} that demonstrated the ability of deep Boltzmann machines to distinguish mysticete species.  Deep CNNs and RNNs have been used for bat species identification\cite{mac2018bat}, whale species identification\cite{thomas2019marine}, detecting and characterizing sperm whale echolocation clicks \cite{bermant2019deep}, and have become one of the dominant types of recognizers for bird species identification since the successful introduction of CNNs in the LifeCLEF bird identification task.\cite{goeau2016lifeclef}

Unsupervised ML has not been used as extensively in bioacoustics, but has several noteworthy applications and large potential.  Much of the work has been to cluster calls into distinct types, with the goal of using objective methods that are repeatable and do not suffer from perceptual bias.  Examples of this include the K-means clustering,\cite{lin2015passive,mccowan1995new} adaptive resonance theory clustering (Deecke and Janik, 2006),\cite{deecke2006automated}  self-organizing maps,\cite{green2011recurring} and clustering graph nodes based on modularity.\cite{frasier2016automated} Clustering sounds to species is also of interest and has been used to investigate toothed whale echolocation clicks in data deficient areas where not all species’ sounds have been well described\cite{frasier2017automated} using Biemann’s (2006) graph clustering algorithm\cite{biemann2006chinese} that shares similarities with bottom up clustering approaches.

There are several repositories for bioacoustic data.  The Macaulay Library at the Cornell Lab of Ornithology (https://www.macaulaylibrary.org) maintains an extensive database of acoustic media with a combination of curated and citizen-scientist recordings.  Portions of the Xeno-Canto collection (https://www.xeno-canto.org) of bird sounds has been used extensively as a competition data set in the CLEF series of conferences.  The marine mammal bioacoustics community maintains the Moby Sound database of marine mammal sounds (https://www.mobysound.org/), which includes many of the data sets used in the Detection, Classification, Localization, and Density Estimation for marine mammals series of workshops.  In addition, there are government databases such as the British Library's sounds library (https://sounds.bl.uk/) which includes animal calls and soundscape recordings.   Many organizations are trying to come to terms with the large amounts of data generated by passive acoustic recordings and some governments are conducting trials of long-term repositories for passive acoustic data such as the United States’ National Center for Environmental Information’s data archiving pilot program (https://www.ngdc.noaa.gov/mgg/pad/).

\begin{figure*}[tb]
\includegraphics[width=0.6\linewidth]{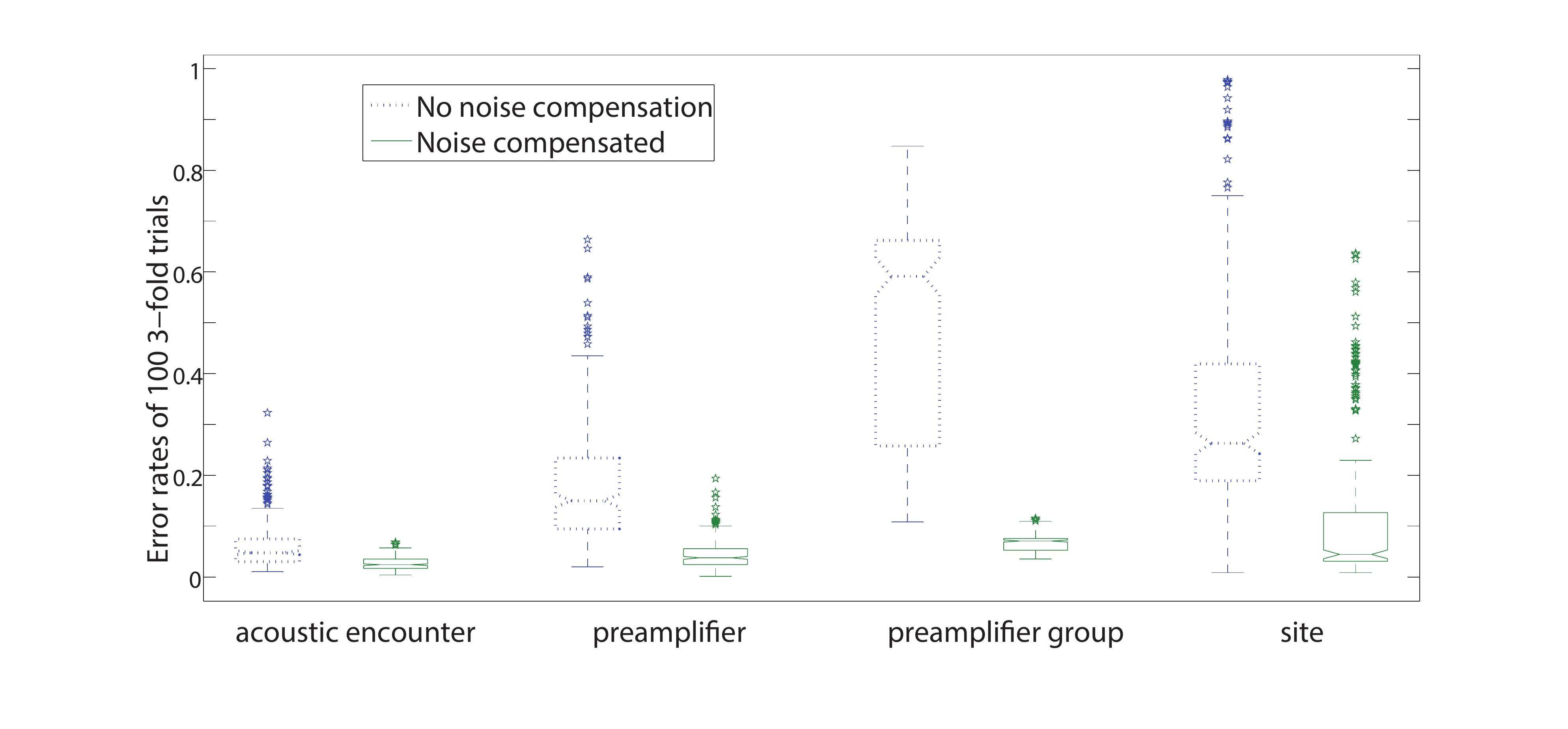}
\caption{\label{fig:cv_split}{(Color online) Effect of cepstral feature compensation on echolocation click classification error rate (for Pacific white-sided and Risso's dolphins). The compensation is performed using local noise estimates and the classification errors are due to environmental and equipment mismatch. The box plots show the error rates estimated by 100 random 3 fold trials.  Train/test boundaries are stratified by varying criteria that illustrate the increase in error rate over mismatch types. First and second sets of plots (blue and green) illustrate the effectiveness of the compensation technique.
 }}
\end{figure*}
In Fig.~\ref{fig:cv_split} we illustrate the effect of recording equipment and sampling site location mismatch on cross validation results in marine mammal call classification. For some problems, changes across equipment or environments can cause severe degradation of performance.  When these types of issues are not considered, performance in the field can vary significantly from what was expected based on laboratory experiments. Each case (acoustic encounter, preamplifier, preamplifier group, and site) specifies a grouping criterion for training/test folds. The acoustic encounter case  are sets of calls from a group of animals when they are within detection range of the data logger. Calls from each encounter are entirely in  training  or  test data.  The preamplifier case adds further restrictions that encounters recorded on the same preamplifier are never split across the training and testing.  The preamplifier group is case stricter yet; clicks from preamplifiers with similar characteristics cannot be split.  The final group indicates that acoustic encounters from the same recording site cannot be split. 

Finally, an ongoing challenge for the use of ML in bioacoustics is managing detection data generated from long-term data sets. Some recent efforts are beginning to organize and store data products resulting from passive acoustic monitoring.\cite{fujioka2014integration,roch2016management} While peripheral to the performance of ML algorithms, the ability to store the scores and decisions of ML algorithms along with descriptions of the algorithms and the parameters used is critical to comparing results and analyzing long-term trends.

\section{Reverberation and Environmental Sounds in Everyday Scenes}\label{sec:reverb}

Humans encounter complex acoustic scenes in their daily life. Sounds are created by a wide range of sources (e.g. speech, music, impacts, scrapes, fluids, animals, machinery), each with its own structure and each highly variable in its own right.\cite{gaver1993world} Moreover, the sound from these sources reverberate in the environment, which profoundly distorts the original source waveform. Thus the signal that reaches a listener usually contains a mixture of highly variable unknown sources, each distorted by the environment in an unknown fashion.

This variability of sounds in everyday scenes poses a great challenge for acoustic classification and inference. Classification algorithms must be sensitive to inter-class variation, robust to intra-class variation, and robust to reverberation---all of which are context dependent. Robust identification of sounds in natural scenes often requires both large training data sets, to capture the requisite acoustic variability, and domain specific knowledge about which acoustic features are diagnostic for specific tasks.

Overcoming these challenges will enable a range of novel technologies. These technologies include, for example, hearing aids which can extract speech from background noise and reverberation, or self-driving cars which can locate a fire-truck siren amidst a noisy street. Some applications which have already been investigated include: inspection of tile properties from impact sounds;\cite{feng2005application} classification of aircraft from takeoff sounds;\cite{marquez2014aircraft} and cough sound recognition in pig farms.\cite{exadaktylos2008real} More examples are given in Table.~2 of Sharan and Moir.\cite{sharan2016overview}  These are all tasks which must deal with the complexities of natural acoustic scenes. Because environmental sounds are so variable and occur in so many different contexts---the very fact which makes them difficult to model and to parse---any ML system that can overcome these challenges will likely yield a broad set of technological innovations. As such, the analysis and understanding of sound scenes and events is an active field of research.\cite{virtanen2018computational}

There is another reason that algorithms which parse natural acoustic scenes are of special interest.  By definition, such algorithms attempt the same challenge that biological hearing systems have evolved to solve---organisms, as well as engineers, desire to make sense of sound and thereby infer the state of the world\cite{bregman1994auditory,wang2006computational}. This convergence of goals means that engineers can take inspiration from auditory perception research. It also raises the possibility that ML algorithms may help us understand the mechanisms of auditory perception in both humans and animals, which remain the most successful systems in existence for acoustical inference in natural scenes.

In the following, we will consider two key challenges of applying ML algorithms to acoustic inference in natural scenes: (1) robustness to reverberation; and (2) classification of a large range of diverse environmental sounds.

\subsection{Reverberation}
\begin{figure}[t]
\includegraphics[width=\reprintcolumnwidth]{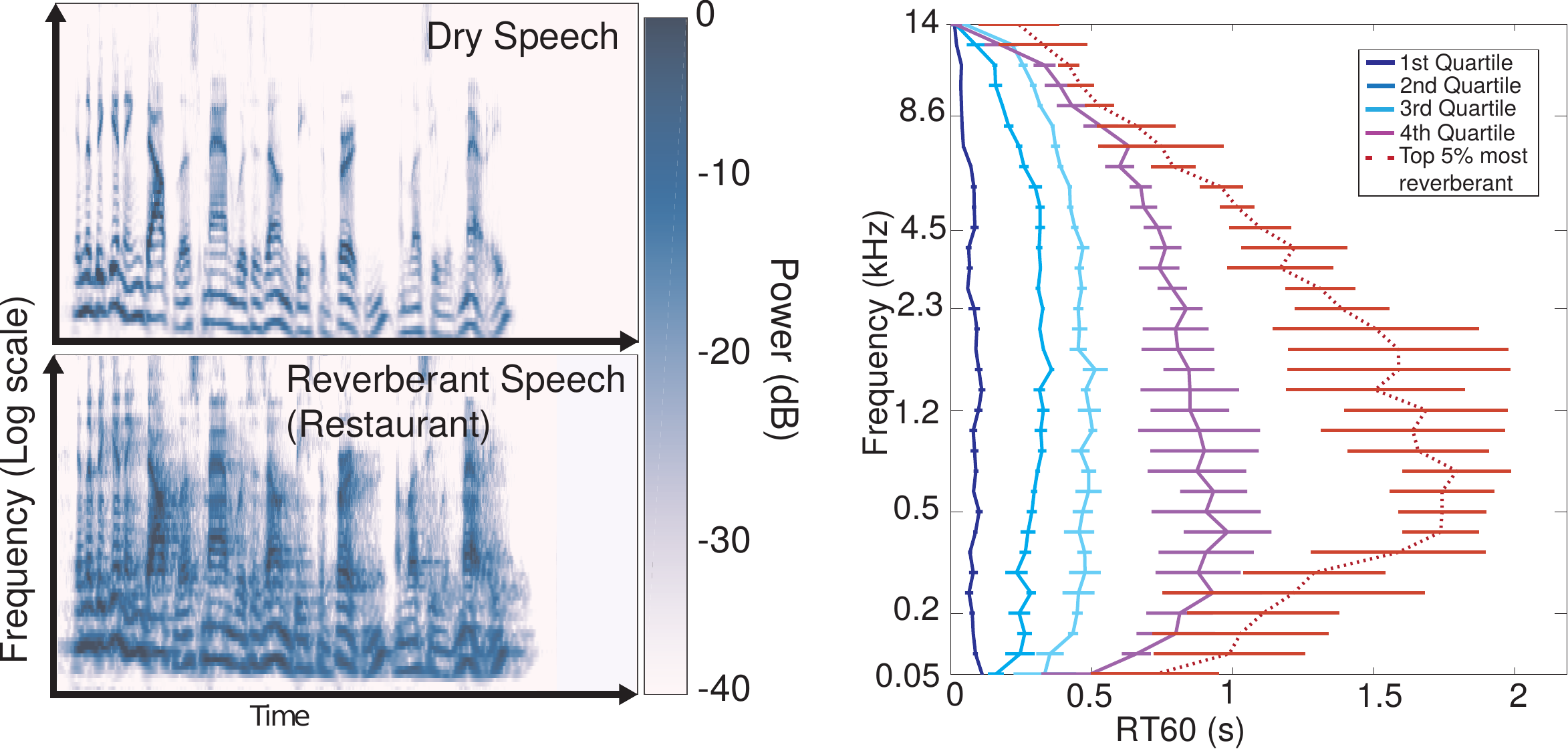}
\caption{\label{fig:JT_Rvrb}{(Color Online) (Left) Cochleagrams of dry and reverberant speech demonstrate the profound distortion that can be induced by natural scenes---in this case a restaurant environment. (Right) Histograms of reverberant decay times (RT60 is the time taken for reverberation to decay 60\,dB) surveyed from natural scenes demonstrate that diverse scenes contain stereotyped IR properties. Humans make use of these regularities to perceive reverberant sounds. (Reproduced from Ref.~\onlinecite{traer2016statistics})}}
\end{figure}

Acoustic reverberation is ubiquitous in natural scenes, and profoundly distorts sounds as they propagate from a source to a listener (Fig.~\ref{fig:JT_Rvrb}). Thus any recorded sound is a product of both the source and environment.  This presents a challenge to source recognition algorithms, as a classifier trained in one environment, may not work when presented with sounds from a different space, or even sounds presented from different locations within the same space. However, reverberation also provides a source of information about the environment,\cite{dokmanic2013acoustic,dokmanic2015listening} and the source-listener distance. Humans can robustly identify sources, source locations, and properties of the environment from reverberant sounds.\cite{zahorik2001loudness} This suggests that the human auditory system can, from a single sound, separately infer multiple causal factors.\cite{traer2016statistics} The process by which this is done is poorly understood, and has yet to be replicated via algorithms.

The effect of reverberation can be described by filtering with the environment Impulse Response (IR),
\begin{equation}
    r_{j}(t)=s(t)*h_{j}(t)
\end{equation}
where $r(t)$ is the reverberant sound, $s(t)$ the source signal, and $h(t)$ the impulse response; the subscript $j$ indexes across microphones in a multi-sensor array.  An algorithm that seeks to identify the source (or IR), must either be robust to variations introduced by natural IRs (or sources), or it must be able to separate the signal into its constituents (i.e. $s(t)$ and $h(t)$). The challenge is that, in general, both $s(t)$ and $h(t)$ are unknown and such a separation is an ill-posed problem.

Presumably, to make sense of reverberant sounds, an algorithm must leverage knowledge about the acoustical structure of sources, IRs, or both.  Natural scenes, despite highly diverse environments, display statistical regularities in their IRs, such as consistent frequency-dependent variation in decay rates (Fig.~\ref{fig:JT_Rvrb}). This regularity partially enables human comprehension of reverberant sounds.\cite{traer2016statistics} If such regularities exist, ML algorithms can in principle learn them, if they receive appropriate training.

One way to address the variability introduced by reverberation is to incorporate reverberant sounds in the training data set. This has been used to improve performance of a deep neural networks (DNNs) trained on speech recognition.\cite{giri2015improving} Though effective in principle, this may require exceptionally large data sets to generalize to a wide range of environments.

A number of datasets with labelled sound sources in a range of reverberant environments have been prepared (REVERB challenge;\cite{kinoshita2017reverb} ASpIRE challenge\cite{harper2015automatic}). Some data sets have focused instead on estimation of room acoustic parameters (ACE challenge\cite{eaton2015ace}). The proceedings of these challenges provide a thorough overview of state-of-the art systems.

Given that the physical process underlying reverberation is well understood and can be simulated, the statistics of reverberation can also be assessed by simulating a large number of rooms. This has been used to train DNNs to reconstruct a spectrogram of dry (i.e. anechoic) speech from a spectrogram of reverberant speech.\cite{han2015learning}

Another approach to addressing reverberation is to derive algorithms which model the effects of reverberation on sound signals. For example, reverberant IRs smooth signals in the time domain, decreasing the signal kurtosis (which is largest for signals containing sparse high-amplitude peaks). Assuming the source signal is sparse, a dereverberation filter can be learned which maximizes the kurtosis of the output signal\cite{wiggins1978minimum} and returns an estimate of the source.\cite{gillespie2001speech,lee2008binaural}  More recent speech dereverberation methods, also employing machine learning methodologies, can be found in \onlinecite{kinoshita2017reverb,ernst2018speech}.

Another feature used is the spatial covariance of a microphone array. The direct-arriving (i.e.~non-reverberant) sound is strongly correlated across two spatially separated microphones, as the signal detected at each channel is the same signal with different time delays. The reverberation, which consists of a summation of many signals incident from different directions\cite{schroeder1962natural} is much less correlated across channels. This can be exploited to yield a dereverberation algorithm,\cite{nakatani2010speech,higuchi2014unified,Schwartz16,jukic2016adaptive,kinoshita2017reverb,braun2018linear,li2019multichannel,schwartz2014online}, and to estimate signal direction-of-arrival.\cite{xiao2015learning} There are also spectral-subtraction based methods to dereverberation, which for example estimate and subtract the late reverberant speech component.\cite{habets2009late,habets2010speech} For a comprehensive review of speech dereverberation methods, please see Ref.~\onlinecite{naylor2010speech}.

In addition to estimating the source signal, it is often desirable to infer properties of the IR from the reverberant signal, and thereby infer characteristics of the environment.\cite{papayiannis2017discriminative}  The most common such property to be inferred is the Reverberation Time (RT), which is the time taken for reverberant energy to decay some amount.  RT can be estimated from histograms of decay rates measured from short windows of the signal.\cite{ratnam2003blind}

The techniques described above have all shown some success in estimating sources or environments from reverberant audio. However, in most cases either the sound sources or the IRs were drawn from a constrained set (i.e only speech, or a small number of rooms). It remains to be seen how well these approaches will generalize to the relative cacophony of everyday scenes.

\subsection{Environmental sounds}
There are many challenges to identifying sources in natural scenes. Firstly, there is the tremendous range of different sound sources. Natural scenes are filled with speech, music, animal calls, traffic sounds, machinery, fluid sounds, electronic devices, and a range of clattering, clanking, scraping and squeaking of everyday objects colliding. Secondly, there is tremendous variability within each class of sound.  The sound of a plate dropped on a floor varies dramatically with the plate, the floor, the height of the drop, and the angle of impact. Thirdly, natural scenes often contain many simultaneous sound sources which overlap and interfere. To recognize acoustic scenes, or the sources therein, an algorithm must simultaneously be sensitive to the differences between different sources and robust to the variation within each source.

The most obvious solution to overcoming the complexity of natural scenes is to train classifiers on large and varied sets of labelled recordings. To this end, a number of public datasets and have been introduced for both source recognition in natural scenes (DCASE challenges;\cite{mesaros2017dcase} ESC;\cite{piczak2015esc} TUT;\cite{mesaros2016tut} Audio set;\cite{gemmeke2017audio} UrbanSound\cite{salamon2014dataset} and scene classification (DCASE; TUT).
Thorough overviews are given for state-of-the-art algorithms in proceedings of these challenges, in Virtanen et al.\cite{virtanen2018computational}, in Sharan and Moir\cite{sharan2016overview} for sound recognition, and in Barchiesi et al.\cite{barchiesi2015acoustic} for scene recognition.

Recently, massive troves of online videos have proven a useful source of sounds for training and testing.  One approach is to use meta-data tags in such videos as ``weak labels".\cite{kumar2016audio}  Even though the labels are noisy and are not time-synced to the actual noise event---which may be sparse throughout the video---this can be mitigated by the sheer size of the training corpus, as millions of such videos can be obtained and used for training and testing.\cite{hershey2017cnn}

Another approach to audiovisual training is to use state-of-the-art image processing algorithms to provide object and scene labels to each frame of the video. These can then be used as labels for sections of audio allowing conventional training of a classifier to recognize sound events from the audio waveform.\cite{aytar2016soundnet} Similarly, a network can be trained to map image statistics to audio statistics and thereby generate a plausible sound for a given image, or image sub-patch.\cite{owens2016ambient}

The synchronicity between object motion (rendered in pixels) and audio events can be leveraged to extract individual audio sources from video. Classifiers which receive inputs from both audio and video channels can be trained to differentiate videos with veridical audio, from videos with the wrong audio or temporally misaligned audio.  Such algorithms learn ``audiovisual features" and can then infer audio structure from pixel patterns alone.  This enables audio source separation with video of multiple musicians or speakers,\cite{zhao2018sound} or identification of where in an image a source is emanating.\cite{owens2018audio,arandjelovic2017look} 

Whether trained by video features or by traditional labels, a sound source classifier must learn a set of acoustic features diagnostic of relevant sources. In principle, the features can be learned directly on the audio waveform. Some algorithms do this,\cite{oord2016wavenet} but in practice, most state-of-the-art algorithms use pre-processing to map a sound to a lower-dimensional representation from which features are learned. Classifiers are frequently trained upon Short-Time-Fourier Transform (STFT) domains, and many variations thereupon with  non-linear frequency decompositions spacings (mel-spaced, Gammatone, ERB, etc).  These decompositions (sometimes termed cochleagrams if the frequency spacing is designed to mimic the sensitivity of the cochlea within the ear) all favor finer spectral resolution at lower frequencies than higher frequencies, which both mirrors the sensitivity of biological audition and may be optimal for recognition of natural sounds.\cite{theunissen2014neural} Beyond the Spectro-temporal domain, algorithms have been presented which learn features upon a wide range of transformations of acoustical data (summarized by Sharan and Moir,\cite{sharan2016overview} Li et al.,\cite{li2017comparison} and Waldekar and Saha\cite{waldekar2018classification}).

Sparse decomposition provides a framework to optimally decompose a waveform into a set of features from which the original sound can be approximately reconstructed.  This has been put to use to optimize source recognition algorithms\cite{sattigeri2014scalable} and, particularly in the form of non-negative matrix factorization (NMF), provides a learned set of features for sound recognition,\cite{cho2005nonnegative} scene recognition,\cite{bisot2016acoustic} source separation,\cite{virtanen2007monaural} or denoising.\cite{wilson2008speech}

\begin{figure*}[t]
\includegraphics[width=1.7\reprintcolumnwidth]{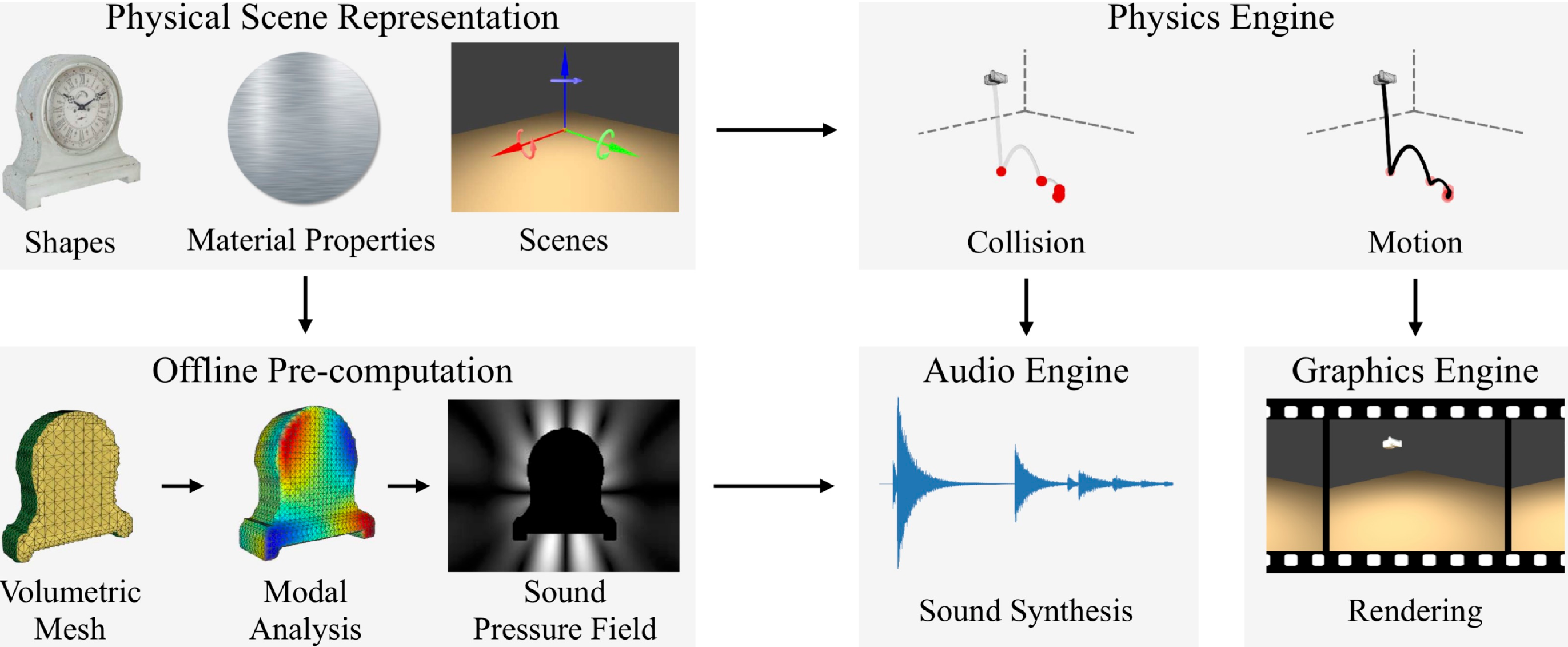}
\caption{\label{fig:JT_GnMdl}{(Color Online) Arbitrarily large datasets of contact sounds can be synthesized via a physical model. Vibrational IRs are pre-computed for a set of synthetic objects, using a boundary element model (BEM). A physics engine is then used to simulate the motion of rigid bodies after initial impulses.  Both sound and video can be computed, and the simulated audio is automatically labelled by the physical parameters: object mass, material, velocity, force of impact, etc.  (Reproduced from Ref.~\onlinecite{zhang2017generative})}}
\end{figure*}

Another approach to choosing acoustic features for classification, is to consider the generative processes by which environmental sounds are created. In many cases, such as impacts of rigid-body objects, the physical processes by which sound is created are well characterized and can be simulated from physical models.\cite{bonneel2008fast} Although full physical simulations are impractically slow for inference by a generative model, such models allow impact audio to be simulated rather than recorded\cite{zhang2017generative,sterling2018isnn} (Fig.~\ref{fig:JT_GnMdl}). This allows the creation of arbitrarily large data sets over which classification algorithms can be trained. The 20K audio-visual data set \cite{zhang2017generative} contains orders of magnitude more labelled impact sounds (with associated videos) than earlier data sets.

Such physical synthesis models allow the training of classifiers which may move beyond recognizing broad sound classes and be able to judge fine-grained physical features such as material, shape or size of colliding objects. Humans can readily make such distinctions\cite{lemaitre2012auditory,giordano2006material} though how they do so is not known.  In principle, detailed and flexible judgments can be made via a generative model which explicitly encodes the relevant causal factors (i.e. the physical parameters we hope to infer, such as material, shape, size, mass, etc.). Such generative models have been used to infer objects and surfaces from images,\cite{yuille2006vision} vocal tract motion from speech,\cite{roweis2005automatic} simple sounds from simulated scenes,\cite{cusimanoauditory} and the motion of objects from the impact sounds made as they bounced and scraped across surfaces.\cite{langlois2014inverse} However, as high-resolution physical sound synthesis is computationally expensive and slow, it is not yet clear how to apply such approaches to more realistic environmental scenes.

Given that the structure of natural sounds are determined by the physical properties of moving objects, audio classification can be aided by video information. Video provides, in addition to class labels as described above, information about the materials present in a scene, and the manner in which objects are moving.  Owens et al.\cite{owens2016visually} recorded a large set of videos of a drum stick striking objects in everyday scenes.  The sounds produced by collision were projected into a low-dimensional feature space where they served as ``labels" for the video data set. A neural network was then trained to associate video frames with sound features, and could subsequently synthesize plausible sounding impacts for silent video of colliding objects.

\subsection{Towards human-level interpretation of environmental sounds and scenes}
As we have described above, recent developments in ML have enabled significant progress in algorithms that can recognize sounds from everyday scenes. These have already enabled novel technologies and will no doubt continue to do so. However, current state-of-the-art systems still do not match up to human perception in many inference tasks.

Consider, for example, the sound of an object (e.g. a coin, a pencil, a wine glass, etc.) dropped on a hard surface.  From this sound alone, humans can identify the source, make guesses about how far and how fast it moved, estimate the distance and location of both the initial impact and the location of settling, distinguish objects of different material or size, and judge the nature of the scene from reverberation. In contrast, current state-of-the-art systems are considered successful if they can distinguish the sound of a basketball bouncing from a door slammed shut or the bark of a dog. They identify but do not \textit{interpret} the sound the way that humans do. Interpreting natural sounds at this level of detail remains an unsolved engineering problem, and it is not known how humans do this intuitively. It is possible that developments in ML hearing of natural scenes and studies of biological hearing will proceed together, each informing and inspiring the other, to yet make a machine that ``hears the world" like a human to parse and interpret the rich environmental sounds present in everyday scenes.

\section{conclusion}
In this review we have introduced ML theory, including deep learning (DL), and discussed a range of applications of ML theory in acoustics research areas. While our coverage of the advances of ML in the field of acoustics is not exhaustive, it is apparent that ML has enabled many recent advances. We hope this article can serve as inspiration for future ML research in acoustics. It is observed that large, publicly available data sets (e.g. Refs.~\onlinecite{Hada1409:Multichannel,mesaros2017dcase,mesaros2016tut,lollmann2018locata,kinoshita2017reverb,harper2015automatic,eaton2015ace}) have encouraged innovation across acoustics field. ML in acoustics has enormous transformative potential, and its benefits can increase with open data.

Despite their limitations, ML-based methods provide good performance relative to conventional processing in many scenarios. However, ML-based methods are data-driven and require large amounts of representative training data to obtain reasonable performance. This can be seen as an expense of accurately modeling complex phenomena, as ML models often have very high capacity. In contrast, standard processing methods often have lower capacity, but are based on training-free statistical and mathematical models. 

Based on  this review, we foresee a  transformation of acoustic processing from  hand-engineering, basic-intuition-driven modeling to a more data-driven ML paradigm. Though the benefits of ML in acoustics cannot be fully realized without building-upon the indispensible physical intuition and theoretical developments within well-established sub-fields, such as array processing. Thus development of ML theory in acoustics should be done without forgetting the physical principles describing our environments.

\begin{acknowledgments}
This work was supported by the Office of Naval Research, Grant No. N00014-18-1-2118.
\end{acknowledgments}

\bibliography{biblio_rev2}
\end{document}